\documentclass[12pt,a4paper,twoside]{book}

\usepackage[]{epsfig,color}
\usepackage{amsmath,amssymb,amsfonts,amsfonts,amsthm}
\usepackage[nice]{nicefrac}
\usepackage{latexsym}
\usepackage{wrapfig}
\usepackage{fancyhdr}
\usepackage{colordvi,alltt}
\usepackage{multirow}
\usepackage{tabularx,threeparttable}
\usepackage{fancybox}
\usepackage{calc}
\usepackage[avantgarde]{quotchap}    %
\usepackage{paralist}        %
\usepackage{multibib}                %
\usepackage{cite}                    %
\usepackage{float}                   %
\usepackage{makeidx}                 %
\usepackage{hyperref}
\usepackage{breakurl}
\usepackage{pstricks,pst-node,pst-text,pst-3d,pst-plot,pst-eucl}
\usepackage{pst-xkey}

\usepackage{subfigure}
\usepackage{enumitem}
\usepackage{rotating}

\usepackage{ifthen}

\usepackage{xspace}        %
\usepackage{setspace}

\usepackage{datetime}

\psset{unit=1cm,xunit=1cm,yunit=1cm}

\newcites{links}{Links}

\floatstyle{boxed}

\renewcommand{\chaptermark}[1]{\markboth{{\it \chaptername\ \thechapter.\ #1}}{}}

\DeclareGraphicsExtensions{.png}
\graphicspath{{.}{fig/}{fig/randomization/}{fig/world/}{fig/points/}{fig/segments/}{fig/geo_traits/}{fig/spherical_voronoi/}{fig/3bisectors/}}

\newtheorem{theorem}{Theorem}[chapter]
\newtheorem{lemma}[theorem]{Lemma}

\newtheorem{observation}[theorem]{Observation}
\newtheorem{corollary}[theorem]{Corollary}
\newtheorem{definition}[theorem]{Definition}
\newtheorem{conjecture}[theorem]{Conjecture}

\newenvironment{remark}[1][Remark]{\begin{trivlist}
\item[\hskip \labelsep {\bfseries #1.}]}{\end{trivlist}}

\newcommand{\Index}[1]{#1\index{#1}}
\makeindex

\makeatletter
\providecommand*{\toclevel@algorithm}{0}
\makeatother

\makeatletter
\renewcommand{\p@subfigure}{\thefigure\space}
\makeatother

\def\capStyle#1{#1}
\newcommand{\captionfonts}{\small\sf}

\makeatletter  %
\long\def\@makecaption#1#2{%
  \vskip\abovecaptionskip
  \sbox\@tempboxa{{\captionfonts #1: #2}}%
  \ifdim \wd\@tempboxa >\hsize
         {\captionfonts #1: #2\par}
         \else
         \hbox to\hsize{\hfil\box\@tempboxa\hfil}%
         \fi
         \vskip\belowcaptionskip}
\makeatother   %

\newdate{submitdate}{18}{5}{2009}
\newdateformat{monthdate}{%
\monthname[\THEMONTH] \THEYEAR}

\newcommand{\calelize}[1]{\ensuremath{\mathcal #1}}
\newcommand{\calA}{\calelize{A}}
\newcommand{\calC}{\calelize{C}}
\newcommand{\calE}{\calelize{E}}
\newcommand{\calF}{\calelize{F}}
\newcommand{\calG}{\calelize{G}}

\newcommand{\calM}{\calelize{M}}
\newcommand{\calS}{\calelize{S}}
\newcommand{\calW}{\calelize{W}}

\newcommand{\providelength} [1]{
  \ifthenelse{\isundefined{#1}}
             {\newlength{#1}}
             {}}

\newcommand{\floor}[1]{\lfloor #1 \rfloor}
\newcommand{\ceil}[1]{\lceil #1 \rceil}

\def\ccode#1{{\small{\texttt{#1}}}}
\def\concept#1{\textsf{\em #1}}

\newcommand{\ie}{i.\,e.,\xspace}
\newcommand{\eg}{e.\,g.,\xspace}
\newcommand{\etal}{{\it et~al}.\xspace}

\newcommand{\cpp}{\textsc{C\raise.08ex\hbox{\tt ++}}\xspace}
\newcommand{\cgal}{{\sc Cgal}\xspace}
\newcommand{\core}{{\sc Core}\xspace}
\newcommand{\leda}{{\sc Leda}\xspace}
\newcommand{\gmp}{{\sc Gmp}\xspace}
\newcommand{\stl}{{\sc STL}\xspace}
\newcommand{\boost}{{\sc Boost}\xspace}
\newcommand{\vrml}{{\sc Vrml}\xspace}

\newcommand{\disks}{\ensuremath{{\mathcal D}}\xspace}
\newcommand{\annulus}[1]{\ensuremath{{N_{#1}}}}
\newcommand{\anwid}[1]{\ensuremath{{W_{\annulus{#1}}}}}
\newcommand{\anouter}[1]{\ensuremath{{\mathcal O}_{#1}}}
\newcommand{\aninner}[1]{\ensuremath{{\mathcal I}_{#1}}}

\newcommand{\anpointorg}{Neighboring center}
\newcommand{\anpoint}{\MakeLowercase{\anpointorg}}
\newcommand{\andirection}{\ensuremath{\vec{d}}\xspace}

\newcommand{\mobius}{M\"o\-bi\-us\xspace}
\newcommand{\bez}{B\'ezier\xspace}
\newcommand{\FPFS}{farthest-point farthest-site\xspace}
\newcommand{\FPFSabrv}{{FPFS}\xspace}
\newcommand{\neighbor}{site\xspace}

\newcommand{\R}[1]{\ensuremath{{\mathbb R}\rule{0.3mm}{0mm} ^ {#1}}}
\newcommand{\unitsphere}{\ensuremath{{\mathbb S}\rule{0.3mm}{0mm} ^ 2}\xspace}
\newcommand{\distsym}{\ensuremath{\rho}\xspace}
\newcommand{\distance}[2]{\ensuremath{\distsym(#1, #2)}}
\newcommand{\vornr}[1]{\ensuremath{{\rm Vor}(#1)}}
\newcommand{\vor}[1]{\ensuremath{{\rm Vor}_\rho(#1)}}
\newcommand{\region}{\ensuremath{{Reg}}}
\newcommand{\env}[1]{\ensuremath{{\calE}_{#1}}}
\newcommand{\mindiag}[1]{\ensuremath{{\calM}_{#1}}}

\newcommand{\daniex}{\ensuremath{\{(i,i)\}_{i=1}^{n/2} \cup
    \{(-i,i)\}_{i=1}^{n/2}}}

\newcommand{\arrtraits}{\concept{ArrangementTraits\_2}\xspace}
\newcommand{\overlaytraits}{\concept{OverlayTraits}\xspace}
\newcommand{\envelopetraits}{\concept{EnvelopeTraits\_3}\xspace}

\newcommand{\arrangement}{\ccode{Arrange\-ment\_2}\xspace}
\newcommand{\aos}{\ccode{Arrangement\_\-on\_\-surface\_2}\xspace}
\newcommand{\envelope}{\ccode{Envelope\_3}\xspace}
\newcommand{\overlay}{\ccode{overlay}\xspace}

\newcommand{\envvor}{\ccode{Envelope\-Voronoi\_2}\xspace}
\newcommand{\vorconcept}{\concept{Envelope\-Voronoi\-Traits\_2}\xspace}
\newcommand{\apolloniustraits}{\ccode{Apollonius\_\-traits\_2}\xspace}
\newcommand{\fpfstraits}{\ccode{Farthest\_\-point\_\-farthest\_\-site\_\-traits\_2}\xspace}
\newcommand{\mobiustraits}{\ccode{CGAL::Mo\-bi\-us\_\-diagram\_\-traits\_2}\xspace}
\newcommand{\site}{\ccode{Site\_2}\xspace}

\newcommand{\arrlineartraits}{\ccode{Arr\_lin\-e\-ar\_traits\_2}\xspace}
\newcommand{\cgalarrlineartraits}{\ccode{CGAL::Arr\_lin\-e\-ar\_traits\_2}\xspace}
\newcommand{\lazyker}{\ccode{CGAL::La\-zy\_\-kernel}\xspace}
\newcommand{\circker}{\ccode{CGAL::Circular\_kernel\_2}\xspace}

\newrgbcolor{annulus-color}{0.000 0.690 0.000}
\newrgbcolor{circles-color}{0.690 0.000 0.690}
\newlength{\annuluslinewidth} 

\begin{document}
\begin{titlepage}
\begin{center}
  \psfig{figure=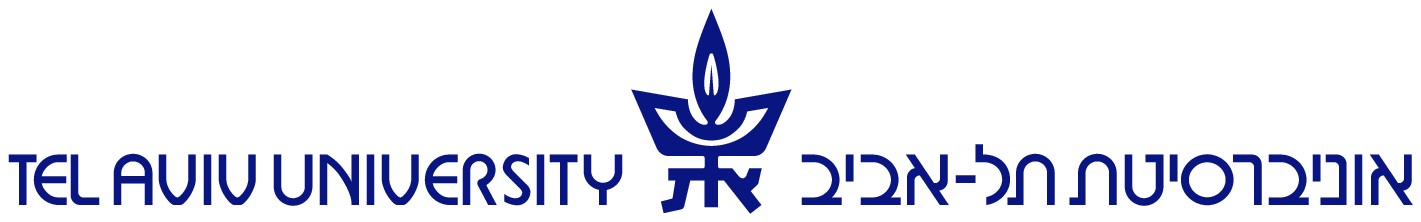,silent=}\\\vspace{0.2ex}
  {\large {\sc Raymond and Beverly Sackler}}\\\vspace{0.2ex}
  {\large {\sc Faculty of Exact Sciences}}\\\vspace{0.2ex}
  {\large {\sc The Blavatnik School of Computer Science}}\\\vfill

  {\Huge\bf Constructing Two-Dimensional}\\\vspace{0.3cm}
  {\Huge\bf Voronoi Diagrams via}\\\vspace{0.3cm}
  {\Huge\bf Divide-and-Conquer of Envelopes}\\\vspace{0.3cm}
  {\Huge\bf in Space}\\\vspace{3cm}

  {\large Thesis submitted in partial fulfillment of the requirements for the M.Sc.}\\\vspace{0.5cm}
  {\large degree in the School of Computer Science, Tel-Aviv University}\\\vspace{0.5cm}
  {\large  by}\\\vspace{0.5cm}
  {\LARGE\bf Ophir Setter}\\\vspace{3cm}
  {\large This work has been carried out at Tel-Aviv University }\\\vspace{0.2ex}
  {\large under the supervision of Prof. Dan Halperin}\\\vfill
  {\large \monthdate\displaydate{submitdate}}

  \author{Ophir Setter}
\end{center}
\end{titlepage}

\thispagestyle{empty}
\frontmatter

\newpage
\subsection*{Acknowledgements}

Many people had great influence on this thesis and its author during
the research period.
I deeply thank my advisor, Prof.~Dan Halperin, for his help in
guidance, support, and encouragement, and for introducing me the field
of applied computational geometry.

\smallskip
\noindent I wish to thank Efi Fogel and Eric Berberich for
fruitful collaboration and for sharing priceless knowledge.
Special thanks are given to Efi for his warm hospitality during 
fruitful Friday afternoons and for providing the basis for the player
software, which enabled the creation of the 3D figures of this thesis.
Special thanks are given to Eric for his admirable motivation and
for sharing his insights through many rich discussions.
I also thank Prof.~Micha Sharir for his cooperation and help in
theoretical parts of the thesis.

\smallskip
\noindent I would also like to thank all other members of the applied
computational geometry lab at the computer science school of Tel-Aviv
University who provided support and useful suggestions.
Special thanks are given to Ron Wein and to Michal Meyerovitch.

\smallskip
\noindent I wish to thank all members of the algorithms group at the
Max-Planck-Insitut f{\"u}r Informatik in Saarbr{\"u}cken, Germany, for
introducing and helping with the field of computational algebra, and
for their hospitality.
Special thanks are given to Michael Hemmer, to Eric Berberich, and to
Michael Kerber.

\bigskip
\noindent Work on this thesis has been supported in part by the Israel
Science Foundation (grant no. 236/06), by the German-Israeli
Foundation (grant no. 969/07), and by the Hermann Minkowski--Minerva
Center for Geometry at Tel Aviv University.

\newpage ~~~ \newpage

\section*{Abstract}

We present a general framework for computing two-dimensional Voronoi
diagrams of different classes of sites under various distance
functions.
The framework is sufficiently general to support diagrams embedded on
a family of two-dimensional parametric surfaces in \R{3}.
The computation of the diagrams is carried out through the
construction of envelopes of surfaces in 3-space provided by \cgal{}
(the Computational Geometry Algorithm Library).
The construction of the envelopes follows a divide-and-conquer
approach.
A straightforward application of the divide-and-conquer approach for
computing Voronoi diagrams yields algorithms that are inefficient in
the worst case.
We prove that through randomization the expected running time becomes
near-optimal in the worst case.
We show how to employ our framework to realize various types of
Voronoi diagrams with different properties by providing
implementations for a vast collection of commonly used Voronoi
diagrams.
We also show how to apply the new framework and other existing tools
from \cgal to compute minimum-width annuli of sets of disks, which
requires the computation of two Voronoi diagrams of two different
types, and of the overlay of the two diagrams.
We do not assume general position. Namely, we handle degenerate input,
and produce exact results.

\tableofcontents
\listoffigures

\newpage ~~~ \newpage
\mainmatter

\chapter{Introduction}
\label{chap:intro}
In layman's terms the Voronoi diagram of a given set of objects is the
subdivision of the space into regions where each region consists of
points that are closer to one particular object than to all others of
the given set of objects.
Voronoi diagrams are intuitive structures and are even found in
various forms in nature.
The concept has reappeared in diverse fields of science
\begin{wrapfigure}{r}{150pt}
  \vspace{-13pt}
  \includegraphics[width=150pt]{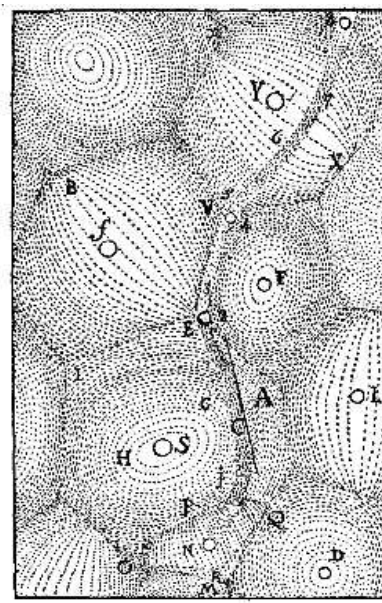}
  \vspace{-20pt}
  \caption[Voronoi diagram figure from Descartes research] %
  {A Voronoi diagram figure from a solar system research by
    Descartes (1644).\footnotemark}
  \vspace{-17pt}
\end{wrapfigure}
throughout history, often receiving a different new name:
Wigner-Seitz zones (chemistry and  physics), domains of action
(crystallography), Thiessen polygons (geography),
etc.\footnotetext{The figure is taken from:
  \url{http://e.simwe.com/228342/viewspace-6935}}

Most text books cite a 1644 solar system research by
Descartes~\cite{d-pp-44} as the first documented application of
Voronoi diagrams, even though they were not explicitly defined there.
There is no controversy, however, that %
Dirichlet was the first to formally define the concept of
\emph{Voronoi diagrams}, and that the initial extensive studies on the
subject were conducted by him and by Georgy
Voronoi~\cite{ak-vd-00,obsc-stcavd-00}.
This clarifies the fact that the two leading aliases for the diagrams
currently are \emph{Voronoi diagrams} and
\emph{Dirichlet tessellations}.
The other most common term refers to the dual diagram to the Voronoi
diagram --- the \emph{Delaunay triangulation} (or tessellation).
Despite the fact that %
Voronoi was the first to conceive the dual diagram to the Voronoi
diagram, %
Delaunay was the first to directly define it, earning an alias after
his name.

Shamos and Hoey introduced Voronoi diagrams to the field of Computer
Science~\cite{sh-cpp-75}.
They used Voronoi diagrams to improve running times of algorithms for
problems, which had been considered unrelated, such as smallest
enclosing circle and various closest-point problems.
Since then, Voronoi diagrams were thoroughly investigated, and were
used to solve many geometric problems.

The concept of Voronoi diagrams, that is, the division of a space into
maximally-connected cells, where each cell consists of points that
are closer to a particular site than to any other site 
of a given collection of sites,
was extended beyond the scope of point sites and the Euclidean
metric~\cite{ak-vd-00,bwy-cvd-06,obsc-stcavd-00}.
In fact, the original diagram referred to by Descartes (mentioned
above) seems more like, what nowadays is called, a weighted Voronoi
diagram than a standard Voronoi diagram of points.

The most straightforward generalization of the standard Voronoi
diagram is to various kinds of geometric sites while staying in the
Euclidean space.
These diagrams include Voronoi diagrams of line segments, Voronoi
diagrams of circular arcs, Voronoi diagrams of disks in the plane, and
Voronoi diagrams of ellipses in the plane.

Different applications impose different types of distance functions to
the sites, inducing diverse types of Voronoi diagrams.
Among those are:
power diagrams of disks,
multiplicatively-weighted Voronoi diagrams, and Voronoi diagrams with
respect to the $L_p$ (Minkowski) and Karlsruhe (Moscow)
metrics~\cite{obsc-stcavd-00}.
Two particularly interesting types of diagrams are the $L_1$-diagram
of points that can have two-dimensional cells induced by multiple
sites (two-dimensional \emph{bisectors}), and the
multiplicatively-weighted Voronoi diagram that can be of quadratic
size in the number of input sites.
Voronoi diagrams were also defined in various ambient spaces.
For example, Voronoi diagrams defined over different two-dimensional
surfaces, such as the Voronoi diagram of points on a sphere, the
Voronoi diagram of points on a cone, and the power diagram on a
sphere.
In higher dimensions the main research efforts concentrated on Voronoi
diagrams of point sites and power diagrams, neglecting other types of
sites.
For example, though the complexity of the Euclidean Voronoi diagram of
lines with fixed orientations in three-dimensions was investigated by
Koltun and Sharir~\cite{ks-tdevd-03}, a complete combinatorial and
algebraic description of the diagram of three lines was given by
Everett~\etal~\cite{elle-vdtl-07} only recently.
In this thesis we concentrate on Voronoi diagrams defined over certain
two-dimensional parametric surfaces in 3-space.

Klein unified some classes of planar Voronoi diagrams under a
generalized framework by introducing \emph{abstract Voronoi diagrams},
which are defined in terms of their bisector
curves~\cite{k-cavd-89,kmm-ricavd-93}.

Every type of nearest-\neighbor Voronoi diagram defines a complementary 
farthest-\neighbor Voronoi diagram.
The farthest-\neighbor Voronoi diagram is useful on its own, and one can find
applications where both (nearest and farthest) Voronoi diagrams are needed;
see Chapter~\ref{chap:min-annulus}.
Various farthest-\neighbor Voronoi diagrams have nearly-linear time
construction algorithms~\cite{cg-ps-85,afw-fsgvd-88}.
Mehlhorn, Meiser, and Rasch~\cite{mmr-fsavd-01} used Klein's
terminology and proved that farthest abstract Voronoi diagrams are
of linear size and can be computed with a randomized algorithm in
$O(n\log n)$ expected running time.

\section*{Algorithms}
Numerous approaches for computing Voronoi diagrams were developed.
Shamos and Hoey used the divide-and-conquer paradigm to
obtain the first optimal $\Theta(n \log n)$-time construction algorithm for
the Voronoi diagram of points in the plane~\cite{sh-cpp-75}.
The algorithm partitions the set of points into two sets of
roughly equal size by a vertical line, computes the respective right
and left Voronoi diagrams, and finally, carefully merges
the two diagrams together.
Their main achievement was to prove that there is a polygonal line that
``stitches'' the two diagrams together (in the merge step), 
and that it can be found in $O(n)$ time,
yielding a $\Theta(n \log n)$-time overall algorithm.
A similar approach was used by Klein~\cite{k-cavd-89} to supply a
$\Theta(n \log n)$-time algorithm for abstract Voronoi diagrams.
Guibas and Stolfi described the algorithm in the context of the
Delaunay graph (the dual to the Voronoi graph) as a major application
for their quad-edge data structure~\cite{gs-pmgscv-83}.
Dwyer improved the expected running time of the divide-and-conquer
algorithm for various point distributions 
to~$O(n \log \log n)$~\cite{d-fdcacd-87}.

A different divide-and-conquer approach was recently proposed by
Aichholzer~\etal \cite{aaahjp-dcvdr-09a}.
They employ a divide-and-conquer medial-axis algorithm on an augmented
domain to compute the Voronoi diagrams of various types of sites, such
as polygonal sites, circular disks, and spline curves.
The combinatorial structure of the Voronoi diagram is computed without
the construction and manipulation of bisector curves.
However, being based on a medial-axis algorithm, their approach
supports only diagrams induced by the Euclidean metric.

The ubiquitous sweep-line paradigm, introduced by Bentley and
Ottmann~\cite{bo-arcgi-79}, was adapted by Fortune for the
construction of Voronoi diagrams of points in the
plane~\cite{f-savd-87}.
The sweep technique proved useful also for constructing other types of
Voronoi diagrams, such as order-$k$ Voronoi
diagrams~\cite{r-okvdsa-91} and Voronoi diagrams of circles in the
Euclidean plane~\cite{jkmkh-saevdc-06}.

Another important and popular technique is the
\emph{incremental construction}~\cite{gs-cdtp-78}, which together with
\emph{randomization}~\cite{cs-arscg-89} yield an $O(n\log n)$
algorithm for constructing Voronoi diagrams~\cite{gks-ricdvd-92}.
This technique was used to attain algorithms for various types
of Voronoi diagrams, and allowed relaxation of certain requirements
on bisector curves in the definition of abstract Voronoi diagrams,
while keeping an optimal time complexity~\cite{ky-vdpco-03,kmm-ricavd-93}.

Lower or upper envelopes of surfaces constitute a fundamental
structure in computational geometry. They are frequently used to solve
various problems including: hidden surface removal,
computing Hausdorff distances, and more~\cite{as-ata-00,sa-dssga-95}.
Agarwal~\etal presented an efficient and simple divide-and-conquer
algorithm for constructing envelopes in three dimensions~\cite{ass-olea-96}.
The theoretical worst-case time complexity of constructing the
envelope of $n$ ``well-behaved'' surfaces in three dimensions using the
divide-and-conquer algorithm is $O(n^{2+\varepsilon})$.\footnote{A
  bound of the form $O(f(n) \cdot n^{\varepsilon})$ means that the
  actual upper bound is $C_{\varepsilon}f(n) \cdot n^\varepsilon$, for
  any $\varepsilon > 0$, where $C_{\varepsilon}$ is a constant that
  depends on $\varepsilon$, and generally approaches infinity as
  $\varepsilon$ goes to $0$.} 
This near-quadratic running time can arise also in cases of envelopes
of linear complexity.
This is also an upper bound, almost tight in the worst case, on the
combinatorial complexity of the envelope.

Edelsbrunner and Seidel~\cite{es-vda-86}
observed the connection between Voronoi diagrams in \R{d} and lower
envelopes of the corresponding distance functions to the sites in
\R{d+1}, yielding a very general approach for computing Voronoi
diagrams.
For example, consider the Voronoi diagram of a set of points in the plane, 
then, the connection is as follows: 
for each point site $p = (x_p, y_p)$ we consider the paraboloid 
$P_p(x, y) = (x - x_p)^2 + (y - y_p)^2$. The \emph{minimization diagram}
of the lower envelope of the paraboloids, which is the vertical
projection of the lower envelope onto the plane, corresponds to the Voronoi
diagram of the points.
Other algorithms include
Yap's algorithm for segments and circular
arcs~\cite{y-avdssc-87}, an optimal algorithm for the construction of
weighted Voronoi diagrams~\cite{ae-oacwvd-84}.
The interested reader is referred to the book by
Okabe~\etal~\cite{obsc-stcavd-00} and to the survey by
Aurenhammer and Klein~\cite{ak-vd-00} for more information.

\section*{Software}
The Computational Geometry Algorithms Library
(\cgal{})\citelinks{cgal-link} is an open-source \cpp{} library of
efficient and reliable geometric algorithms.\footnote{Throughout the
  thesis a number in brackets (e.g.,~\citelinks{gmp-link}) refers to
  the link list on page~\pageref{sec:linkbib}, and an alphanumeric
  string in brackets (e.g.,~\cite{fsh-eiaga-08}) is a standard
  bibliographic reference.}
It follows the exact geometric computation
paradigm~\cite{y-rgc-04,yd-ecp-95} to achieve robustness with exact
results.
\cgal contains implementations of algorithms for computing the dual
Delaunay graphs to standard Voronoi diagrams, Apollonius diagrams, and
segment Voronoi diagrams~\cite{bdpty-tic-02,ek-padaai-06,k-reisvd-04}.
Moreover, Voronoi diagrams of ellipses can be computed using the
same \cgal{} framework~\cite{ett-vdec-08}.
Other exact implementations include the construction of segment
Voronoi diagrams in \leda{}~\cite{bms-hcvdls-94} and the
implementation of the randomized algorithm for constructing abstract
Voronoi diagrams in \leda{}~\cite{s-eiavd-94}.

Approximated
alternatives include the {\sc Vroni} code for computing
two-dimensional Voronoi diagrams of
points and line segments~\cite{h-vearec-01}, the use of
the Graphics Processing Unit (GPU) to visualize Voronoi
diagrams~\cite{hklmc-fcgvd-99, n-itvdg-08}, and more.

A large number of the implementations for constructing Voronoi
diagrams use the incremental construction paradigm.
The time complexity achieved by the incremental construction relies on
the fact that most changes applied to the diagram are local with
respect to the location of the inserted site.
This assumption usually implies that the diagrams are of linear size.
Typically, the time complexity of constructing a Voronoi diagram that has
\emph{linear complexity}, using the above algorithms, is \emph{nearly linear}.

One of the main packages included in \cgal{} is the \aos package,
which supports construction and maintenance of arrangements of bounded
or unbounded curves embedded on certain two-dimensional parametric
surfaces in three-dimensions and different operations on
them~\cite{cgal:wfzh-a2-08,wfzh-aptaca-07,bfhmw-smtdas-07}.
The \aos package is robust when using exact arithmetic and handles all
degenerate input.

\cgal{} contains a robust and efficient implementation of the
divide-and-conquer algorithm mentioned earlier for constructing
envelopes of surfaces in three dimensions~\cite{cgal:mwz-e3-08,m-rgeces-06}.
The implementation, provided in \cgal{}'s \envelope{} package,
relies heavily on arrangements and algorithms from the \aos package.
As mentioned above, the divide-and-conquer algorithm achieves a worst-case
near-quadratic running time.
This fact poses an obstacle when attempting to utilize this algorithm
for the construction of Voronoi diagrams that have linear complexity, as we aim
for algorithms that run in near-linear time.

\section*{Contribution of the Thesis}

\begin{figure}[t] %
  \centering
  \providelength{\subfigwidth}\setlength{\subfigwidth}{110pt}
  \newlength{\spacewidth}\setlength{\spacewidth}{0pt}
  \subfigure[]{\label{fig:various-a}\includegraphics[width=\subfigwidth]{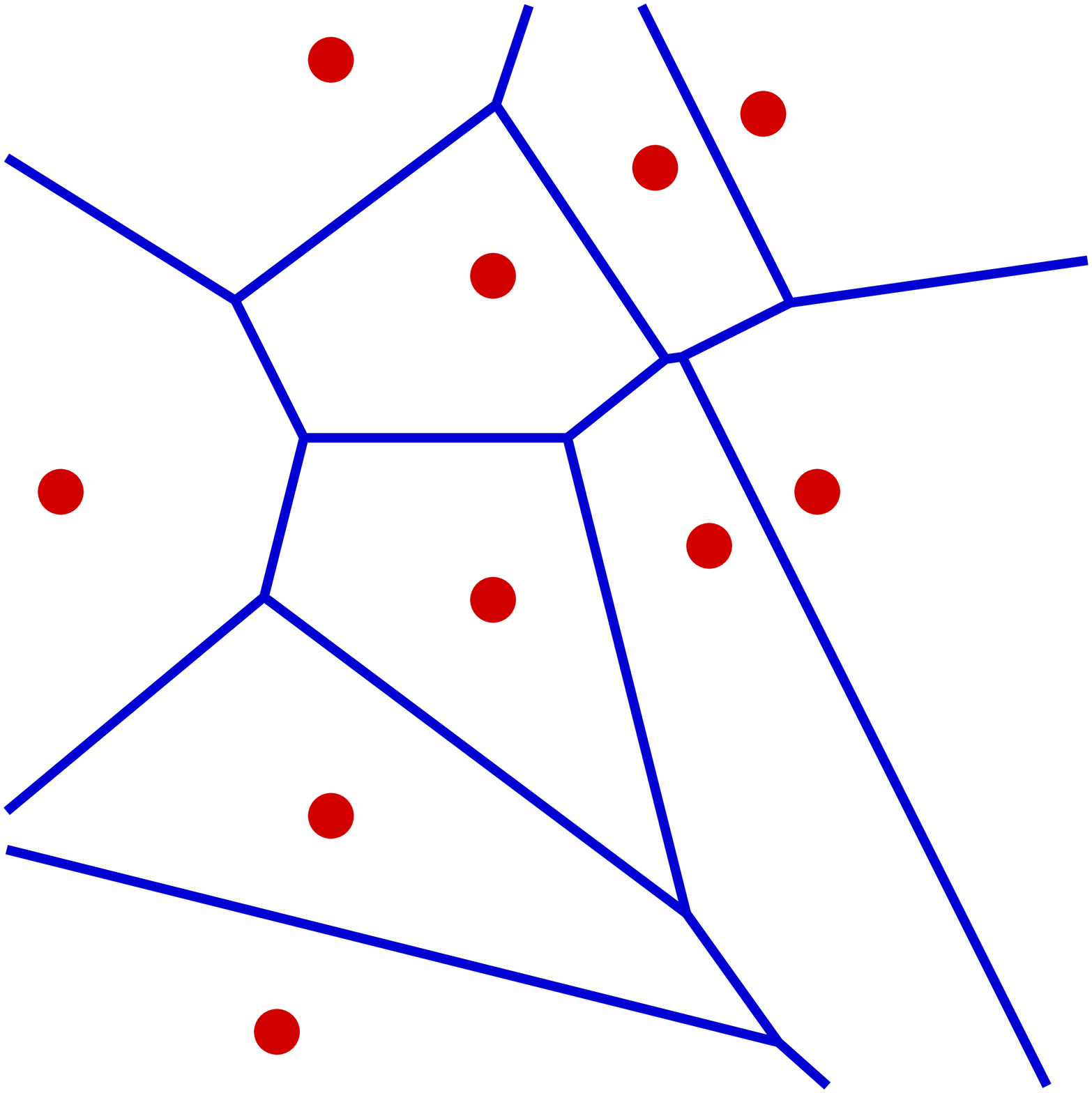}}
  \hspace{\spacewidth}
  \subfigure[]{\label{fig:various-b}\includegraphics[width=\subfigwidth]{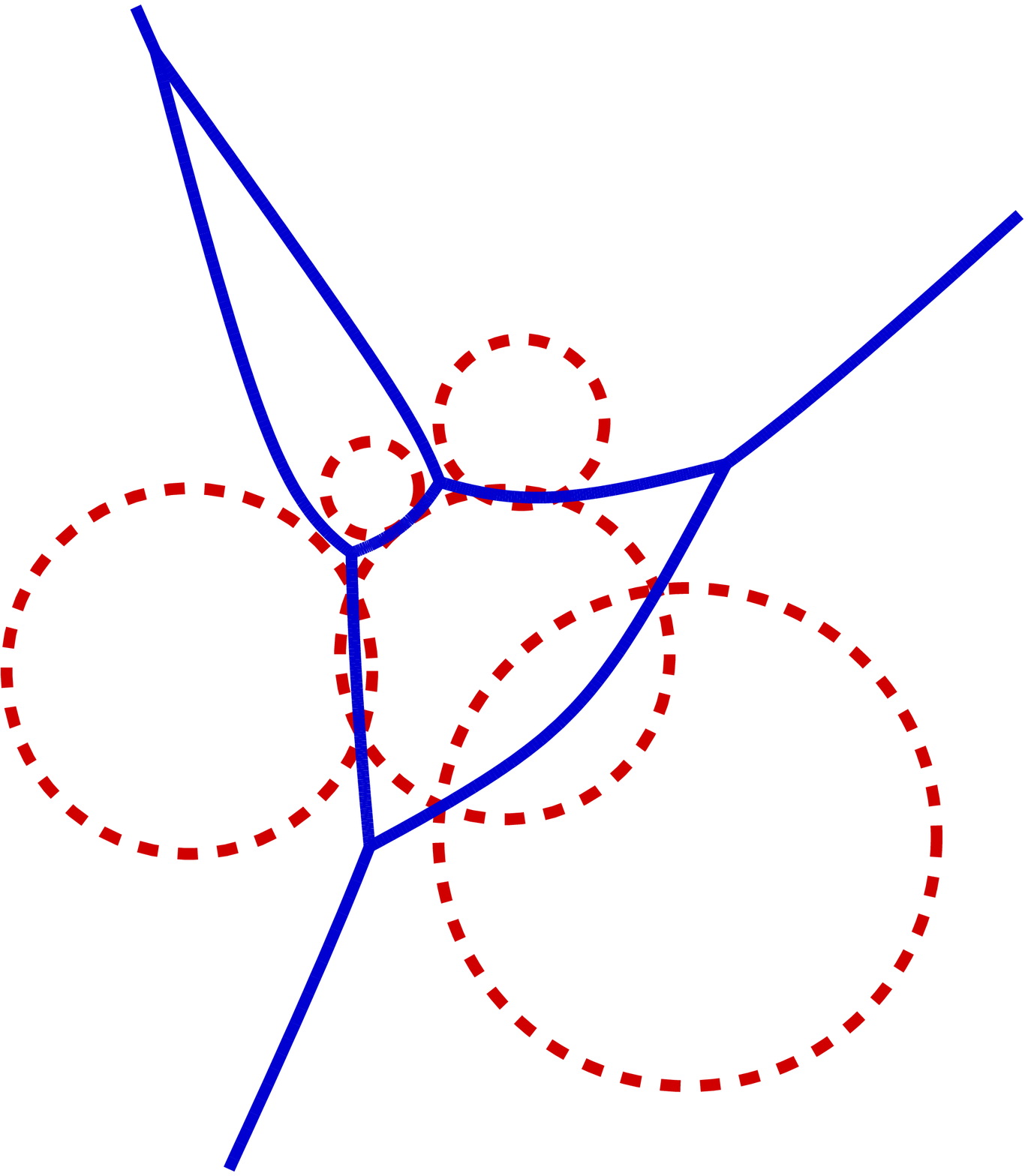}}
  \hspace{\spacewidth}
  \subfigure[]{\label{fig:various-c}\includegraphics[width=\subfigwidth]{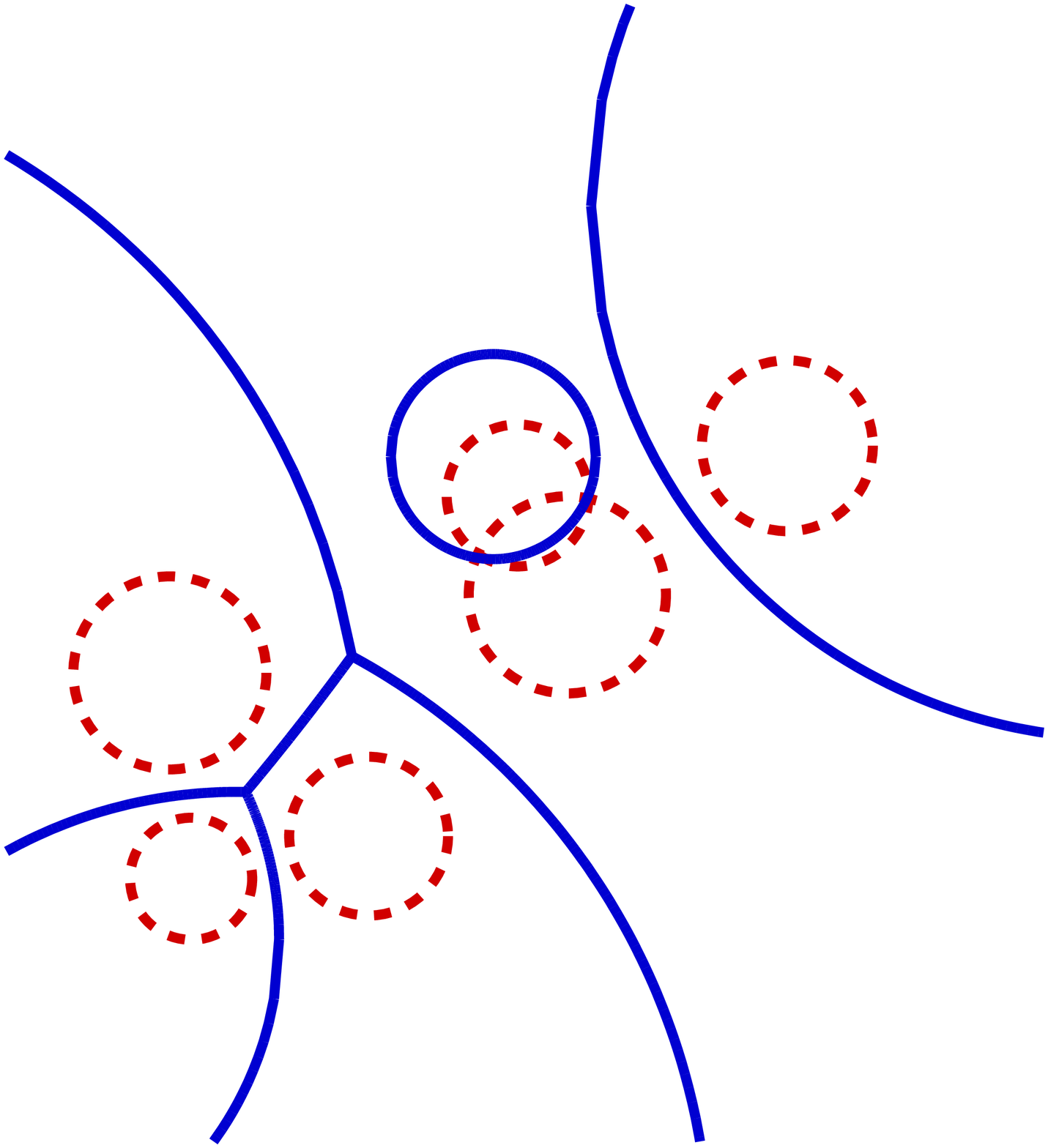}}
  \hspace{\spacewidth}
  \subfigure[]{\label{fig:various-d}\includegraphics[width=\subfigwidth]{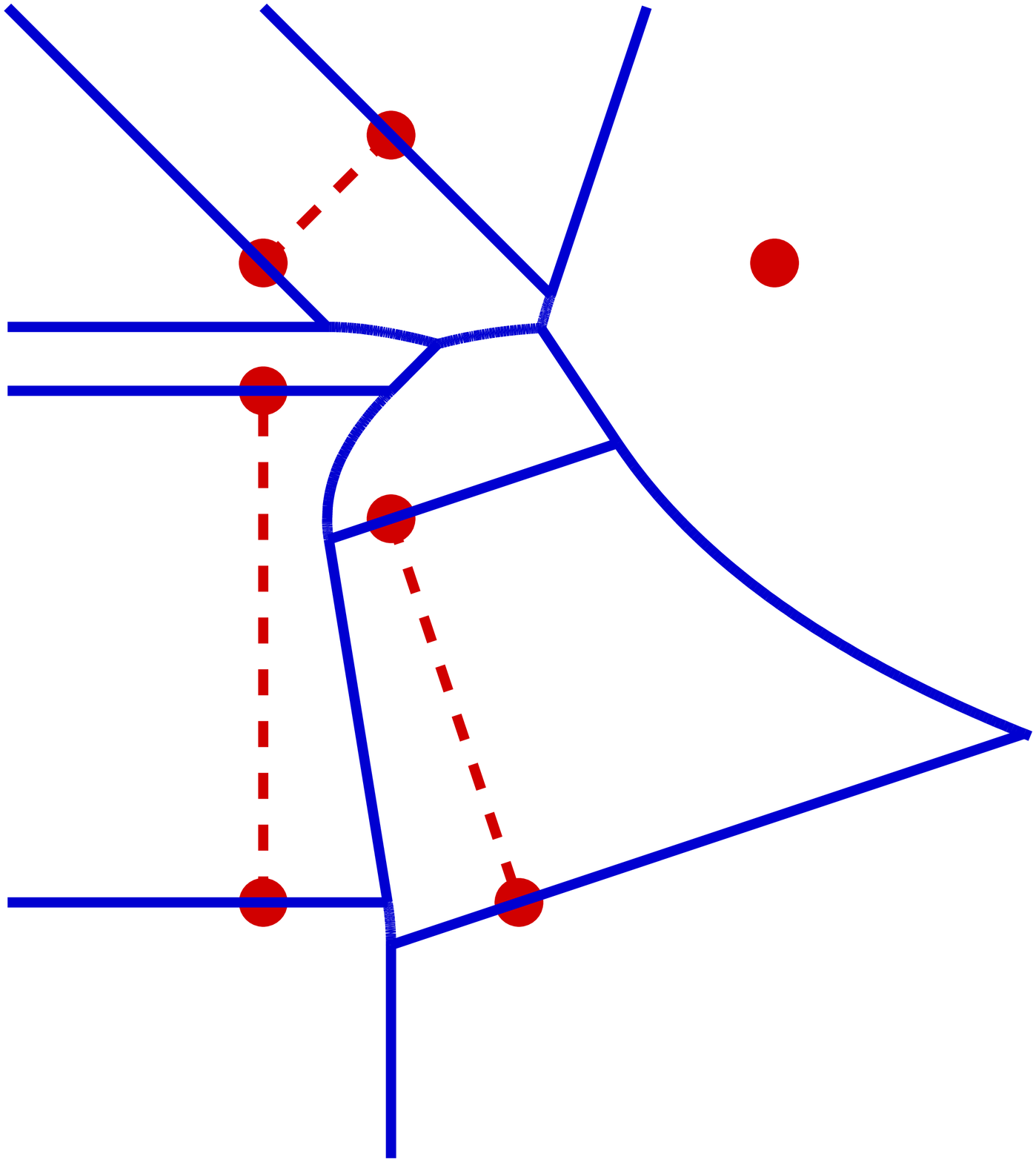}}

  \caption[Various Voronoi diagrams]{\capStyle{Various Voronoi
      diagrams computed with our software.
      (For the parameters of sites in each diagram, see
      Table~\ref{tab:supp-vd}.)
      \subref{fig:various-a} A standard Voronoi diagram (point sites
      with $L_2$ metric).
      \subref{fig:various-b} An additively-weighted Voronoi
      (Apollonius) diagram with disk centers as sites and disk radii
      as weights.
      \subref{fig:various-c} A \mobius{} diagram with disk centers as
      sites.  The distance from every point on the boundary of a
      disk to its corresponding site is zero.
      \subref{fig:various-d} A Voronoi diagram of segments 
      and points.
      The sites in \subref{fig:various-b}, \subref{fig:various-c}, and
      \subref{fig:various-d} are illustrated with dashed curves.
      The figures in this thesis are best viewed in color.
  }}
  \label{fig:various}
\end{figure}

We present a general framework for constructing various two-dimensional
Voronoi diagrams, exploiting the efficient, robust, and
general-purpose envelope code of \cgal{}.
We have extended the \envelope{} package to work together
with the new \aos{} package and created spherical geodesic Voronoi
diagrams based on our self-developed code for constructing
arrangements of geodesic arcs on the sphere.
The work in this context was presented at the $24^{th}$ European
Workshop on Computational Geometry~\cite{fsh-eiaga-08}
and in the multimedia session of the $24^{th}$ Annual Symposium on
Computational Geometry~\cite{fsh-agas-08};
for more information on the computation of arrangements of geodesic
arcs on the sphere and applications see Efi Fogel's
thesis~\cite{f-mscaag-08}.

\begin{table}[h]
\begin{threeparttable}
  \caption[Supported Types of Voronoi diagrams]{\capStyle{Types of Voronoi 
      diagrams currently supported by our implementation, and their bisector 
      classes.}}
  \label{tab:supp-vd}

    \small
    \begin{tabularx}{\textwidth}{|p{3cm}|p{4cm}|p{3.4cm}|X|}
      \hline
      \multicolumn{1}{|c|}{{\bf Name}} &
      \multicolumn{1}{c|}{{\bf Sites}} & 
      \multicolumn{1}{c|}{{\bf Distance function}} &
      \multicolumn{1}{c|}{{\bf Class of bisectors}}\\ \hline \hline

      \emph{\mbox{Standard Voronoi} diagram} & 
      points $p_i$ & \vspace{-6pt}$||x - p_i||$ &
      \multirow{2}{*}{lines} \\ \cline{1-3}

      \emph{Power diagram} &  disks (with center
      $c_i$ and radius $r_i$) & 
      \vspace{-6pt}$\sqrt{(x - c_i)^2 - r_i^2}$ & \\ \hline

      \emph{2-point \mbox{triangle-area} Voronoi diagram} &  pairs of points
      $\{p_i, q_i\}$ & 
      area of $\triangle xp_iq_i$ & pairs of lines \\ \hline

      \emph{Apollonius \newline diagram}
      & points $p_i$ and weights $w_i$
      & \vspace{-6pt}$||x - p_i|| - w_i$
      & hyperbolic arcs \\ \hline

      \emph{\mobius diagram}  &  points $p_{i}$ with
      scalars $\lambda_{i}, \mu_{i}$ & \vspace{-6pt}$\lambda_{i}(x -
      p_{i})^{2} - \mu_{i} $ & circles and lines
      \\ \hline
      
      \emph{Anisotropic \mbox{diagram}} &
      points $p_{i}$, with
      positive definite matrices $M_{i}$, 
      and scalars $\pi_{i}$ & \vspace{-6pt}$(x -
      p_{i})^{t} M_{i} (x - p_{i}) - \pi_{i}$ & conic arcs \\ \hline
      
      \emph{Voronoi diagram of linear objects} & 
      interior-disjoint points, segments, rays, or lines &
      Euclidean distance & piecewise algebraic curves
      composed of
      line segments and parabolic arcs \\ \hline
      
      \emph{\mbox{Spherical Voronoi} diagram} & points on a
      sphere & geodesic distance & \multirow{2}{3.4cm}{arcs of great 
        circles (geodesic arcs)}
      \\ \cline{1-3}
      
      \emph{\mbox{Power diagram on} a sphere} & circles on a
      sphere & ``spherical'' power distance\tnote{a} &
      \\ \hline
	      
    \end{tabularx}

  \begin{tablenotes}
  \item[a] {Given a point $p$ and a circle with center~$q$ and
    radius~$r$ on the sphere, the spherical power ``proximity''
    between $p$ and the circle is defined to be 
    $\frac{\cos{d(p, q)}}{\cos{r}}$ where $d(p, q)$ is the geodesic
    distance between $p$ and $q$~\cite{s-lvds-02}.}
  \end{tablenotes}
\end{threeparttable}
\vspace{-10pt}
\end{table}

Chapter~\ref{chap:background} gives the necessary background and basic
definitions.
Chapter~\ref{chap:env-to-vd} describes the adaptation of the
divide-and-conquer algorithm for envelopes to Voronoi diagrams
embedded on two-dimensional parametric surfaces in 3-space, and the
elimination of the above complexity obstacle using randomization in
the divide step.
We describe the software interface between the construction of Voronoi
diagrams and the envelope code of \cgal{}.
An analysis by Micha Sharir for the expected time complexity of
constructing lower envelopes in this randomized divide-and-conquer
setup can be found in Section~\ref{sec:micha-proof}.
Chapter~\ref{chap:impl} presents details about our implementation of
a large set of Voronoi diagrams with different properties using our
framework.
The section demonstrates the generality of the framework and gives
information on techniques we have applied to speed-up the exact (and
costly) computation in practice.
Table~\ref{tab:supp-vd} summarizes the types of diagrams that are
currently supported by our implementation.
In Chapter~\ref{chap:exper}, we thoroughly discuss the advantages and
the limitations of our framework. We also show experimental results,
demonstrating the randomization effect on the running time.
We present an application of our framework to solve the problem of
computing a minimum-width annulus of a set of disks in the plane,
which exploits the generality and the flexibility of the framework in
Chapter~\ref{chap:min-annulus}; a short introduction to the
minimum-width annulus problem is given in
Section~\ref{sec:min-annulus:intro}.
The solution requires the computation of two Voronoi diagrams of two
different types, and of the overlay of the two diagrams.
We presented an extended abstract of this work of computing a
minimum-width annulus of disks at the $25^{th}$ European Workshop on
Computational Geometry~\cite{sh-ecmwad-09}.
An extended abstract of the thesis will be presented on the~$6^{th}$
annual International Symposium on Voronoi Diagrams in science and
engineering (ISVD)~\cite{ssh-ctdvd-09}.

The major strength of our approach is its completeness, robustness,
and generality, that is, the ability to handle degenerate input, 
the agility to produce exact results, and the capability to construct
diverse types of Voronoi diagrams.
The code is designed to successfully handle degenerate input,
while exploiting the synergy between generic programming and exact 
geometric computing, and the 
divide-and-conquer framework to construct Voronoi diagrams.
Theoretically, the randomized divide-and-conquer envelope approach for
computing Voronoi diagrams is efficient and it is asymptotically
comparable to other (near-)optimal methods.
However, the method uses constructions of bisectors and Voronoi
vertices as elementary building blocks, and they must be exact, which
makes the concrete running time of our exact implementation
inferior to those of existing implementations of various dedicated
(specific diagram type) implementations.

\begin{wrapfigure}[7]{r}{210pt}
  \vspace{-20pt}
  \centerline{
    \begin{tabular}{cc}
      \includegraphics[width=100pt]{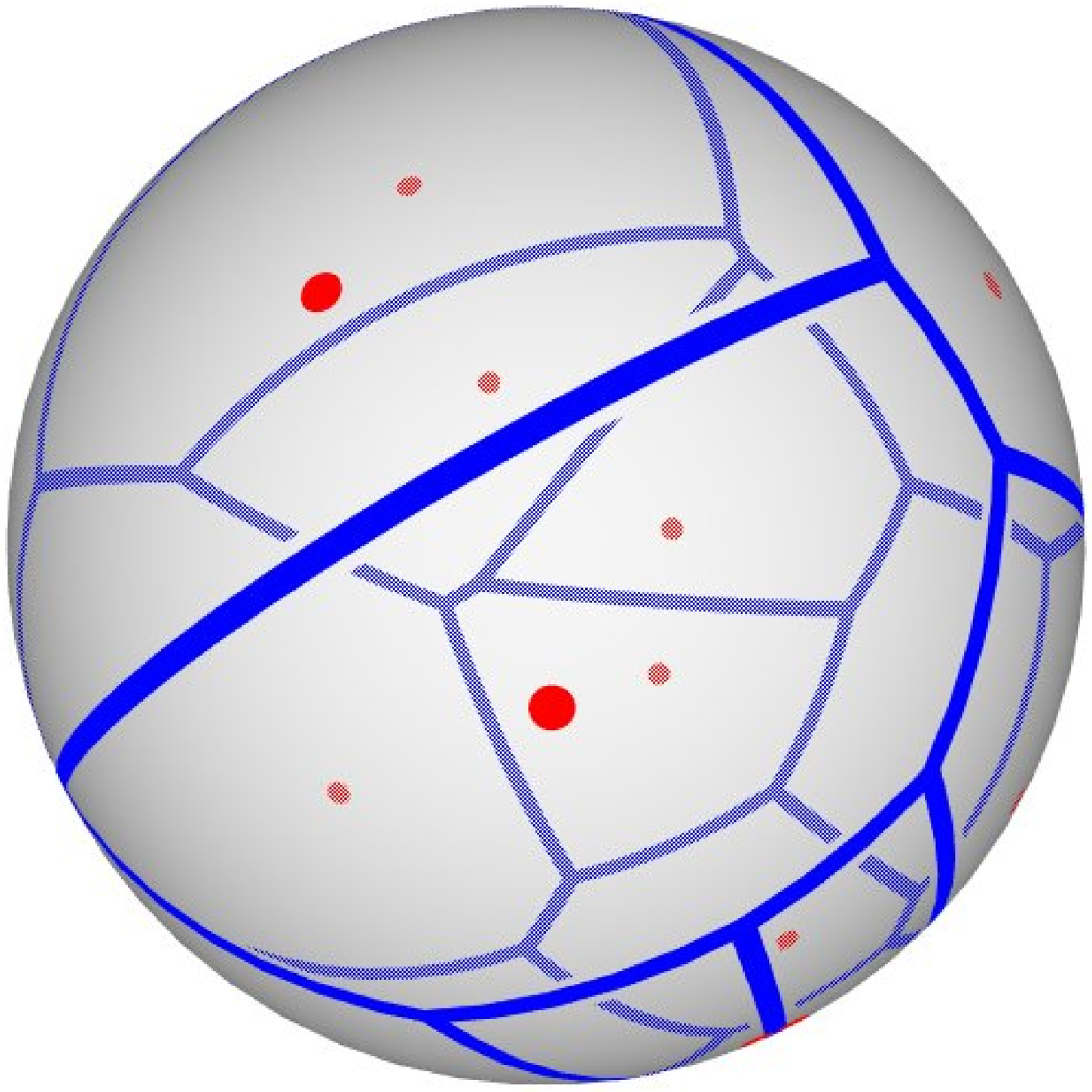} &
      \includegraphics[width=100pt]{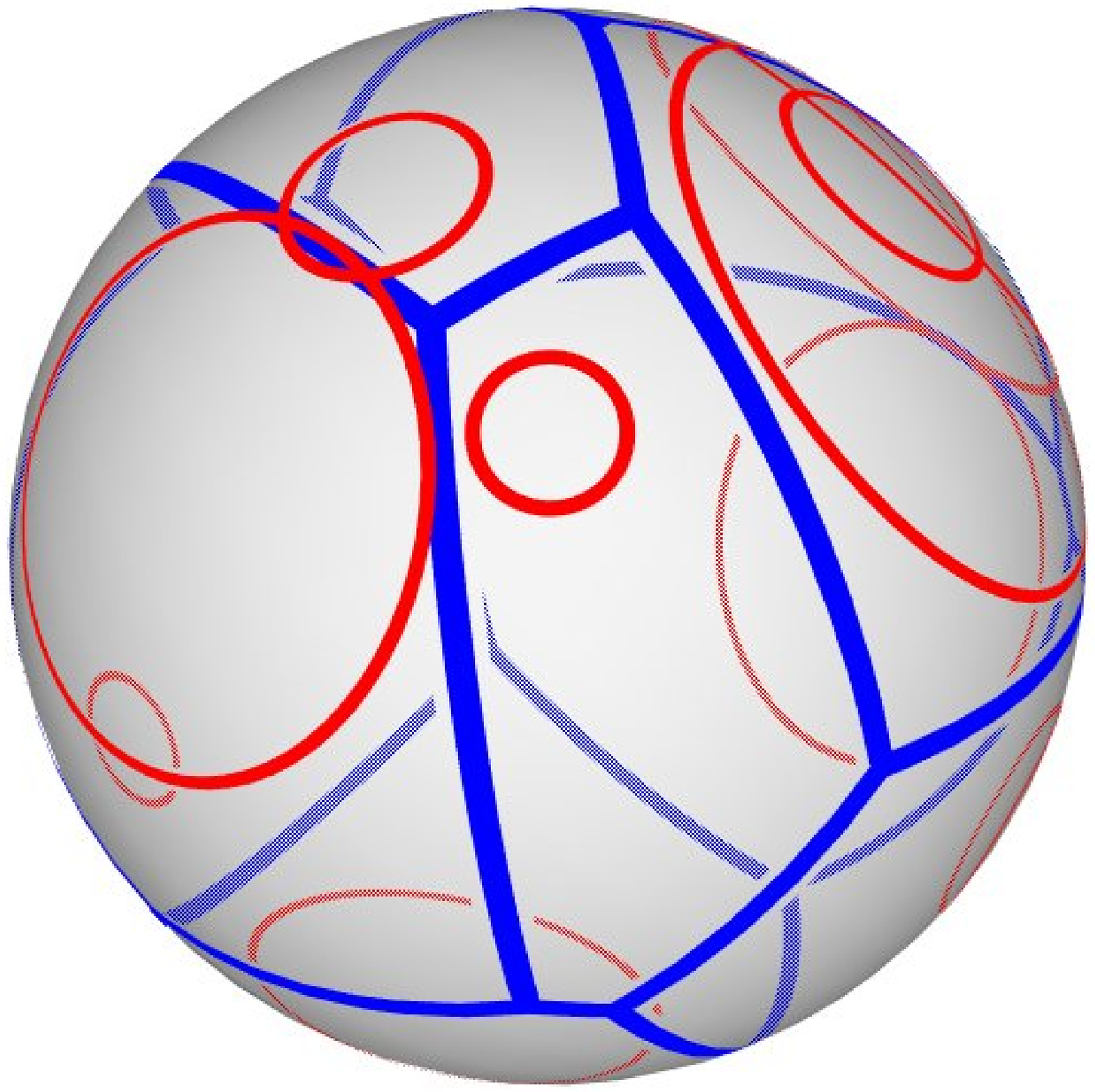} \\
    \end{tabular}
  }
\end{wrapfigure}
Our software can support practically any kind of Voronoi
diagrams, provided that the user supplies a set of basic
procedures for manipulating a small number of sites and their
bisectors (see Section~\ref{sec:env-to-vd:impl} and
Chapter~\ref{chap:impl} for details on how to use the framework to
compute new types of Voronoi diagrams.)
Figure~\ref{fig:various}
illustrates several types of planar Voronoi diagrams computed by our software.
The figure above shows two types of
Voronoi diagrams on the sphere computed with our software; its left
part shows a spherical 
Voronoi diagram of 14 points and its right part shows a spherical
power diagram of 10 circles. Both diagrams are composed of geodesic
arcs.\footnote{The figure and other 3D figures in this thesis were
  created using an interactive viewer for an extended \vrml{} format
  called \emph{player}, which is based on a Scene Graph Algorithm
  Library called \emph{SGAL}.}

\cgal had a significant impact on this thesis.
All the software components involved in this thesis are based on \cgal
and are developed according to its guidelines.
The developed software adheres to the generic programming paradigm and
follows the exact geometric computation paradigm, similar to other
existing \cgal components.
Nonetheless, this thesis also had an influence on \cgal.
The results of this thesis contributed to \cgal in the form of
improving existing components and in developing new components that
are planned to be integrated into a future public release of \cgal.
This includes, for example, contributing to the development of the new
\aos package, extending the \envelope package to support envelopes
embedded on two-dimensional surfaces, and enhancing existing traits
classes for the arrangement package and the arrangement package
itself.
The code presented in this thesis is packed into a package, in the form
of a \cgal package, named \envvor.

\chapter{Preliminaries}
\label{chap:background}

This chapter provides background material and definitions required for
the understanding of this thesis.
Sections~\ref{sec:background:vd} and~\ref{sec:background:envelopes}
provide general definitions for Voronoi diagrams and describe the
divide-and-conquer algorithm for constructing envelopes of surfaces in
three-dimensions.
Basic software components and technical background, the
implementations included in this thesis are built upon, are reviewed
in Sections~\ref{sec:background:cgal}
and~\ref{sec:background:cgal-envelopes}.

\section{Voronoi Diagrams}
\label{sec:background:vd}
Let $O = \{o_1, \ldots, o_n\}$ be a set of $n$ objects, referred to as
\emph{Voronoi sites}, in an ambient space~\calS{}.
Let $\distsym : O \times \calS \to \R{}$ be a distance function between
Voronoi sites and points in the space.
The \emph{Voronoi diagram} \vor{O} of the set $O$ with respect to the
distance function \distsym is defined to be the partition of the space
\calS{} into maximally connected cells, where each cell consists of
points that are closer to one particular site (or a set of sites) than
to any other site.
Formally, every point $p \in \calS$ lies in a cell corresponding to a set of
sites $I \subseteq O$ if, and only if, $\rho(o_i, p) < \rho(o_j, p)$
for every $o_i \in I, o_j \notin I$, and $\rho(o_i, p) = \rho(o_j, p)$
for every $o_i, o_j \in I$.
Likewise, the \emph{farthest Voronoi diagram} is the partition of the
space into maximally connected cells, where each cell consists of
points that are farther from one site than from the other sites.

All Voronoi diagrams can be defined by adjusting the parameters $O$,
$\rho$, and \calS{} as required.
Defining the standard Voronoi diagram, for example, amounts to the
selection of a set of points in the plane as the set of sites, the
selection of the plane itself as the ambient space, and the selection
of the Euclidean distance between two points in the plane as the
distance function.
The power diagram on the sphere, yet another example, is defined by
choosing $O$ to be a set of circles on the unit sphere, \calS{} to be the
sphere, and $\rho$ to be the spherical power distance.

In certain cases, the distance to a site may depend on
various parameters associated with the site.
For example, \mobius{} diagrams' distance depends on two positive
scalars, and anisotropic Voronoi diagrams' distance depends on
one positive definite matrix and one positive scalar.
Table~\ref{tab:supp-vd} lists some of the more prevalent
two-dimensional Voronoi diagrams together with their respective types
of sites, ambient spaces, and distance functions.

The \emph{bisector} $B(o_i, o_j)$ of two Voronoi sites $o_i$ and $o_j$ is
the locus of points that have an equal distance to both sites, that is 
\begin{displaymath}
  B(o_i, o_j) = \{p \in \calS \mid \rho(o_i, p) = \rho(o_j, p)\}.
\end{displaymath}

From here on we refer only to ambient spaces that are two-dimensional,
parameterizable, and orientable (\eg a plane, a sphere, a torus, etc.)

The above definition is the more classical definition conceived from
the need to divide the space into areas of influence or dominance.
The sites are, of course, the dominating entities and the distance
functions correspond to the measure of dominance
of each site on points of the ambient space.
In addition to this application-inspired definition an alternative
implementation-oriented definition arose.

Considering the plane as the ambient space, Voronoi diagrams were
defined through their bisecting curves instead of the distance
function.
Such diagrams that also comply to additional restrictions are referred to
as \emph{abstract Voronoi diagrams}~\cite{k-cavd-89}.
In this definition, a set $B$ is called a bisecting curve if, and only
if, $B$ is homeomorphic to the open interval $(0, 1)$ and closed as a
subset of \R{2}.
A bisecting curve $B$ partitions the plane into two unbounded
areas.
For each pair of sites $o_i$ and $o_j$ we assume that 
$B(o_i, o_j) = B(o_j, o_i)$ is a bisecting curve, and denote by
$D(o_i, o_j)$ and $D(o_j, o_i)$ the two areas obtained in the
partition induced by the bisecting curve. (One of the 
areas is known to be $D(o_i, o_j)$ and one is known to be $D(o_j,
o_i)$.) The Voronoi diagram \vornr{O} is defined as follows:
\begin{eqnarray}
  \region{}(o_i, o_j) & = & \left\{ \begin{array}{ll}
    D(o_i, o_j) \cup B(o_i, o_j), & \text{if $i < j$}\\
    D(o_i, o_j) & \text{if $i > j$}
  \end{array} \right.
  \\
  \region{}(o_i, O) & = & \bigcap_{o_j \in O, j \neq i}
  \region{}(o_i, o_j)
  \\
  \vornr{O} & = & \bigcup_{o_i \in O}\partial\region{}(o_i, O)
\end{eqnarray}

Abstract Voronoi diagrams do not cover the entire variety of Voronoi
diagrams discussed in this thesis. For example, \mobius diagrams and
anisotropic diagrams are \emph{not} abstract Voronoi diagrams.
In addition, the classical definition of abstract Voronoi diagrams
does not include the cases of different ambient spaces, such as
two-dimensional parametric surfaces.

\section{Divide-and-Conquer Algorithm for Envelopes}
\label{sec:background:envelopes}
Given a set of bivariate functions (partially) defined over a
two-dimensional domain~\calS{},
$\calF{} = \{f_1, \ldots, f_n\}$,
$f_i:~\calS{}~\to~\R{}$, its
\emph{lower envelope} $\env{\calF}:~\calS{}~\to~\R{}$ is defined to be
their point-wise minimum:
\begin{displaymath} %
  \env{\calF}(x) = \min_{i}f_i(x).
\end{displaymath}

The \emph{minimization diagram} \mindiag{\calF} of \calF{} is the
subdivision of~\calS{} into maximal relatively-open connected cells,
such that the function (or the set of functions) that attains the
lower envelope over a specific cell of the subdivision is the same.
Changing ``$\min$'' to ``$\max$'' in the definition above
results in the corresponding definitions for \emph{upper envelope} and 
\emph{maximization diagram}.

Agarwal, Schwarzkopf, and Sharir presented a simple and efficient
divide-and-conquer algorithm for the construction of envelopes of
bivariate functions defined over the plane~\cite{ass-olea-96}.
The algorithm is essentially an application of the overlay of
two-dimensional minimization diagrams.
They showed that the combinatorial complexity of such an overlay of
two envelopes of $n$ ``well-behaved'' functions is
$O(n^{2+\varepsilon})$; see~\cite{as-ata-00} for a full description of
the assumptions on such functions.
An alternative proof was given by Koltun and Sharir~\cite{ks-ptoe-03}.
As acceptable in the field of computational geometry they assumed
that the input is given in general position.
This assumption creates a gap between the theoretical
divide-and-conquer algorithm for constructing envelopes in
three-dimensions and its practical anticipated implementation.

Meyerovitch presented an implementation for the above algorithm for
constructing envelopes of functions defined over the
plane~\cite{m-rgeces-06,m-rgeces-thesis-06}, which handles all
inputs, including all degenerate situations; see more details in
Section~\ref{sec:background:cgal} below.
Following is a description of the algorithm in the context of
lower envelopes construction.
The description is easily adapted to the case of upper envelopes as well.
The input for the algorithm is a set \calF{} of bivariate functions,
which could be partially defined, and the result is the minimization
diagram \mindiag{\calF} of the set of functions.

The envelope of a single function comprises the function itself.
Hence, the minimization diagram is constructed by projecting
the boundary of the domain of the function onto~\R{2}.
In the case where the set of functions consists of more than one function
we partition the set into two sets of functions
of roughly equal size $\calF{}_1$ and $\calF{}_2$, and compute their respective
minimization diagrams \mindiag{1} and \mindiag{2}, recursively. 
Every feature --- a vertex, an edge, or a face --- of \mindiag{1} and
\mindiag{2} is labeled with the set of functions that attain the lower
envelope over it.
We merge \mindiag{1} and \mindiag{2} into the final minimization
diagram~\mindiag{\calF}.
The merge operation is composed of three stages:
\begin{enumerate}

\item {\bf Overlaying \mindiag{1} and \mindiag{2}}
  each represented as a planar arrangement to obtain a new planar
  arrangement.
  We use a sweep-based overlay algorithm, during the execution of
  which new features are created depending on existing features of
  both input diagrams;
  for example, a new vertex is created by overlaying two
  intersecting edges from \mindiag{1} and \mindiag{2}, respectively.
  The newly created features are labeled with references to the
  functions attaining the lower envelopes over both input minimization
  diagrams.

\item {\bf Constructing the minimization diagram over every feature}
  (the vertices, the edges, and the faces) of the overlay.

  If the overlaid features from both minimization diagrams that induce
  the subject feature reference no functions then the feature is
  labeled with the empty set.
  If only one feature references a set of functions then the subject
  feature is labeled with this set of functions.

  The case where \emph{both} input features reference sets of functions
  makes our task a non-trivial one, as we may have to split the
  feature into several pieces 
  (if it is an edge or a face).
  All referenced functions originating from a single minimization diagram
  are identified over the subject feature, so we can consider one
  representative function from each of the diagrams (\mindiag{1} and
  \mindiag{2}).
  The projected intersection between these two functions 
  induces the split of the feature into fragments.
  Each of the resulting fragments is then labeled with the correct set
  of functions according to the value obtained by the functions over
  it (the set of functions originating from \mindiag{1} or \mindiag{2}
  in case one set obtains a lower value, or the union of both sets in
  case they obtain an equal value).

\item {\bf Removing redundant features}.
  Neighboring faces in the refined overlay and their connecting edges can be
  labeled with identical sets of functions.
  If this is the case we compute the union of the redundant faces by
  removing edges and vertices, which yields the final minimization
  diagram.
\end{enumerate}

Assuming that all functions are ``well-behaved''
(see, \eg~\cite{as-ata-00}), the complexity of
the algorithm is dependant on the complexity of the overlay of the two
minimization diagrams. Therefore, the theoretical worst-case
time-complexity for constructing envelopes using the
divide-and-conquer algorithm is $O(n^{2+\varepsilon})$.
The actual implementation in~\cgal, presented in
Section~\ref{sec:background:cgal-envelopes}, contains speed-ups that
expedite the practical running time.

\section{CGAL and the \texttt{Arrangement\_on\_surface\_2} Package}
\label{sec:background:cgal}
Providing robust, efficient, and general implementations of
computational geometry algorithms is a notoriously difficult task.
Two prime issues bring up the majority of difficulties.

The first is the general hardship of implementing geometric algorithms
while considering all kinds of degenerate input and boundary situations.
Assuming that the given input is in general position is used to avoid
these marginal cases in many geometric algorithms in theory.
The assumption of general position suggests that special or
``coincidental'' inputs be discarded;
for example, three lines in the plane are assumed not to intersect at
a single point.
This discards many cases that appear in practical applications and
real-world problems, and creates a large gap between
computational-geometry algorithms in theory and their
implementation.

The second issue is related to robustness and rounding
errors. Geometric algorithms in theory generally follow a
computational
model named ``real RAM''~\cite{cg-ps-85}. They assume that all
numerical computations are performed with unlimited precision, and
require constant time per operation.
In reality this model cannot be realized. Numbers represented by
machines either have limited precision, or require more than constant
time per operation, as native types comprise a fixed number of bits.
Inaccurate numerics can impose inconsistent predicates and
constructions, forming unstable geometric
algorithms~\cite{kmpsy-cerpg-08}.
Developing a robust and efficient geometric algorithm under these
constraints is a very challenging task even for a qualified
professional.

\cgal{}, the Computational Geometry Algorithms Library, was launched
in 1996 as a collaborative effort of several research
institutes in Europe and Israel to provide easy access to efficient,
robust, and reliable geometric algorithms and data structures for
academic and industrial use.

\cgal{} provides various geometric data structures and algorithms like
convex hull algorithms, Delaunay triangulations, Voronoi diagrams,
Boolean operations on polygons and polyhedra, arrangements of curves,
Minkowski sums of polygons, alpha shapes, search structures, and
more~\cite{cgal:eb-08}\citelinks{cgal-link}.
\cgal{} is used in various fields in academia and industry, such as
computer graphics, scientific visualization, computer-aided
design, bioinformatics, motion planning, and more.

\cgal{} overcomes the above difficulties by adhering to the exact
geometric computation paradigm~\cite{yd-ecp-95}, and by relying on
computation with exact number types to achieve robustness.
\cgal{} adheres to the generic programming paradigm (see below) to
achieve maximum flexibility without compromising efficiency.

Generic programming is a programming discipline in which concrete
algorithms are gradually lifted over specific required types by
describing them in terms of polymorphic abstract
types~\cite{a-gps-99}.
The types are provided later at instantiation time of the algorithm as
parameters.
This approach empowers the programmer with the ability to write
dynamic and general programs on abstract types, at the expense of code
tangibility.
Collections of requirements from abstract polymorphic types are
referred to as \emph{concepts}, and specific types used to instantiate
the algorithms are referred to as \emph{models}.

Templates (or \emph{template programming}) are \cpp{} language
constructs that were designed to support the generic programming paradigm.
They have been found to be extremely useful, providing \cpp{}
programmers with the ability to write static code-generators and
perform static computations (meta-programming), %
which improves the run-time of non-templated \cpp{} compiled code
while maintaining and even enhancing flexibility.
In fact, \cpp{} can be regarded as a two-level language where
each level is Turing-complete;
the first level is the code-generating statically-expanding compile-time
consuming template declarations and meta-programs, and the second
level is the ``standard'' non-templated runtime-consuming code.
For an in-depth discussion of \cpp{} templates we refer the reader
to ``C++ Templates'' by~Vandevoorde and~Josuttis~\cite{vj-ct-02} and
to ``Modern C++ Design'' by~Alexandrescu~\cite{a-mcd-01}.

\cgal{} is divided into packages, where each package provides an
efficient implementation of a geometric algorithm (or a family of
algorithms) or a useful data-structure, and related functionalities.
\cgal{} packages are organized in three parts.
Geometric data-structures and algorithms, as just mentioned, are one
part.
These data structures and algorithms operate on geometric objects
like points and segments, and perform geometric tests on them.
The objects and predicates are regrouped in \cgal{} ``Kernels'',
which constitute another part of \cgal{}.
The third part of \cgal{} is the ``Support Library'' which offers
fundamental utilities used throughout \cgal{}; for example, various
extensions for the \stl~\citelinks{stl-link},
\boost~\citelinks{boost-link} and QT libraries.

The support library also contains classes that represent numbers,
referred to as ``number types,'' which are used as parameters to \cgal
kernel classes.
Depending on the problem (and the input) to be handled, the number
types provide a trade-off between efficiency and accuracy.
For example, in some cases the user is able to instantiate a
geometric kernel with a built-in number-type of \cpp that represent a
discrete (bounded) subset of the rational numbers, while in other
cases he/she has to use a number type that supports all operations in
unlimited precision over the rationals, such as the rational number
type {\tt CGAL::Gmpq} based on \gmp --- Gnu's Multi Precision
library~\citelinks{gmp-link}.

A leading package of \cgal{} --- both in terms of size (lines of code)
and the extent of usage --- is the \aos{} package.
Given a set \calC{} of curves embedded on a given two-dimensional
surface, the arrangement \calA{}(\calC{}) is the subdivision of the
surface into cells, induced by the curves of \calC{}.
Arrangements are defined more generally~\cite{bfhmw-smtdas-07}.
However, we restrict ourselves here to 2D arrangements, which are
supported by \cgal{}.
The \aos package is based on the earlier \arrangement package that
supported planar arrangements of bounded and unbounded
curves~\cite{fwh-cfpeg-04,wfzh-aptaca-07}.
The \aos package enables the user to construct and maintain
arrangements embedded on certain two-dimensional orientable parametric
surfaces~\cite{bfhmw-smtdas-07}.
In addition to the ability to construct arrangements, the package
supports various operations on arrangements, including traversing an
arrangement, performing point-location queries on an arrangement, and
overlaying two arrangements~\cite{cgal:wfzh-a2-08}.

The arrangement package of \cgal{} achieves robustness and exact
results when using exact arithmetic types, and handles all kinds of
degenerate situations. 
It supports two algorithmic frameworks, that is the sweep-line
framework and the zone-computation framework.
Both are used in various geometric algorithms. For example, the former
is used for Boolean-set operations between linear
or general polygons in the plane, and the latter is used for inserting
a curve into an existing arrangement of curves.

The arrangement package follows the generic programing paradigm
through the use of \emph{traits} classes, which enable the separation
of the geometric and the topological aspects of the computation.
Models that describe behaviors are referred to as traits
classes~\cite{m-tnutt-97}.
A traits class is passed as a parameter to a templated method or a
class template and should provide certain predefined types and
methods that enable the operation of a specific algorithm.
For example, in our case, a geometry traits class for the \aos{}
class (see below) should contain types that represent points and
$x$-monotone curves, a method to compare the $x$-coordinates of two
points, etc.
This decouples the implementation of the algorithms contained in the
\aos{} package from the specific geometric computations, and
enables the user of the package to create different types of
arrangements for different classes of curves.

The main class of the package --- \aos{} --- is parameterized with two
template classes, a \emph{geometry-traits} class and a
\emph{topology-traits} class.
The geometry-traits class controls the geometric aspects of the
arrangement, namely, it defines associated geometric types (points,
curves, and $x$-monotone curves) for the specific family of curves and
provides the arrangement with required geometric operations and
predicates on those types (\eg intersecting two $x$-monotone curves,
determining whether a point lies below or above a curve, etc.).
The topology-traits class, as its name suggests, is responsible for the
topological (abstract, graph-like) representation of the arrangement,
namely, keeping the correct relations between the arrangement's cells
(faces, edges, and vertices) and their neighboring cells with respect
to the embedding surface.

The package is designed for maximum efficiency and flexibility, where
flexibility refers to both adaptability and extensibility.
In other words, the arrangement package was designed to have the
ability to be incorporated into existing user code and the ability to
be enhanced with additional code.

\section{Exact Construction of Envelopes in CGAL}
\label{sec:background:cgal-envelopes}
The \envelope{} package, which implements the algorithm mentioned in
Section~\ref{sec:background:envelopes} for computing the lower (or the
upper) envelope of a set of surfaces in
three-dimensions~\cite{cgal:mwz-e3-08}, is strongly built-upon the
\arrangement{} package.
There is one difference between the description of the algorithm in
Section~\ref{sec:background:envelopes} and the implementation of
the \envelope{} package, that is, the support of \envelope{} in
constructing the envelope of general three-dimensional surfaces by the
decomposition of the surfaces into bivariate (partially defined)
functions.
The package decouples the topology-related computation from the
geometry-related computation, making its code generic and easy to
reuse and adapt.

The code insures stability, namely, it handles all possible
degenerate situations in the case of general surfaces;
among those are vertical surfaces, overlapping (and partially
overlapping) surfaces, and a common intersection point of
more than three surfaces.
While insuring stability, the number of calls to the exact (and slow)
geometric predicates is minimized by propagating pre-computed
information about the structure of the envelope to neighboring cells
in the merge step of the algorithm.

The \envelope{} package defines a new concept for a traits class for
computing envelopes of surfaces in three-dimensions.
The \envelopetraits refines the \arrtraits concept, adding types and
predicates that are used to compute envelopes.
Every model of the \envelopetraits has to supply the
ability of constructing an arrangement from the projected intersection
curves and boundary curves of the given surfaces.
Available traits classes for the \envelope{} package include traits
classes for computing the envelopes of triangles, spheres, planes, and
quadrics~\cite{bm-ceq-07,m-rgeces-06}.

The implementation mainly makes use of two operations supported by the
arrangement package:
\begin{inparaenum}[(i)]
\item sweep-based overlay operation, which is used to overlay two minimization
  diagrams, and
\item zone computation-based insertion operation, which is used to
  insert projected intersection curves (of the surfaces)
  that partition cells of the refined arrangement.
\end{inparaenum}
The new \aos{} package extends the aforementioned operations, that is,
the sweep-line and zone-computation, to support
arrangements on two-dimensional parametric surfaces.
Thus, we extended the \envelope{} code to work together with the new
\aos{} package, and handle minimization diagrams that are embedded on
two-dimensional parametric surfaces.

Though computing lower envelopes of functions defined over
two-dimensional orientable parametric surfaces has its own
significance, we concentrate on describing how this ability is
exploited to compute Voronoi diagrams on surfaces;
see Section~\ref{sec:env-to-vd:impl} for more details and 
Section~\ref{sec:impl:sphere-vd} for a concrete example of computing
Voronoi diagrams on the sphere.

\chapter{From Envelopes to Voronoi Diagrams}
\label{chap:env-to-vd}
This chapter describes the adaptation of the algorithm for
constructing envelopes
of bivariate functions to the computation of general Voronoi diagrams.
Section~\ref{sec:env-to-vd:dc-vd} describes the way the algorithm works in
the context of Voronoi diagrams.
Section~\ref{sec:env-to-vd:theo} discusses theoretical aspects of the
algorithm. It shows that through randomization, the expected running
time of the algorithm is near-optimal in the worst case.
Section~\ref{sec:env-to-vd:impl} provides comprehensive
details on the software interface that allows the computation of general 
two-dimensional Voronoi diagrams through the construction of envelopes in
\cgal.

\section{Divide-and-Conquer Algorithm for Voronoi Diagrams}
\label{sec:env-to-vd:dc-vd}

There is a strong connection between Voronoi diagrams in
$d$-dimensions and envelopes of functions in $(d+1)$-dimensions,
which was first observed by Edelsbrunner and Seidel~\cite{es-vda-86}.
Their main revelation was the relation between Euclidean Voronoi
diagrams of point-sets in~\R{d} (and their higher-order Voronoi
diagrams) to arrangements of hyperplanes in \R{d+1}, which yielded a
very general approach for computing various types of Voronoi
diagrams.

We demonstrate the connection between Voronoi diagrams and envelopes
in the planar case.
This connection can be easily established also for
two-dimensional parametric surfaces as the embedding spaces.
Let $O = \{o_1, \ldots, o_n\}$ be a set of $n$ Voronoi sites in the plane,
and let $\rho : O \times \R{2} \to \R{}$ be a distance function between
Voronoi sites and points in the plane.
Recalling the definitions of Voronoi diagrams and envelopes
from Section~\ref{sec:background:vd} and
Section~\ref{sec:background:envelopes}, it is clear that if we define 
$f_i: \R{2} \to \R{}$ to be $f_i(x) = \rho(o_i, x)$, for each 
$i = 1, \ldots, n$, then the minimization diagram of $\{f_1, \ldots,
f_n\}$ corresponds to the Voronoi diagram of $O$.
Likewise, the respective maximization diagram corresponds to the
farthest Voronoi diagram of $O$.

As we aspire to use envelopes to compute Voronoi diagrams, we
``translate'' below the terms used in envelopes construction to terms 
used in Voronoi diagrams computation.
This translation will be useful in Section~\ref{sec:env-to-vd:impl}
where we define the interface for computing Voronoi diagrams.
The interface is simpler than the given \cgal interface for computing 
envelopes.
This allows the user of our framework to create a new type of Voronoi
diagrams by providing certain functions and types without the
knowledge of the underlying envelope algorithm.

Each Voronoi site is transformed into a bivariate function defined over
the whole two-dimensional domain, as opposed to envelopes of general surfaces
that can be partially defined.
The bisector of two Voronoi sites is the locus of points that have an
equal distance to both sites (see Section~\ref{sec:background:vd}),
thus, the projection of the intersection between two functions that
correspond to two Voronoi sites is the bisector of the two sites.

\begin{figure} [h] %
  \centering
  \setlength{\spacewidth}{35pt}
  \newlength{\figwidth}\setlength{\figwidth}{120pt}
  
  \subfigure[]{\includegraphics[width=\figwidth]{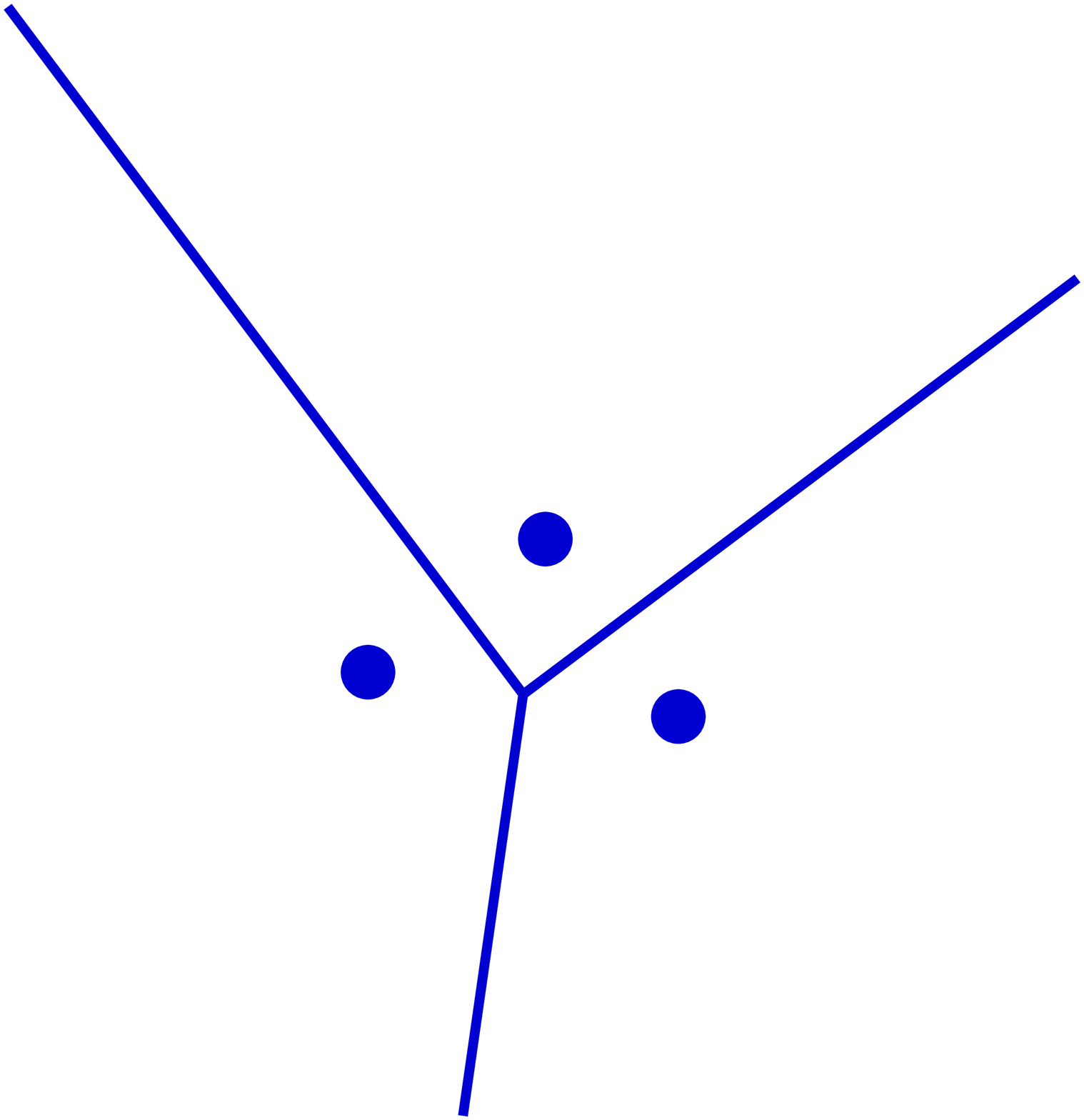}}
  \hspace{\spacewidth}
  \subfigure[]{\includegraphics[width=\figwidth]{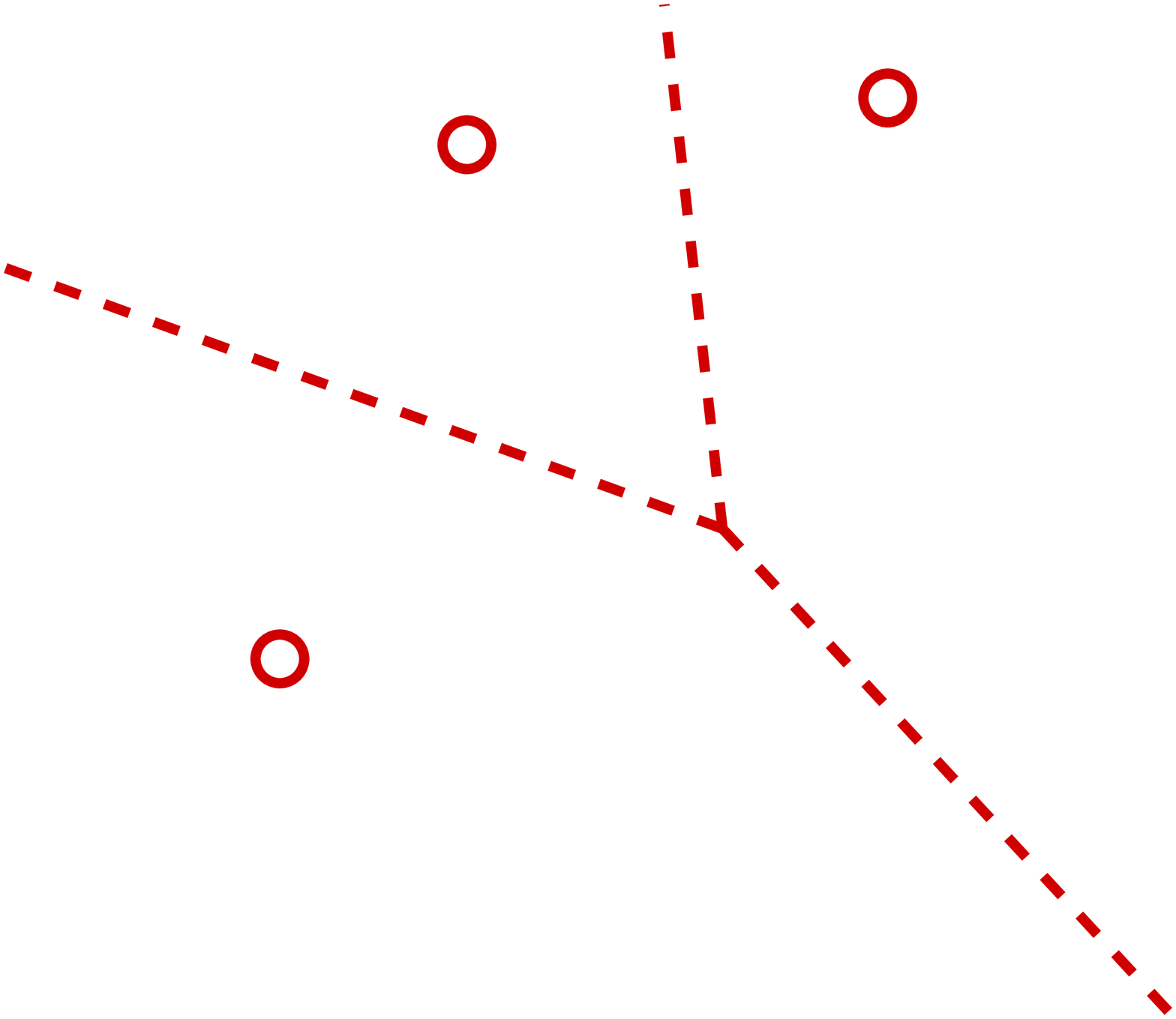}}
  \hspace{\spacewidth}
  \subfigure[]{\includegraphics[width=\figwidth]{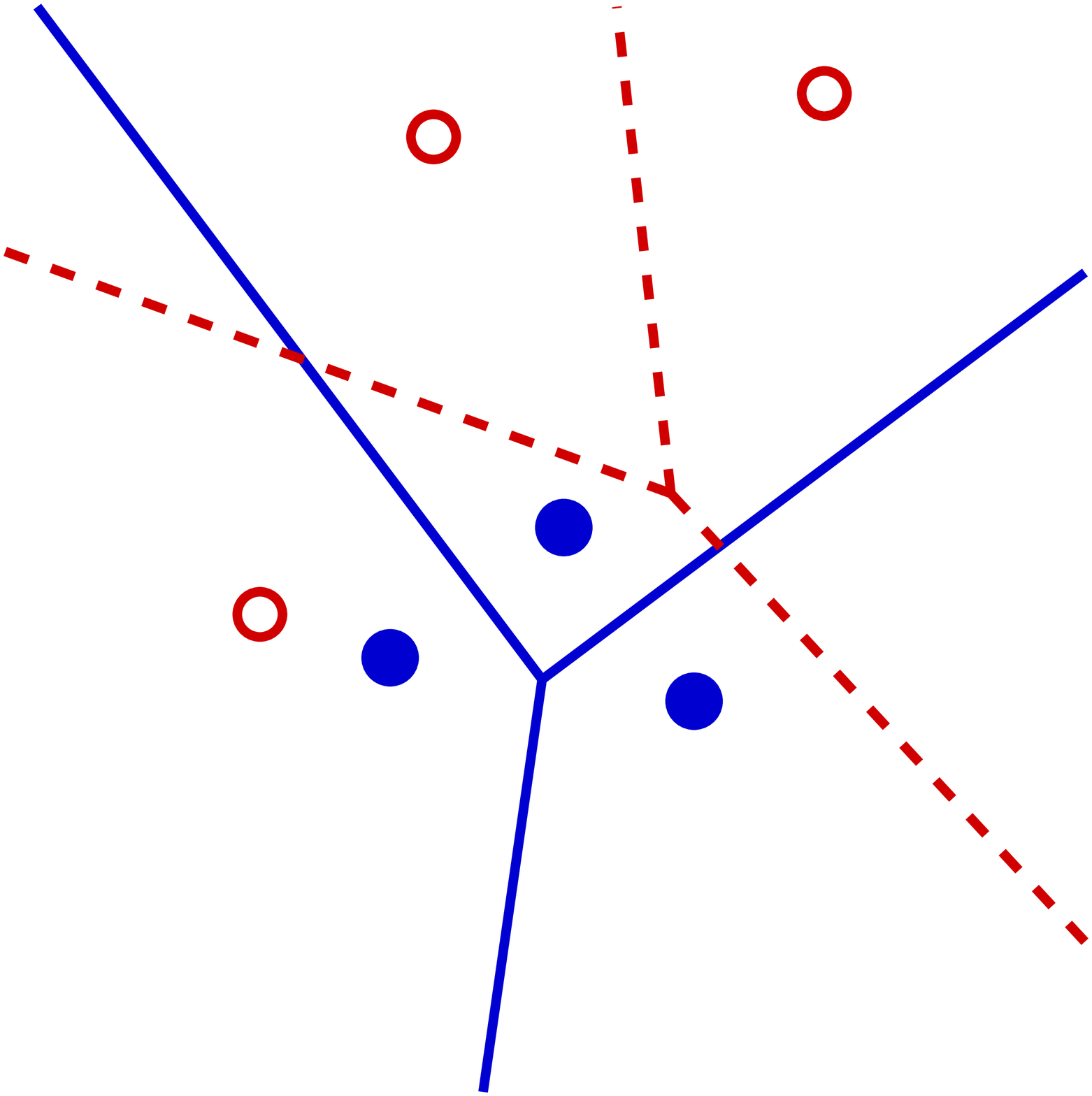}}
  
  \hspace{\spacewidth}
  \subfigure[]{\includegraphics[width=\figwidth]{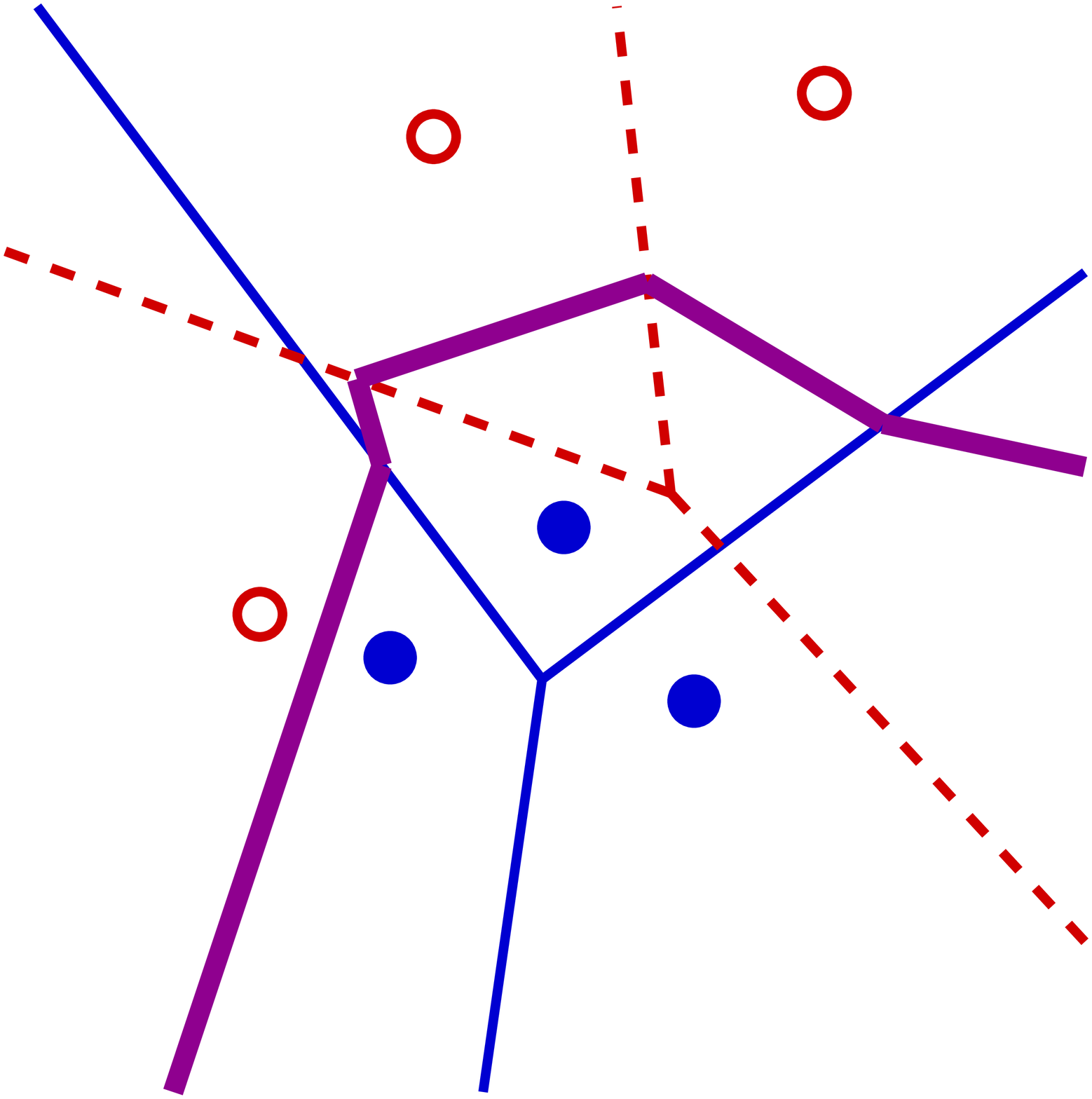}}
  \hspace{\spacewidth}
  \subfigure[]{\includegraphics[width=\figwidth]{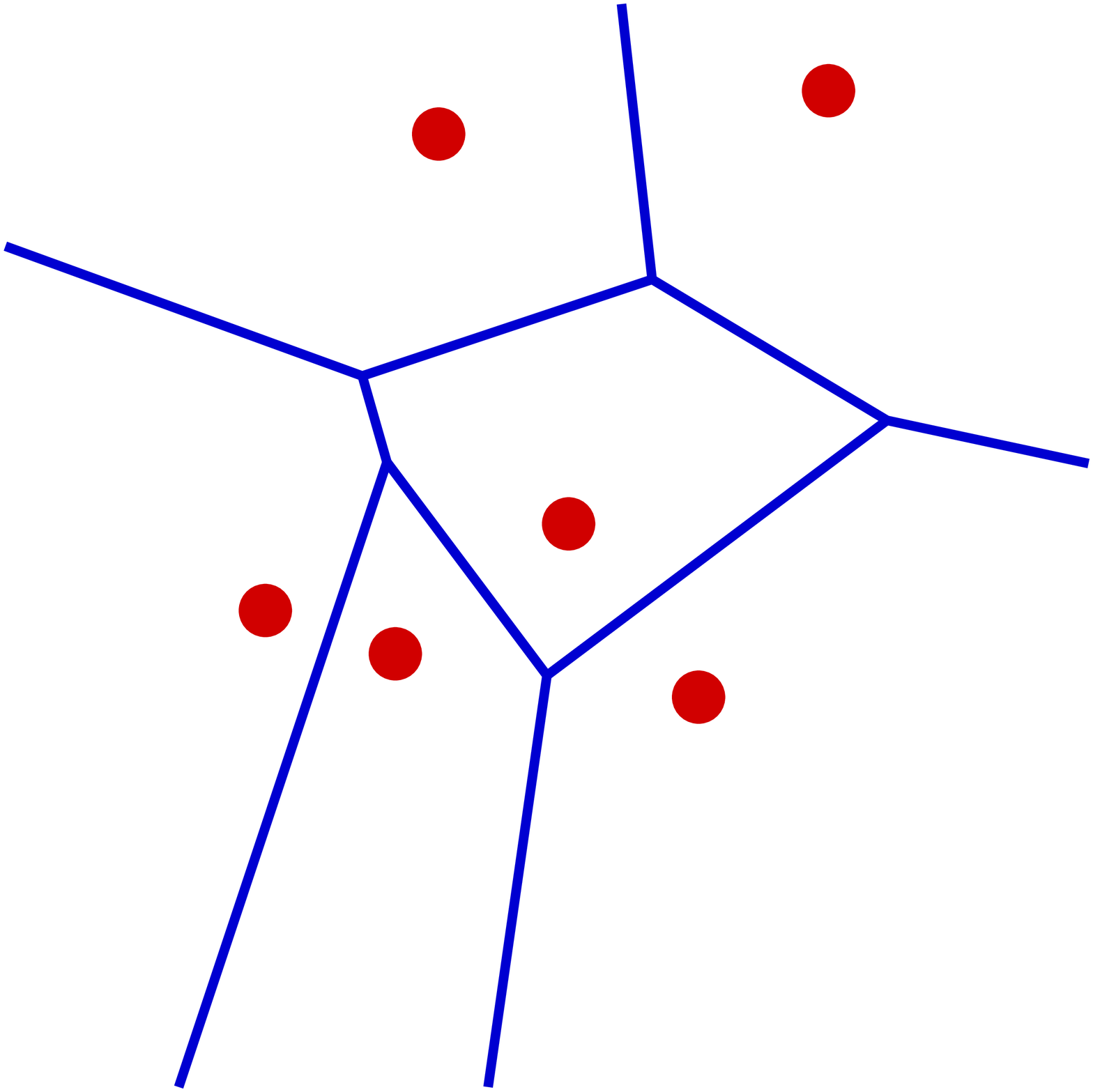}}
  
  \vspace{-10pt}
  \caption[Merging two Voronoi diagrams.]{\capStyle{The merge step of the 
      divide-and-conquer algorithm for computing Voronoi diagrams.
      (a)~The first Voronoi diagram, \vor{S_1}.
      (b) The second Voronoi diagram, \vor{S_2}.
      (c) The overlay of the two diagrams.
      (d) The refined overlay.
          Each face is partitioned to regions that are closer to sites 
          from $S_1$ and region that are closer to sites from $S_2$.
      (e) The final diagram obtained after the removal of redundant features
          from the refined overlay.}}
  
  \label{fig:env_to_vd:merge}
\end{figure} 

The adapted algorithm for Voronoi diagrams computation follows.
We split the set of sites into two disjoint subsets $S_1$ and
$S_2$ of (roughly) equal size, construct their respective Voronoi
diagrams \vor{S_1} and \vor{S_2}, recursively, and then merge the two
diagrams to obtain \vor{S}.

The merge step begins with overlaying the two diagrams. 
For each face $f$ of the overlay, all
its points have a fixed pair of nearest sites $s_1$ and $s_2$ from
$S_1$ and $S_2$, respectively, where the bisector between $s_1$ and
$s_2$ (restricted to $f$) partitions $f$ into its portion of points
nearer to $s_1$ and the complementary portion of points nearer to
$s_2$. This results with portions of the final Voronoi cells. Each
feature of the refined overlay is labeled with the sites nearest to
it.

Finally, redundant features are removed and subcells of the same cell
are stitched together, to yield the combined final diagram.
Figure~\ref{fig:env_to_vd:merge} illustrates the process of merging
two Voronoi diagrams of points (a red one and a blue one), to yield
the final Voronoi diagram of the unified set of points.

\section{Theoretical Aspects}
\label{sec:env-to-vd:theo}

\begin{figure}
  \begin{center}
    \includegraphics[width=\textwidth]{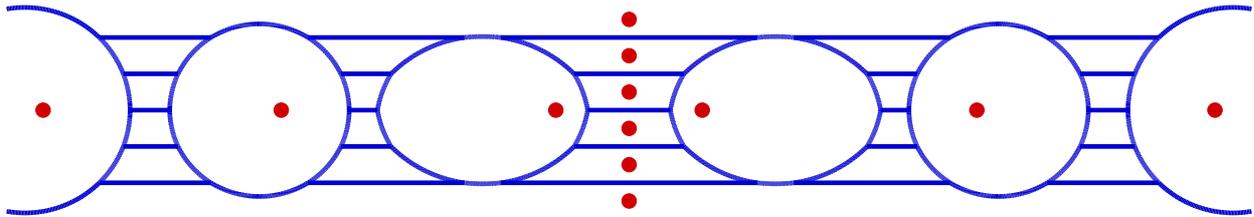}
  \end{center}
  \caption[Worst-case quadratic multiplicatively-weighted Voronoi
    diagram.]{\capStyle{A worst-case quadratic-size example of a
      multiplicatively-weighted Voronoi diagram with 12 sites, based
      on an example of Aurenhammer and Edelsbrunner~\cite{ae-oacwvd-84}.
      The diagram was computed with our software.}}
  \label{fig:env-to-vd:mobius-worst}
\end{figure}

Recall that the asymptotic worst-case time complexity of the
divide-and-conquer envelope algorithm (under the natural assumption
that the functions have ``constant description
complexity'') is $O(n^{2+\varepsilon})$, for any $\varepsilon > 0$.
Indeed, there are planar Voronoi diagrams that
obtain quadratic complexity, and for which this
construction is nearly worst-case optimal.
For example, Figure~\ref{fig:env-to-vd:mobius-worst} illustrates the
worst-case behavior of multiplicatively-weighted Voronoi diagrams,
based on an example by Aurenhammer and
Edelsbrunner~\cite{ae-oacwvd-84}.
However, for cases where the complexity of the diagram is sub-quadratic
(for most of the cases, linear), we would like the algorithm to obtain a
sub-quadratic (or near-linear) running time.

The complexity of the merge step of the algorithm 
directly depends on the complexity of the overlay of the two
sub-diagrams.
(The cost of the best general algorithm for constructing the overlay
is larger by a logarithmic factor than the combined complexity of the
input diagrams and of the overlay.)
Careless partition of the input sites into two subsets can
dramatically slow down the computation. For example, consider the
following point-set input to the standard $L_2$-diagram in the plane,
\daniex. If we partition the set into two subsets, to the left and to
the right of the $y$-axis, then in the final merge step, the overlay of
the two sub-diagrams has $\Theta(n^2)$ complexity. Hence the algorithm
runs in $\Omega(n^2)$ time, whereas the complexity of the final diagram
is only $\Theta(n)$; see Figure~\ref{fig:exper:randomization} for an
illustration and Section~\ref{sec:advantages:optimality} for more
details.

Micha Sharir has shown that if the partitioning of the sites into two
subsets is done \emph{randomly}, then the expected complexity of the
overlay is comparable with the maximum complexity of the diagram for
essentially any kind of sites and distance functions, and for any
possible input.
Sharir's proof is given in Section~\ref{sec:micha-proof}. Here we cite
the theorem and point to relevant consequences of it.

\begin{theorem}
  \label{theo:env-to-vd:random}
  Consider a specific type of two-dimensional Voronoi diagrams, so that
  the worst-case complexity of the diagram of any set of at most $n$
  sites is $F(n)$. 
  Let $S$ be a set of $n$ sites. If we randomly split $S$
  into two subsets $S_1$ and $S_2$, by choosing at random for each
  site, with equal probability, the subset it belongs to, then the expected
  complexity of the overlay of the Voronoi diagram of $S_1$ with the
  Voronoi diagram of $S_2$ is $O(F(n))$.
\end{theorem}

As we aim to compute Voronoi diagrams of a large variety of types, we use
a sweep-line based algorithm that exhibits good practical performance,
and incurs a mere logarithmic factor over the optimal computing time,
namely $O(F(n)\log n)$.
In particular we use the overlay operation provided by the \aos
package.

\begin{corollary}
  \label{cor:env-to-vd:exp-comp}
  For a specific type of two-dimensional Voronoi diagrams, so that the
  worst-case complexity of the diagram of any set of at most $n$~sites
  is~$O(n)$, the divide-and-conquer envelope algorithm computes it in
  expected $O(n\log^2n)$~time.
  If the worst-case complexity~$F(n)$ is $\Omega(n^{1 + \varepsilon})$
  then the expected running time is $O(F(n) \log n)$. 
\end{corollary}

When the diagram is a convex subdivision, one can carry out the merge
step more efficiently, in linear $O(F(n))$ expected time using the
procedure described by Guibas and Seidel~\cite{gs-ccrs-87}. In
particular, we have:
\begin{corollary}
  The $L_2$-Voronoi diagram of $n$ points in the plane, or the power
  diagram of $n$ disks in the plane, can be computed
  using the randomized divide-and-conquer envelope algorithm in
  expected optimal $O(n \log n)$ time.
\end{corollary}

\section{Robust Implementation with CGAL}
\label{sec:env-to-vd:impl}
This section describes the software interface between the
computation of Voronoi diagrams and the construction of
envelopes.
The reduced and convenient interface consists of several functions,
each operating on a small number of user-defined types
(Voronoi sites or bisector curves).
A user wishing to add a new type of diagrams does not have to know the
algorithmic details of constructing minimization diagrams.
The section contains technical details on 
the types and functions that need to be supplied by the user of the
framework in order to implement a new type of Voronoi diagrams.
We assume in this section, as well as in Chapter~\ref{chap:impl}
below, some familiarity of the reader with the \cpp 
programming language~\cite{s-cpl-97} and the generic programming
paradigm~\cite{a-gps-99}.

The \envelope{} package of \cgal is general and handles surfaces having
two-dimensional intersections, hence we can compute Voronoi
diagrams composed of two-dimensional bisectors.\footnote{An example of
  such a diagram is the Voronoi diagram of points with respect to the
  $L_1$-metric, in which two point sites can have a two-dimensional
  bisector.}
In the case of two-dimensional bisectors, the merge step of the
algorithm is almost identical to the case of one-dimensional
bisectors, only that each face of the overlay can be split and labeled
with multiple Voronoi sites.
For the computation of Voronoi diagrams having two-dimensional
bisectors we advise the user to directly use the \envelope
package~\cite{cgal:mwz-e3-08}.
(Section~\ref{ssec:impl:plane-vd:segs} gives a detailed example of
implementing a Voronoi diagram with two-dimensional bisectors.) 

Nevertheless, most types of Voronoi diagrams have only one-dimensional
bisectors, and are generally simpler than envelopes of general
functions. For example, abstract Voronoi diagrams
(Section~\ref{sec:background:vd}) require that each side of a bisector
will be dominated by one of the sites.
Given a dominant site on one side of the bisector, the other site is
the dominant site on the other side.
We reduced and simplified the interface required for the
implementation of new types of Voronoi diagrams with one-dimensional
bisectors.
We require the user of our framework to define a set of geometric
types and operations that will be used by the algorithm. This way
the user can adapt the algorithm to compute the desired type of
Voronoi diagrams. 
Our algorithm is parameterized with a \emph{traits class}~\cite{m-tnutt-97}.
A traits class should provide certain predefined types and methods,
and is passed as a parameter to a class template. In our case, the
algorithm is the class template, which accepts a traits class that
encapsulates the geometric types and the geometric operations that the
algorithm requires.
The concept of the traits class for our
Voronoi diagrams construction algorithm is called \vorconcept{}.
Table~\ref{tab:env-to-vd:concept} summarizes the requirements of
\vorconcept for types and function objects (also referred to as
  ``functors'').

The first step in creating a new type of Voronoi diagrams is to define
the embedding surface of the diagram.
The creator of the traits class picks the embedding surface of the Voronoi
diagram by defining the \emph{Topology\_traits} type, which
encapsulates the topology of the surface on which the diagram is
embedded.
The topology traits class is a standard requirement by the \aos
package~\cite{bfhmw-smtdas-07}.
Currently implemented topology-traits classes in the \aos{} package
include topology traits classes for the bounded or unbounded plane,
for elliptic quadrics, for ring Dupin cyclides that generalize tori,
and a specially tailored topology-traits class for the
sphere~\cite{fsh-eiaga-08}.

\begin{table}
  \caption[Required types and functors by the \vorconcept{}
    concept.]{Required types and functors by the \vorconcept{}
    concept.  
    \ccode{Comparison\_result} is an existing \cgal type.}
  \label{tab:env-to-vd:concept}

  \vspace{10pt}
  \begin{center}
    \begin{sideways}
      \small
      \begin{tabularx}{\textheight - 1.5cm}{|p{5.2cm}|p{4cm}|p{4cm}|X|}
        \hline
        \multicolumn{1}{|c|}{\emph{Name}} & 
        \multicolumn{1}{c|}{\emph{Input}} &
        \multicolumn{1}{c|}{\emph{Output}} &    
        \multicolumn{1}{c|}{\emph{Description}} \\ \hline \hline

        \site{} & --- & ---
        & A type that represents a Voronoi site. \\ \hline

        \ccode{Construct\_bisector\_2} & Two \site{} objects &
        An output iterator with values of type of \ccode{X\_monotone\_curve\_2}
        &
        Returns \ccode{X\_monotone\_curve\_2} objects that together form
        the bisector of the two input sites. \\ \hline

        \ccode{Compare\_distance\_above\_2} & 
        Two \site{} objects and an \ccode{X\_monotone\_curve\_2} &
        \ccode{Comparison\_result}
        &
        Determines which of the given Voronoi sites is closer to the
        area ``above'' the given $x$-monotone curve, where ``above'' is
        the area that lies to its left when the curve is traversed from
        its $xy$-lexicographically smaller end to its
        $xy$-lexicographically larger end. \\ \hline

        \ccode{Compare\_distance\_at\_point\_2} & 
        Two \site{} objects and a \ccode{Point\_2}  & 
        \ccode{Comparison\_result} & 
        Determines which of the given Voronoi sites is closer to the
        given point. \\ \hline

        \ccode{Compare\_dominance\_2} & Two \site{} objects & 
        \ccode{Comparison\_result} & 
        Determines which of the sites dominates the other in case that
        there is no bisector between the two sites. \\ \hline

        \ccode{Construct\_point\_on\_x\_mono\-tone\_2} & 
        \ccode{X\_monotone\_curve\_2} & 
        \ccode{Point\_2} & 
        Constructs an interior point on the given $x$-monotone curve. \\ \hline

      \end{tabularx}
      
    \end{sideways}
  \end{center}
\end{table}

The \vorconcept{} concept refines the \concept{ArrangementTraits\_2}
concept, defined in the \aos{} package.
Models of the \concept{ArrangementTraits\_2} concept are
geometry-traits classes that enable the \aos package to robustly
construct, maintain, traverse, and query two-dimensional arrangements.
The process of computing Voronoi diagrams in our approach requires
these predicates and operations for the creation and manipulation of
bisector curves of pairs of Voronoi sites.
A Voronoi site is represented by the user-defined \site{} type.

Given two \site{} variables and an output iterator, the
\ccode{Construct\_bisector\_2} functor returns a sequence of objects
of type \ccode{X\_monotone\_curve\_2} that
together form the bisector of the two Voronoi sites.
If the bisector between the two sites does not exist, the function
returns an empty sequence.\footnote{There are cases where there is no
  bisector between two Voronoi sites. 
  For example, two Apollonius sites where one is completely contained
  inside the other have no bisector.}

Other required functors are \emph{proximity predicates}.
Each proximity predicate is given a set of points $P$ in the domain
(\eg an edge) and two Voronoi sites, and should indicate which of
the sites dominates $P$.
The \ccode{Compare\_distance\_above\_2} functor accepts two site
objects and an $x$-monotone curve, which is part of their bisector,
and indicates which site dominates the region above the $x$-monotone
curve, where ``above'' is defined to be the region to the left of the
$x$-monotone curve when it is traversed from the
$xy$-lexicographically smaller endpoint to the $xy$-lexicographically
larger endpoint.
The framework utilizes the fact that each of the sites dominates one
side of the bisector to implement the ``below'' version of the
functor.
If there is no bisector between the two sites then the
\ccode{Compare\_dominance\_2} functor is used to indicate which of the
sites dominates the other.

The \ccode{Compare\_distance\_at\_point\_2} functor is a general
proximity predicate that indicates which site (of two sites) dominates
a given point in the two-dimensional domain.
The functor is used together with the
\ccode{Construct\_point\_on\_x\_monotone\_2} functor that constructs
an interior point on a given $x$-monotone curve.

After the user created a model class that satisfies all the
requirements of the \vorconcept{} concept,
he/she can call the function \ccode{CGAL::voronoi\_2} to compute the
Voronoi diagram of a sequences
of sites, or the function \ccode{CGAL::farthest\_voronoi\_2} to
compute the respective farthest-\neighbor Voronoi diagram.
The \ccode{Voronoi\_2\_to\_Envelope\_3\_adaptor} class is used to
adapt the Voronoi traits class to a full \envelope{} traits
class~\cite{cgal:mwz-e3-08}.

\section{Randomizing for Optimality}
\label{sec:micha-proof}

For completeness, we conclude this chapter with the proof by Sharir of
Theorem~\ref{theo:env-to-vd:random} (see
Section~\ref{sec:env-to-vd:theo}) and a generalization of it.

\textbf{Theorem~\ref{theo:env-to-vd:random}.}
\textit{  Consider a specific type of two-dimensional Voronoi diagrams, so that
  the worst-case complexity of the diagram of any set of at most $n$
  sites is $F(n)$. 
  Let $S$ be a set of $n$ sites. If we randomly split $S$
  into two subsets $S_1$ and $S_2$, by choosing at random for each
  site, with equal probability, the subset it belongs to, then the expected
  complexity of the overlay of the Voronoi diagram of $S_1$ with the
  Voronoi diagram of $S_2$ is $O(F(n))$.
}

\begin{proof}

  Each vertex of the overlay is either a vertex of \vor{S_1}, a
  vertex of \vor{S_2}, or a crossing between an edge of \vor{S_1} and an
  edge of \vor{S_2} (non exclusive disjunction).
  The number of vertices of the first two kinds is $O(F(n))$, so it
  suffices to bound the expected number of crossings between edges
  of \vor{S_1} and of \vor{S_2}. Such a crossing is
  a point $u$ that is defined by four sites, $p_1,q_1\in S_1$
  and $p_2,q_2\in S_2$, so that $u$ lies on the Voronoi edge $b(p_1, q_1)$
  of \vor{S_1} that bounds the cells of $p_1$ and
  $q_1$, and on the Voronoi edge $b(p_2, q_2)$ of \vor{S_2} that
  bounds the cells of $p_2$ and $q_2$. Without loss of
  generality, assume that $\distance{u}{p_1} = \distance{u}{q_1} \le
  \distance{u}{p_2} = \distance{u}{q_2}$ (the inequality is strict if we
  assume general position).

  A simple but crucial observation is that $u$ must also lie
  on the Voronoi edge between the cells of $p_1,q_1$ in the
  {\em overall} diagram \vor{S}. Indeed, if this were
  not the case then there must exist another site $s\in S$ so
  that $u$ is nearer to $s$ than to $p_1,q_1$. But then $s$
  cannot belong to $S_1$, for otherwise it would prevent $u$
  from lying on the Voronoi edge of $p_1,q_1$. For exactly
  the same reason, $s$ cannot belong to $S_2$ --- it would then
  prevent $u$ from lying on the Voronoi edge of $p_2,q_2$ in
  that diagram. This contradiction establishes the claim.

  Define the {\em weight} $k_u$ of $u$ to be the number of
  sites $s$ satisfying
  \begin{displaymath}
  \distance{u}{p_1} = \distance{u}{q_1} < \distance{u}{s} <
  \distance{u}{p_2} = \distance{u}{q_2}.
  \end{displaymath}
  Clearly, all these $k_u$ sites must be assigned to $S_1$.

  In other words, for any crossing point $u$ between two
  Voronoi edges $b(p_1,q_1)$, $b(p_2,q_2)$, with weight $k_u$
  (with all the corresponding $k_u$ sites being farther from
  $u$ than $p_1,q_1$ and nearer than $p_2,q_2$), $u$ appears
  as a crossing point in the overlay of \vor{S_1},
  \vor{S_2} if and only if the following three
  conditions (or their symmetric counterparts, obtained by
  reversing the roles of $S_1,S_2$) hold: (i) $p_1,q_1\in
  S_1$; (ii) $p_2,q_2\in S_2$; and (iii) all the $k_u$ sites
  that contribute to the weight are assigned to $S_1$. This
  happens with probability ${\displaystyle
    \frac{1}{2^{k_u+3}}}$.

  Hence, if we denote by~$N_w$ and~$N_{\le w}$ the number of crossings
  of weight~$w$ and the number of crossings of weight at most~$w$,
  respectively, the expected number of crossings in the overlay is
  \begin{equation} \label{exp}
    \sum_{w\ge 0} \frac{N_w}{2^{w+3}} = O\left( \sum_{w\ge 0}
    \frac{N_{\le w}}{2^{w}} \right) ,
  \end{equation}
  where the right-hand side is obtained by substituting $N_w = N_{\le w}
  - N_{\le w - 1}$, and by a simple rearrangement of the sum.

  We can obtain an upper bound on $N_{\le w}$ using the
  Clarkson-Shor technique~\cite{cs-arscg-89, s-cstre-03}. 
  Specifically, denote by $N_w(n)$ and $N_{\le w}(n)$ the maximum
  value of $N_w$ and $N_{\le w}$ taken over all sets of $n$ sites,
  respectively. Then, since a crossing is defined by four sites, we
  have
  \begin{displaymath}
    N_{\le w}(n) = O\left(w^4N_0(n/w)\right) .
  \end{displaymath}
  Note that if a crossing $u$, defined by $p_1,q_1,p_2,q_2$,
  has weight $0$ then $p_1,q_1,p_2,q_2$ are the four nearest
  sites to $u$. The number of such quadruples is thus upper
  bounded by the complexity of the {\em fourth-order} Voronoi
  diagram of some set $S_0$ of $n/w$ sites.

  We claim that the complexity of the fourth-order Voronoi diagram of
  $n$ sites is $O(F(n))$. Indeed, any quadruple $p_1$, $q_1$, $p_2$,
  $q_2$ of four nearest sites to some point $u$ can be charged to a face
  of the fourth-order diagram (the one whose 
  projection contains $u$). Each such face
  can in turn be charged either to one of its vertices, or to its
  rightmost point, or to a point at infinity on one of its
  edges. Assuming general position, each such boundary point can be
  charged at most $O(1)$ times. Now another simple application of the
  Clarkson-Shor technique shows that the number of these vertices and
  boundary points is $O(F(n))$ --- each of them becomes a feature of the
  (0-order) Voronoi diagram if we remove a constant number of sites,
  which happens with large probability when we sample a constant
  fraction of the sites.

  In other words, we have $N_{\le w}(n) = O(w^4F(n/w)) = O(w^4F(n))$. 
  Substituting this into~(\ref{exp}), we obtain an upper bound of $O(F(n))$
  on the complexity of the overlay, as claimed.
\end{proof}

\begin{remark}
  The analysis in this section can easily be extended to the case of
  the lower envelope of an arbitrary collection of bivariate functions
  (of constant description complexity). As a result, we get the following.
\end{remark}

\begin{corollary}
  Let $\calG{}$ be a collection of $n$ bivariate functions of constant
  description complexity, and let $F(m)$ be an upper bound on the
  complexity of the lower envelope of any subcollection of at most $m$
  functions. Then the expected complexity of the overlay of the
  minimization diagrams of two subcollections $\calG{}_1$ and
  $\calG{}_2$, obtained by randomly partitioning~$\calG$, as above,
  is~$O(F(n))$.
  Consequently, the lower envelope of $\calG$ can be constructed by
  the above randomized divide-and-conquer technique, in expected time
  $O(F(n)\log n)$, provided that $F(n)=\Omega(n^{1+\varepsilon})$, for
  some $\varepsilon > 0$. The expected running time is $O(n\log^2n)$
  when $F(n)=O(n)$.
\end{corollary}

\chapter{Examples and Implementation Details}
\label{chap:impl}

In this chapter we describe the implementation details involved in
realizing diverse types of Voronoi diagrams that can be computed with
our framework.
Essentially, all discussed Voronoi diagram can be categorized by two
aspects:
\begin{inparaenum}[(i)]
  \item the embedding space --- the unbounded plane or the sphere, and
  \item the type of arithmetic used in
    the implementation --- rational arithmetic or higher-degree
    algebraic arithmetic.
\end{inparaenum}
The order of the sections in the chapter is by the first category,
distinguishing between the diagrams embedded in the plane
(Section~\ref{sec:impl:plane-vd}) and the diagrams embedded on the
sphere (Section~\ref{sec:impl:sphere-vd}), and then by the second
category.

We have applied optimizations to try and reduce the running time of
our software as much as possible.
Section~\ref{sec:exper:speed} presents these optimizations.
Our efforts in expediting the computation concentrate on Voronoi
diagrams of points and power diagrams in the plane (Voronoi diagrams
with affine bisectors).

The implementations of the types of Voronoi diagrams presented in this
chapter are just the tip of the iceberg and are aimed to demonstrate
the ability of our framework to compute a wide variety of Voronoi
diagrams with different properties.
Additional types of Voronoi diagrams can be (easily) implemented using
our framework; see Chapter~\ref{chap:conclusion} for suggested future work.

\section{Planar Voronoi Diagrams}
\label{sec:impl:plane-vd}

Voronoi diagrams embedded in the plane are most useful, and may be the 
most investigated type of Voronoi diagrams~\cite{obsc-stcavd-00,ak-vd-00}.
Historically, the intention to use the \envelope package of \cgal for the
computation of general Voronoi diagrams was limited to planar Voronoi
diagrams.
In fact, the idea was conceived before the \aos package of \cgal had
an initial implementation.

\subsection{Voronoi Diagrams with Linear Bisectors}
\label{ssec:impl:plane-vd:linear}

One of the ways to categorize Voronoi diagrams is by the class of its
bisector curves. This section discusses Voronoi diagrams, the
bisectors of which are composed of linear objects.
Linear bisectors enable us to create efficient traits classes based on
the arrangement package of \cgal and the various geometric kernels
supplied by \cgal.

The first type of Voronoi diagrams is characterized by the fact that
all its bisectors are single lines in the plane. 
The second part of this section is an example of the usage of our
framework to compute Voronoi diagrams of two-point sites.
Specifically, we compute the Voronoi diagram induced by the two-point
triangle-area distance function of a set of points~\cite{bdd-psvd-02}.

\subsubsection{Affine Voronoi Diagrams}
\begin{wrapfigure}[23]{r}{6.1cm}
  \vspace{-15pt}
  \centerline{
    \includegraphics[width=6cm]{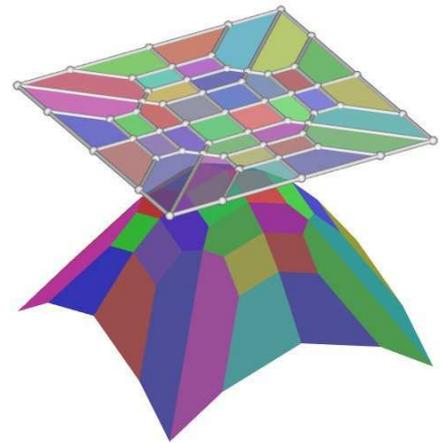}
  }

  \vspace{-10pt}
  \caption[Computing Voronoi diagrams of points using lower envelopes.]
          {\capStyle{Computing the Voronoi diagram of a set of points
              using the lower envelope of a set of planes.
              The minimization diagram (appears above the envelope for
              clarity) of 36~planes, which represent distance
              functions to 36~point sites in a degenerate
              constellation, constitutes the Voronoi diagram of the
              points.
              The lower envelope, as well as the minimization diagram,
              are clipped to within an axis-aligned square.}} 
          \label{fig:impl:lower-env-planes}
\end{wrapfigure}
\emph{Affine Voronoi diagrams} is the class of all Voronoi diagrams
whose sites have affine bisectors.
The power diagram of a set of disks in the plane (defined below) is an
affine Voronoi diagram and is a generalization of the standard Voronoi
diagram of points.
Interestingly, the class of affine Voronoi diagrams is identical
to the class of power diagrams in the plane~\cite[\S2.3.3]{bwy-cvd-06}.
Every abstract Voronoi diagram (Section~\ref{sec:background:vd}) whose
bisectors are lines has a corresponding power diagram that constitutes
it.
Therefore, having a robust implementation for computing power
diagrams of sets of disks is, in some sense, a complete solution in
this context.

\begin{definition}[Planar power diagram]
  The planar power distance is measured from a point $x \in \R{2}$ in
  the plane to a disk $d$ with center $c$ and positive radius $r$, and
  is defined to be:
  $\distance{x}{d} = (x - c) ^ 2 - r ^ 2$.
  The power diagram of a set $D$ of disks in the plane is defined to
  be the nearest-\neighbor Voronoi diagram induced by the power
  distance.
\end{definition}

Using a simple observation, one can construct the lower envelope of a
set of \emph{planes} instead of constructing the lower envelope of the
paraboloids that represent the power distance functions from the disk
sites.
We transform each paraboloid of the form 
$f_i(x) = x^2 -2xc + c^2 - r^2 $ to the plane 
$\pi_i : -2xc + c^2 - r^2$.
In other words, for each point $x \in \R{2}$ and a site we remove the
$x^2$ factor from the power distance.
Notice that each distance function is decreased by the same amount
for a specific point~$x$, and thus, the topological structure of the
lower envelopes of the original paraboloids and the
new linear functions is identical.
See Figure~\ref{fig:impl:lower-env-planes} for an illustration.

The \envelope package of \cgal contains a traits class for
constructing the lower or upper envelope of a set of planes and
half-planes in~\R{3}, named \ccode{CGAL::Env\_plane\_traits\_3}.
However, as it might be expected, the implementation of this traits
class is more complicated than is needed when considering the 
computation of power diagrams only (and not general envelopes).
The reason is that the \ccode{Env\_plane\_traits\_3} traits class
handles cases that cannot occur during the computation of power
diagrams, 
\ie the planes that represent power distance functions are always full
planes and are never vertical.

We implemented an easy-to-use traits class for the computation of
standard Voronoi diagrams of points and power diagrams of disks using
our framework.
The traits class is named \ccode{CGAL::Power\_diagram\_traits\_2}
and is a model of the \vorconcept concept (see
Section~\ref{sec:env-to-vd:impl}).
The implementation of the traits class is much simpler than the
\envelope package's traits class, and provides a better interface for
computing Voronoi diagrams --- the user of the traits class does not
need to construct any plane.
A traits class modeling the \vorconcept does not have to actually
construct the distance functions from the sites;
see Chapter~\ref{chap:env-to-vd}.
The \ccode{Power\_diagram\_traits\_2} traits class does not construct
the three-dimensional planes, but directly implements all the
required predicates and operations.

\begin{figure}[t]
  \centering
  \providelength{\spacewidth}\setlength{\spacewidth}{0pt}
  \providelength{\subfigwidth}\setlength{\subfigwidth}{98pt}

  \subfigure[]{\label{fig:points-on-circle}
    \includegraphics[width=\subfigwidth]{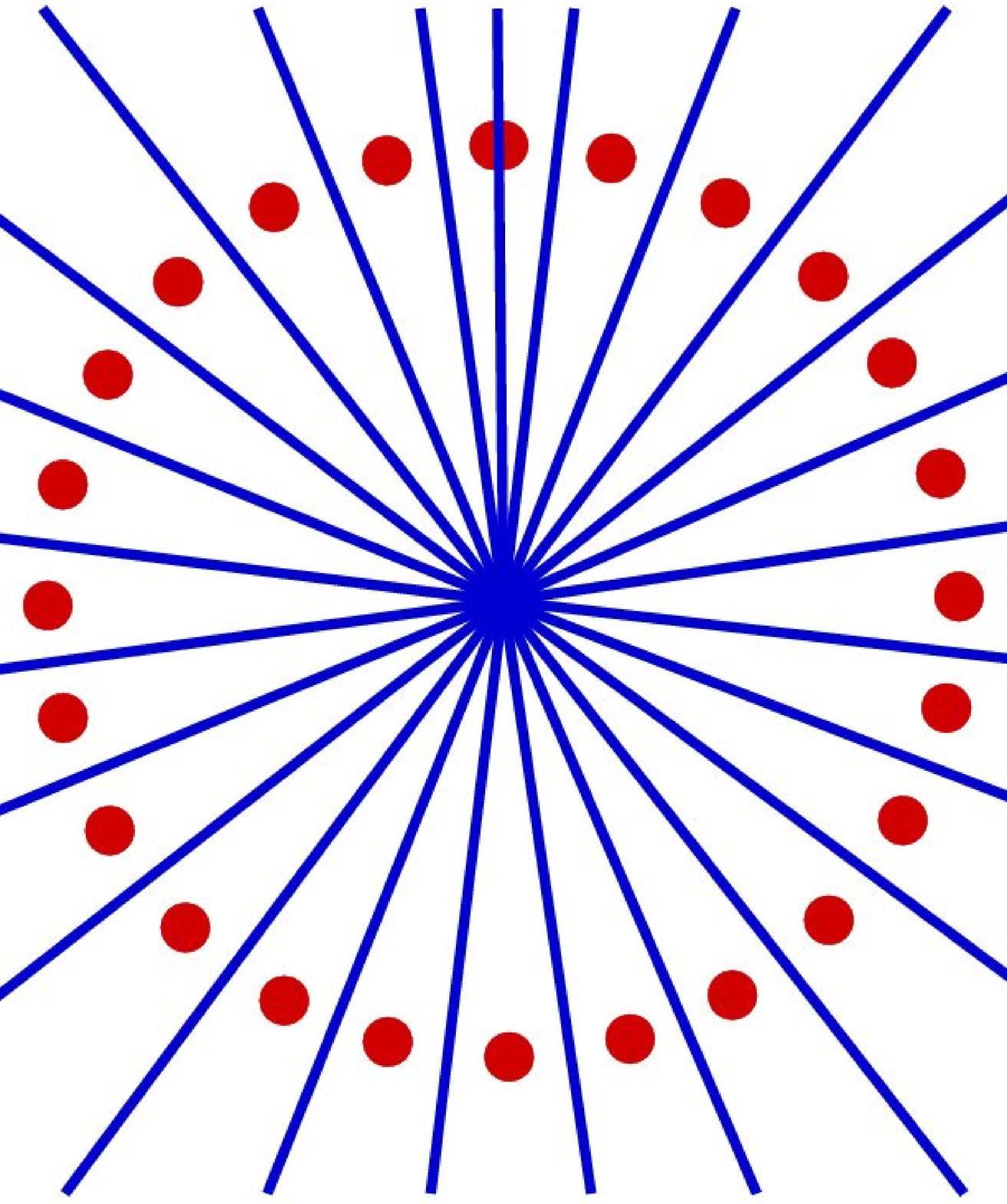}}
  \hspace{\spacewidth}
  \subfigure[]{\label{fig:points-on-grid}
    \includegraphics[width=\subfigwidth]{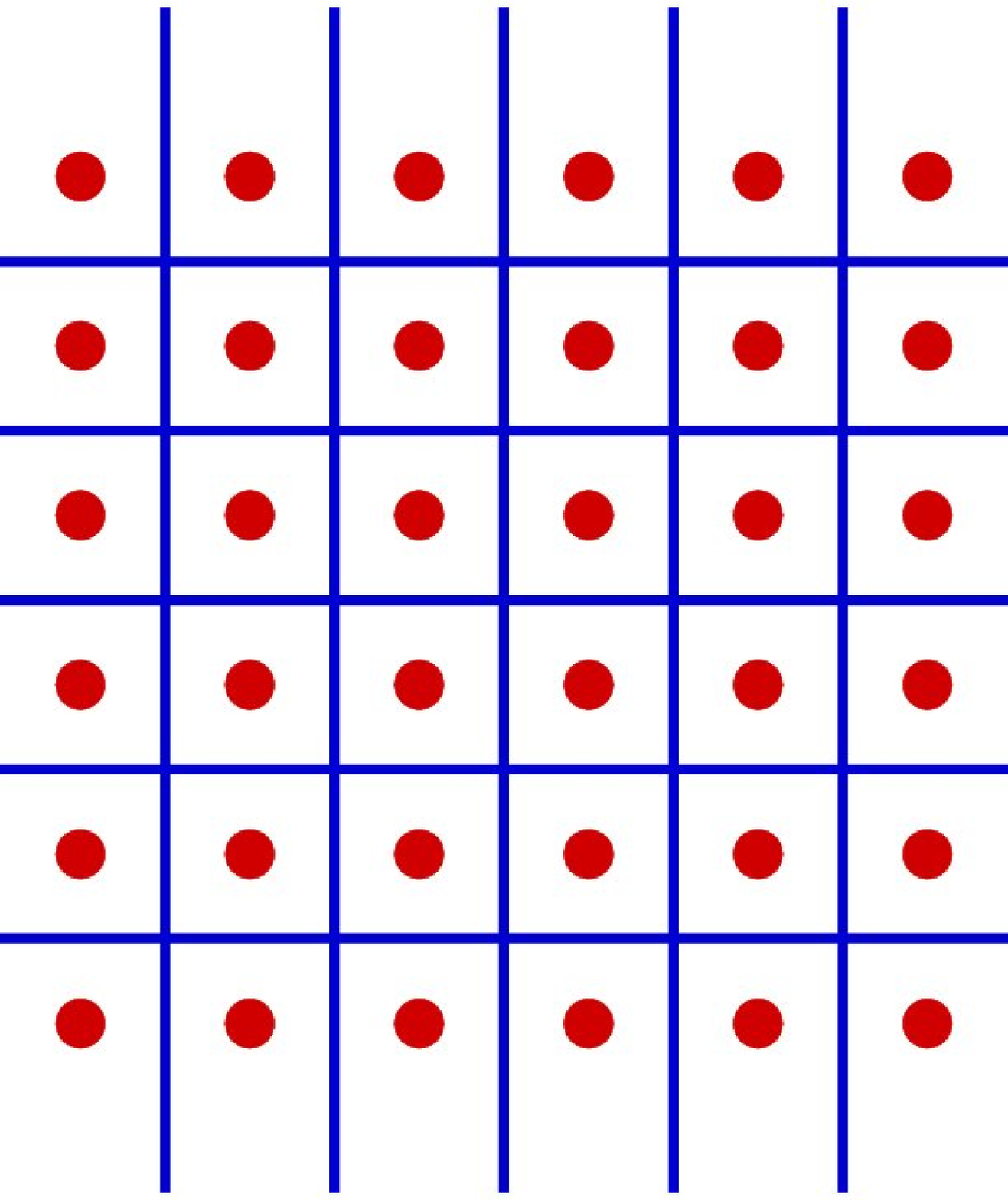}}
  \hspace{\spacewidth}
  \subfigure[]{\label{fig:points-on-cross}
    \includegraphics[width=\subfigwidth]{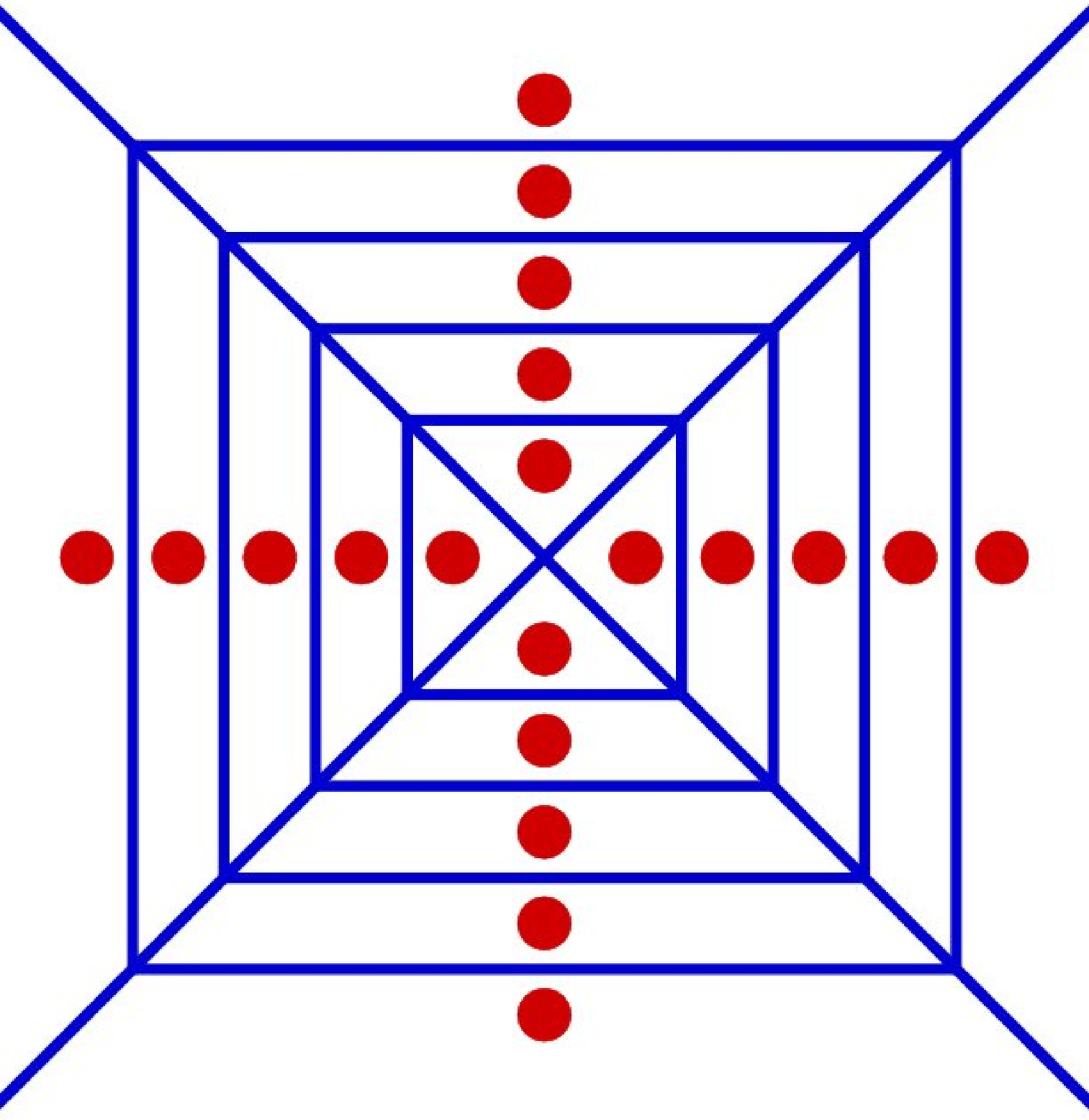}}
  \hspace{\spacewidth}
  \subfigure[]{\label{fig:cgal_vd}
    \includegraphics[height=\subfigwidth]{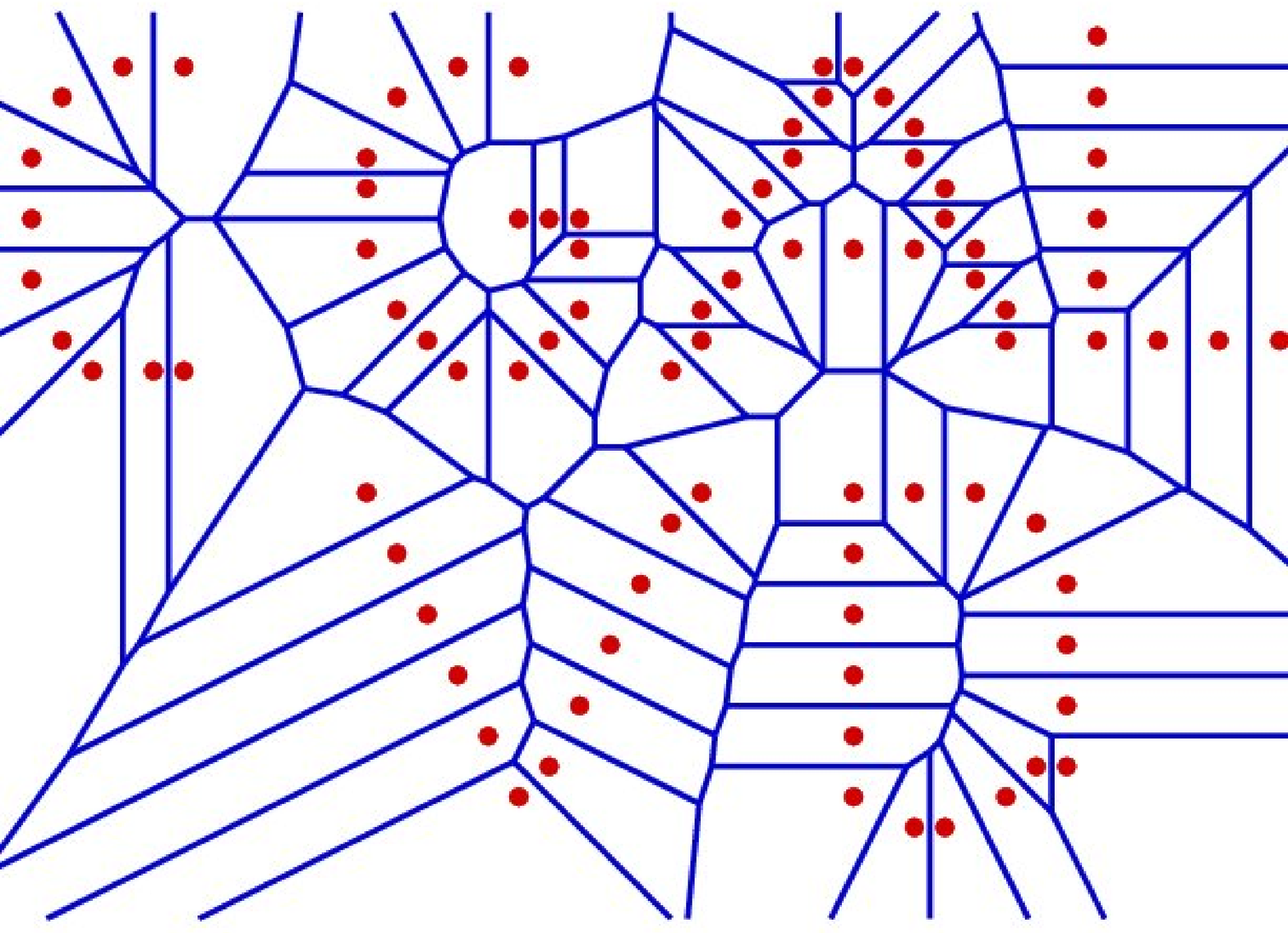}}

  \caption[Voronoi diagrams of points]{Voronoi diagrams of various
    point-sets computed with our framework and our traits class for
    computing power diagrams in the plane.
  \subref{fig:points-on-circle} The Voronoi diagram of points ordered
  exactly on a circle. The Voronoi diagram has a single vertex.
  \subref{fig:points-on-grid} The Voronoi diagram of points ordered on
  a grid.
  \subref{fig:points-on-cross} The Voronoi diagram of points ordered
  on a cross.
  \subref{fig:cgal_vd} The Voronoi diagram of points creating the
  letters of \cgal and VD (an abbreviation for Voronoi Diagrams).
  }
  \label{fig:points}
\end{figure}

Our traits class inherits from the
\cgalarrlineartraits traits
class~\cite{cgal:wfzh-a2-08} that models the \arrtraits concept and
supports arrangements induced by linear objects, which may be bounded
(segments) or unbounded (rays and lines).
The traits class is parameterized by a geometric kernel
class~\cite{fgkss-dccga-00, hhkps-aegk-07}.
The choice of the geometric kernel, which consists of types of
constant size non-modifiable geometric primitive objects (\eg
points, lines, triangles, circles, etc.), and determines the
number-type\footnote{There are several different classes that can
  represent numbers in \cgal.} used and the implementations of all
geometric operations on kernel objects.
In this context, the geometric kernel determines, for example, the
type representing the bisector curves (lines) and the number-type.
The geometric kernel is passed as the underlying kernel for the
\arrlineartraits base class.

The \ccode{Power\_diagram\_traits\_2} class requires the underlying
number-type to only support exact rational arithmetic; as opposed to
number types required in the following
Sections~\ref{sec:impl:plane-vd:algebraic}
and~\ref{ssec:impl:plane-vd:semi}.
Namely, the number type should support the arithmetic operations $+$,
$-$, $*$, and $/$ with unlimited precision over the
rationals. An example for such number type is the rational number type
{\tt CGAL::Gmpq} based on \gmp --- Gnu's Multi Precision
library~\citelinks{gmp-link}.

Disk sites (objects of type \site) are represented as unoriented
circles in the two-dimensional Euclidean plane by the kernel type
\ccode{Kernel::Circle\_2}.
In order to stay in the rational domain and apply only fast rational
arithmetic operations, two restrictions on the disk sites must be
enforced: the coordinates of their centers have to be rational, and
the squares of their radii have to also be rational.

Although the new traits class provides a simpler implementation
and a more convenient interface, it does not provide a significant
increase in performance in comparison with the
\ccode{Env\_plane\_traits\_3} traits class.
The reason is that the main performance hit during the execution of
the algorithm is caused, in fact, by the exact and expensive geometric
operations and predicates that are provided by the kernel.
Section~\ref{sec:exper:speed} provides details on optimizations, which
improve the computation time significantly.
Most of the optimizations there concentrate on minimizing the number
of calls to geometric operations and predicates.

\subsubsection{Two-Site Triangle-Area Voronoi Diagram}
Planar two-point distance functions are defined from a point~$x$ in the
plane to a pair of point site $\{p, q\}$.
Among the known distance functions are 
\begin{inparaenum}[(i)]
\item the sum of distances, which is defined by 
  $\distance{x}{\{p, q\}} = d(x, p) + d(x, q)$ where $d$ is the
  Euclidean distance,
\item the product of distances which is defined by 
  $\distance{x}{\{p, q\}} = d(x, p) \cdot d(x, q)$,
\item the triangle area distance function defined by the area of
  $\triangle xpq$,
\item the triangle perimeter distance function defined by the
  perimeter of~$\triangle xpq$, and
\item the difference of distances defined by
  $\distance{x}{\{p, q\}} = |d(x, p) - d(x, q)|$.
\end{inparaenum}
The combinatorial complexity of the diagrams and the time complexity
of the known algorithms for computing various two-point site Voronoi
diagrams varies~\cite{bdd-psvd-02}.

We describe in this section the implementation of a traits class
named, \ccode{Triangle\_area\_dis\-tance\_traits\_2} that enables the
computation of the Voronoi diagram of points as induced by the
triangle area distance function.
The traits class uses rational arithmetic only and handles
degenerate input.

The triangle distance function between a point $(x, y)$ in the plane
and a pair of points $(p_x, p_y)$ and $(q_x, q_y)$ is defined to
be the area of the triangle formed by these three points, namely,
\begin{displaymath}
  \frac{1}{2} abs \left(
  \left| \begin{array}{ccc}
    x & y & 1 \\
    p_x & p_y & 1 \\
    q_x & q_y & 1 
  \end{array}
  \right| \right) 
  \; .
\end{displaymath}

The combinatorial complexity of the nearest-\neighbor Voronoi diagram
of $n$~points in general position as induced by the triangle distance 
function is~$\Theta(n^4)$, and~$\Theta(n^2)$ for the farthest-\neighbor
Voronoi diagram.\footnote{Remember that there are $\Theta(n^2)$ pairs
  of points, so we actually obtain a quadratic Voronoi diagram in the
  number of sites.}

\ccode{Triangle\_area\_dis\-tance\_traits\_2} is a model of the
\vorconcept concept.
Similar to the \ccode{Power\_diagram\_traits\_2} traits class, our
traits class inherits from the \cgalarrlineartraits traits class,
which models \arrtraits concept.
The class is parameterized by a geometric kernel class whose number
type is required to support exact rational arithmetic for achieving a
robust implementation.
The geometric kernel is passed to the
\arrlineartraits base class as a template
parameter.

The Voronoi sites (of \site type) of our diagram are pairs of
points.\footnote{We use \stl's templated \ccode{std::pair} class,
  instantiated with \ccode{Kernel::Point\_2}.}
The user is accountable for providing all pairs of points as part of
the input.
The distance from a rational point in the plane to a pair of rational
points is rational (due to the rational nature of the distance
function).
The \ccode{Compare\_distance\_at\_point\_2} functor compares the two
rational distances of a point in the plane to two 2-point sites.

The bisector of two sites consists of two rational lines.
Let~$S_1$ and~$L_1$ be the segment connecting the two points of one
site and its supporting line, respectively, and let~$S_2$ and~$L_2$ be
the segment connecting the two points of another site and its
supporting line, respectively.
The intersection of~$L_1$ and~$L_2$ is the intersection of the two
bisector lines. 
If~$L_1$ and~$L_2$ are parallel, then the bisector consists of two
parallel lines that become the same line if~$S_1$ and~$S_2$ are of the
same length.
If~$S_1$ and~$S_2$ are collinear then the bisector is either one
line, or, in case that~$S_1$ and~$S_2$ are of the same length, does
not exist.
We construct the two lines in the \ccode{Construct\_bisector\_2}
functor implementation.
We have to perform an additional operation before outputting the two
lines. The \envelope package code does not tolerate
bisectors whose $x$-monotone parts intersect in their interior.
Before outputting the bisector we have to intersect the two lines and
transform them into 4~interior-disjoint rays.

The last functor that is required by the \vorconcept is the
\ccode{Compare\_distance\_a\-bove\_2} functor.
The bisector lines partition the plane into a maximum of 4~regions. If one
of the points of a Voronoi site is inside a specific region, then the
region is dominated by the Voronoi site of the point.
When crossing a bisector we move from a region that is dominated by
one site to a region that is dominated by the other site
(except for the case where all four points are collinear).
We compute the number of curves we need to cross to get from one of
the points of the first site and from the region above the input curve
to the upper-most face. (If we have a vertical line then we take the
left one.)
If both numbers are even or odd together we return that the region
above the input curve is dominated by the first site. If one is odd
and the other is even then the region is dominated by the second
site.

\subsection{Voronoi Diagrams with Higher-Degree Algebraic Bisectors}
\label{sec:impl:plane-vd:algebraic}

This section describes diagrams where exact rational arithmetic only
is insufficient for computing the diagrams in an exact and robust
manner.
The presented diagrams are curved Voronoi diagrams~\cite{bwy-cvd-06}
requiring a higher-degree algebraic machinery.

\subsubsection{\mobius Diagrams}

\begin{definition}[\mobius diagram]
  \label{def:mobius}
  Let $w$ be a \emph{\mobius site} defined by a triple 
  $(p, \lambda, \mu)$ where $p \in \R{2}$ and $\lambda, \mu \in \R{}$.
  The \emph{\mobius diagram} of a set $\calW = \{w_1, \ldots, w_n\}$ of \mobius
  sites is  the Voronoi diagram induced by the following distance function:
  $\distance{x}{w_i} = \lambda_{i}(x - p_{i})^{2} - \mu_{i}$
\end{definition}

The \mobius diagram is a generalization of the power diagram of disks
(Section~\ref{ssec:impl:plane-vd:linear}) and of the
multiplicatively-weighted Voronoi diagram.
When all $\lambda_i$ are equal, the \mobius diagram becomes
the power diagram of the set of disks centered at $p_i$ with radii
$\sqrt{\frac{\mu_i}{\lambda_i}}$.
When $\mu_i = 0$ for all $i$, the \mobius diagram becomes the 
multiplicatively-weighted Voronoi diagram with $p_i$ as the points 
and $\sqrt{\lambda_i}$ as the weights.

The bisector of two \mobius sites is either a circle or a line (in
case of equal $\lambda$ parameters).
Moreover, every abstract Voronoi diagram whose bisectors are circles
or lines has a corresponding \mobius diagram that
constitutes it~\cite[\S2.4.1]{bwy-cvd-06}
(similar to the relation of power diagrams and affine Voronoi diagrams).

\cgal contains a traits class for the arrangement package that
supports arrangements of circular and linear bounded segments, named
\ccode{Arr\_circle\_segment\_traits\_2}.
The efficient implementation of the traits class, which supports these
curves, is attained by the observation that the coordinates 
of all intersection points between the curves are of 
algebraic degree~2 only, and, therefore, can be represented with
\emph{square-root extension} numbers (which are an extension to the
rational field).

We have created a model class for the \vorconcept concept named
\mobiustraits, which supports the construction
of \mobius diagram using our framework.
The traits class supports rational \mobius sites (with rational $p$,
$\lambda$, and $\mu$).
The \mobiustraits class is based on an extension
we have implemented for the \ccode{Arr\_circle\_segment\_traits\_2},
named \ccode{Arr\_circle\_linear\_traits\_2}, which supports, in addition,
unbounded linear curves (lines and rays).
The fact that \ccode{Arr\_circle\_linear\_traits\_2} harnesses
efficient square-root extension arithmetic does not immediately imply
that an implementation for the \mobius diagram that uses only
square-root extension arithmetic (and not higher-degree algebraic
numbers) is easily obtainable.
Performing operations with square-root extension numbers is limited,
since square-root extension numbers form an algebraic field only if
they are extended by the same root. (Rational numbers are considered
to be present in all extension fields.)
This means that arithmetic operations on two square-root
extension numbers can be easily performed, while staying in the same
algebraic-degree arithmetic, only if they have the same extension.

We compute the distance from a point, which is of type
\ccode{Arr\_circle\_linear\_traits\_2::\-Point\_2}, to a \mobius site
using square-root extension arithmetic.
The $x$ and $y$ coordinates of a point of the
\ccode{Arr\_circle\_linear\_traits\_2} traits class are either both
rational, one is rational and the other is a square-root extension
number, or both are square-root extension numbers with the same
extension.
Therefore, we can perform all needed arithmetic operations for
computing the distance by using square-root extension arithmetic.

Another non-trivial functor is the
\ccode{Construct\_point\_on\_x\_monotone\_2} functor.
In order to construct a point on an edge, assuming that the edge is
not vertical, we construct isolating rational intervals for both of
its endpoints.
We use both intervals to get a rational $c$ between
the endpoints and construct the vertical line $y = c$.
The resulting point is the intersection point between this vertical line and
the original edge.
We perform similar operations in the cases of a vertical curve and an
unbounded curve.

\subsubsection{Anisotropic Voronoi Diagrams}

\begin{definition}[Anisotropic Voronoi diagram]
  Let $s$ be an \emph{anisotropic site} defined by the triple 
  $(p_i, M_i, \pi_i)$ where $p \in \R{2}$, $M \in \R{2\times 2}$ is a
  symmetric positive definite matrix, and $\pi \in \R{}$. 
  The \emph{Anisotropic Voronoi diagram} of a set
  $\calS = \{s_1, \ldots, s_n\}$ of anisotropic sites is the Voronoi
  diagram induced by the following distance function: 
  \begin{displaymath}
    \distance{x}{s_i} = (x - p_i)^t M_i (x - p_i) - \pi_i
  \end{displaymath}
\end{definition}

Anisotropic Voronoi diagrams are a natural generalization of \mobius
diagrams (take the matrix $M$ to be the scalar $\lambda$ from
Definition~\ref{def:mobius} times the identity matrix).
Anisotropic Voronoi diagrams have various applications, \eg
guaranteeing the quality of meshes~\cite{ls-avdgqa-03}.

The bisector of two Anisotropic sites is a full planar quadratic curve
(\ie a circle, an ellipse, a parabola, a hyperbola, or a degenerate
version of one of the previous).
Again, every abstract Voronoi diagram whose bisectors are full
quadratic curves has a corresponding anisotropic diagram that
constitutes it~\cite[\S2.4.2]{bwy-cvd-06}.

We have created the \ccode{Anistropic\_voronoi\_traits\_2} traits class 
that enables the computation of anisotropic Voronoi diagrams using our 
framework.
The traits class is based on a new traits class for the arrangement 
package that enables the construction of arrangements of algebraic plane 
curves~\cite{acs-be-cckva-08,ek-eeaaac-08}.
The remaining required functors are a functor for constructing the 
bisector of two anisotropic sites and functors to answer proximity
queries.

All required proximity predicates are similarly implemented as
follows: we construct a point~$p$ inside the region in question 
(rational points get a higher priority).
Then, we compare the distances from $p$ to the two anisotropic sites 
and decide which of the sites is closer.

Deciding which of the sites is closer is a non-trivial task as we try
to expedite the computation that involves expensive algebraic arithmetic.
We use several techniques to optimize the computation.
First, we try to construct an exact rational point (as opposed to a
point of a higher algebraic degree). In most cases this is possible
(but not in all, \eg 
it is not possible to construct a rational point inside a singleton
that consists of a non-rational point).
Points with rational coordinates induce a rational distance, the
computation of which is easy and fast.
Second, if we are forced to construct a non-rational point,
we first try to use rational interval arithmetic to decide the
predicate~\cite{bbp-iaedf-01}. 
Each point represented by the algebraic plane curves traits class has a
rational bounding-box with which we try and deduce a correct answer. 
Only if we can not use rational arithmetic to decide the predicate
correctly, we resort to more expensive algebraic computations based on
the \core{} library~\cite{klpy-clrngc-99}\citelinks{core-link}.

\subsection{Voronoi Diagrams with Semi-Algebraic Bisectors}
\label{ssec:impl:plane-vd:semi}
\label{ssec:impl:plane-vd:apo}
\label{ssec:impl:plane-vd:segs}

In this section we describe the implementation of two important Voronoi
diagrams, the Apollonius diagram (or additively-weighted Voronoi diagram)
and the Voronoi diagram of linear objects.
The bisectors of the two diagrams are semi-algebraic, \ie composed
of portions of algebraic curves.
The implementation of both diagrams is based on the algebraic plane curve
traits class mentioned in the previous section.
We concentrate on the additional problems introduced by the computation
of Voronoi diagrams with semi-algebraic bisectors.
Those problems include, among others, non-rational distance functions
and complicated boundary conditions.

\subsubsection{Apollonius Voronoi Diagrams}

The Apollonius diagram, also known as the additively-weighted Voronoi
diagram, has applications in many
fields~\cite{obsc-stcavd-00,ada-msawvd-07}.
It is often used as a replacement for the Voronoi diagram of circles
in the plane.
(See Chapter~\ref{chap:min-annulus} for a concrete application of the
Apollonius diagram --- computing a minimum-width annulus of a set of
disks.)

\begin{definition}[Apollonius diagram]
Given a set $D = \{d_1, \ldots, d_n\}$ of disks with respective
centers $p_i$ and radii $r_i$, the \emph{Apollonius diagram}
of the set is the diagram induced by the following distance function:
$\distance{x}{d_i} = ||x - p_i|| - r_i$
\end{definition}

Given two disks in the plane, their Apollonius bisector is either
\begin{inparaenum}[(i)]
\item one branch of a hyperbola, which ``bends'' toward the disk with the 
  smaller radius,
\item a line, which is the bisector of the centers of the disks, in case of
  equal radii, or
\item the empty set in the case where one disk contains the other 
  (there is no bisector as one site dominates the whole plane).
\end{inparaenum}
In the case that the bisector is a branch of a hyperbola, the transverse 
axis of the hyperbola is the line passing through the centers of the disks.

The traits class \apolloniustraits enables the construction of Apollonius
diagrams using our framework, and handles all degenerate cases (including all
types of bisectors).
The implementation of this traits class is similar to the implementation 
of the traits class for the anisotropic diagram (see 
Section~\ref{sec:impl:plane-vd:algebraic} above).
There are two main differences in the implementation of the 
\apolloniustraits class.
The first is the construction of the bisector of two Apollonius sites, which is
not a simple full algebraic curve.
The second is that the distance function between a point in the plane and an
Apollonius site is not ``rational,'' meaning that the distance is not 
necessarily rational even if the given point is.

The construction of a bisector amounts to the following task.
Given the full algebraic hyperbolic curve, pick the correct branch and
return it.
We efficiently implement the functor for constructing the bisector by 
exploiting the \ccode{Make\_x\_monotone\_2}
functor of the algebraic curves traits class.
The \ccode{Make\_x\_monotone\_2} of the traits class not only splits the 
hyperbola into two/four different $x$-monotone curves, but also
fortunately returns them in a lexicographic order.
By examining the coordinates of the centers of the disks and their
radii, we can determine which of the two branches is the bisector.
For example, in the case of a vertical asymptote, we pick
the correct branch by comparing the $x$-coordinates of the centers of
the disks; 
the correct branch is the one on the same side as the disk
of the smaller radius.
Notice, that if one Apollonius site contains another site then there
is no bisector between them and the site with the larger radius
dominates the entire plane.

The distance function from a point in the plane to an Apollonius site
is not rational. Even if a rational point is given, the resulting distance
is not necessarily rational.
Still, comparing the distances of a rational point to two Apollonius sites
is faster than of a non-rational point.
A rational point results in a distance value of algebraic degree~2
where a general point of the traits class may result in a number of
algebraic degree up to~8.
Again, we use the \core{} library for comparing 
numbers of high algebraic-degree in an exact manner.

\subsubsection{Voronoi Diagrams of Linear Objects}
\begin{figure}[h]
  \centering
  \providelength{\subfigwidth}\setlength{\subfigwidth}{110pt}

  \subfigure[]{\label{fig:linear:grid}
    \includegraphics*[angle=-90,totalheight=\subfigwidth]{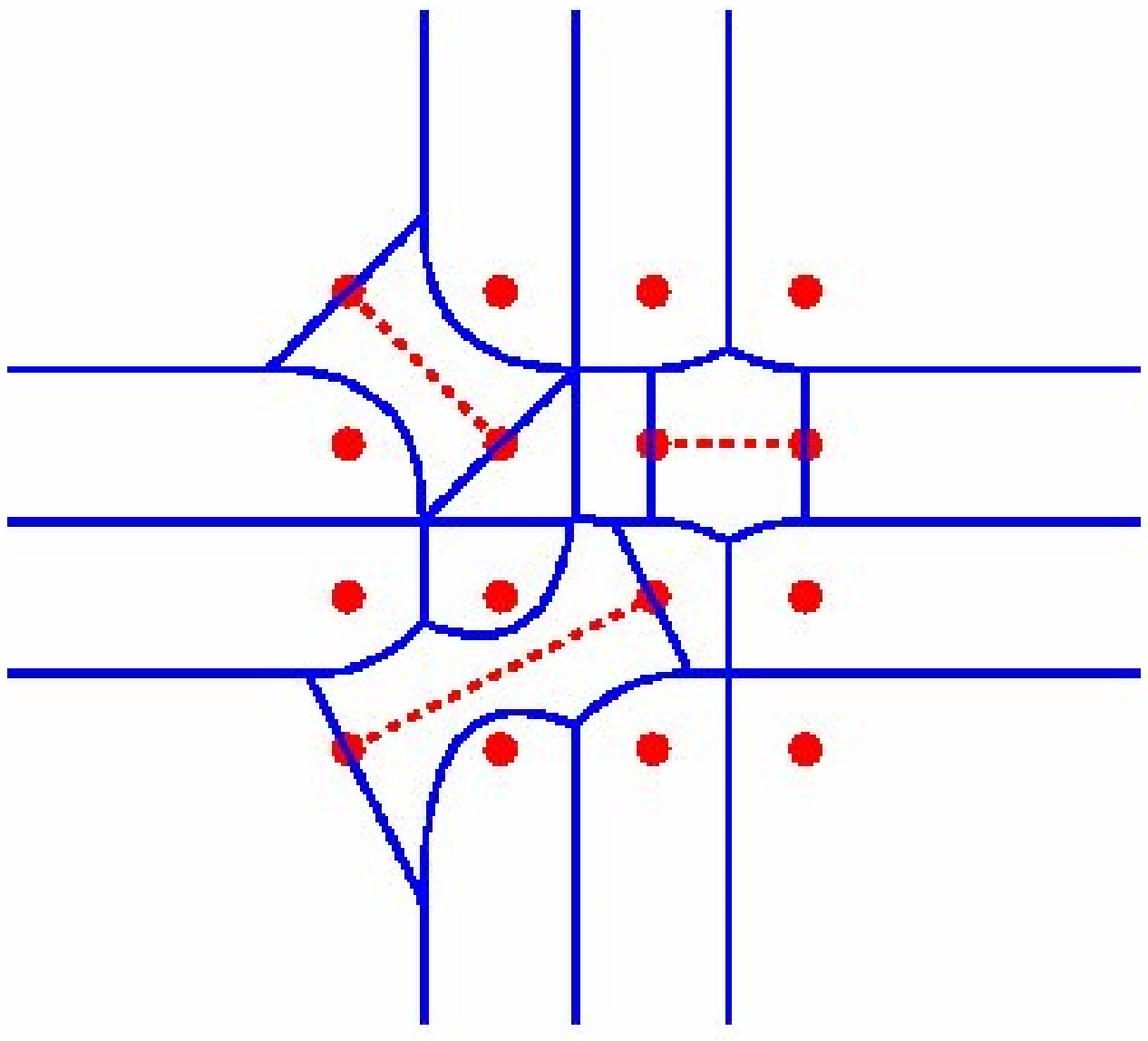}}
  \subfigure[]{\label{fig:linear:grid-and-bound}
    \includegraphics*[angle=-90,totalheight=\subfigwidth]{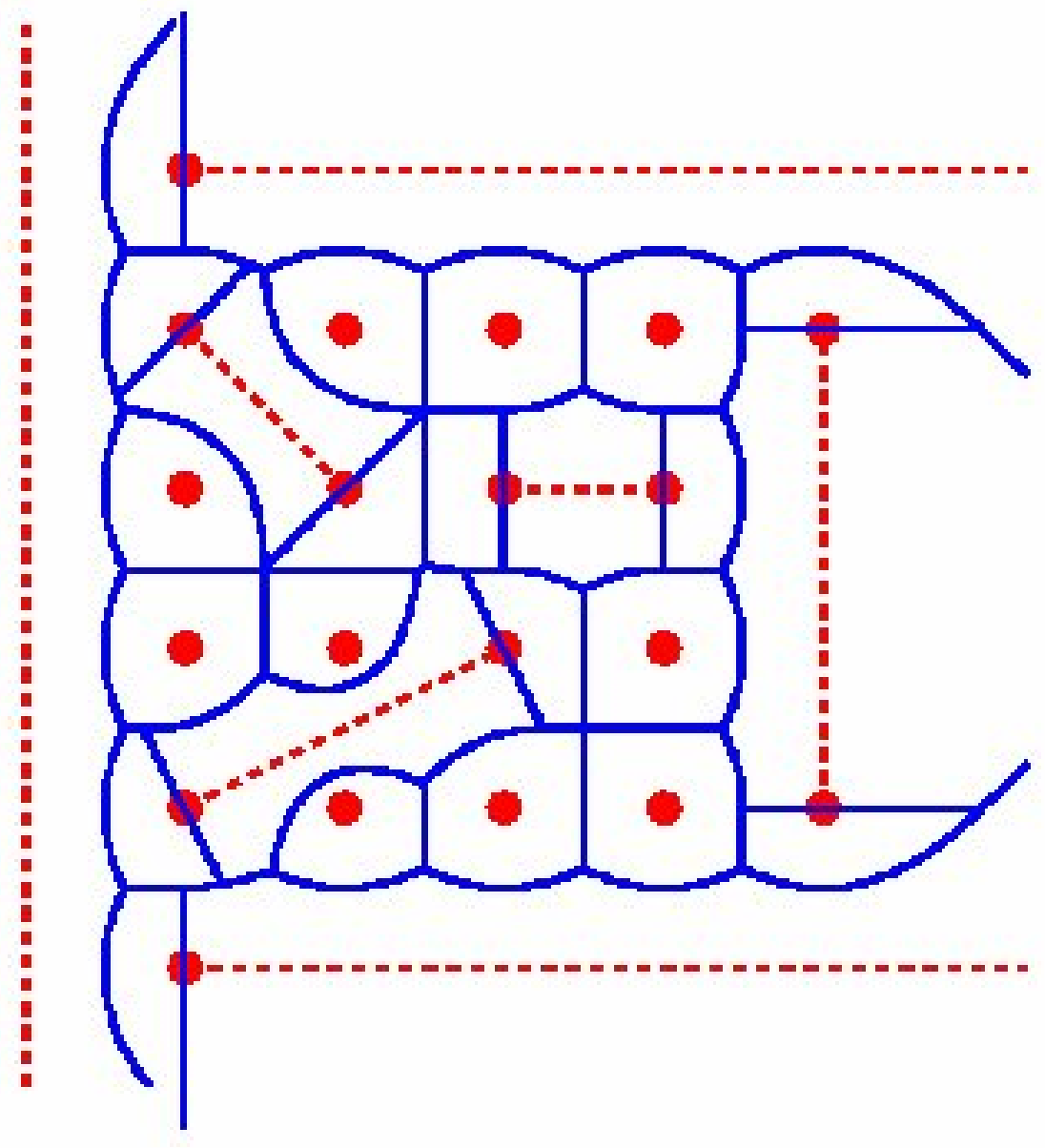}}
  \subfigure[]{\label{fig:linear:ray}
    \includegraphics*[angle=-90,totalheight=\subfigwidth]{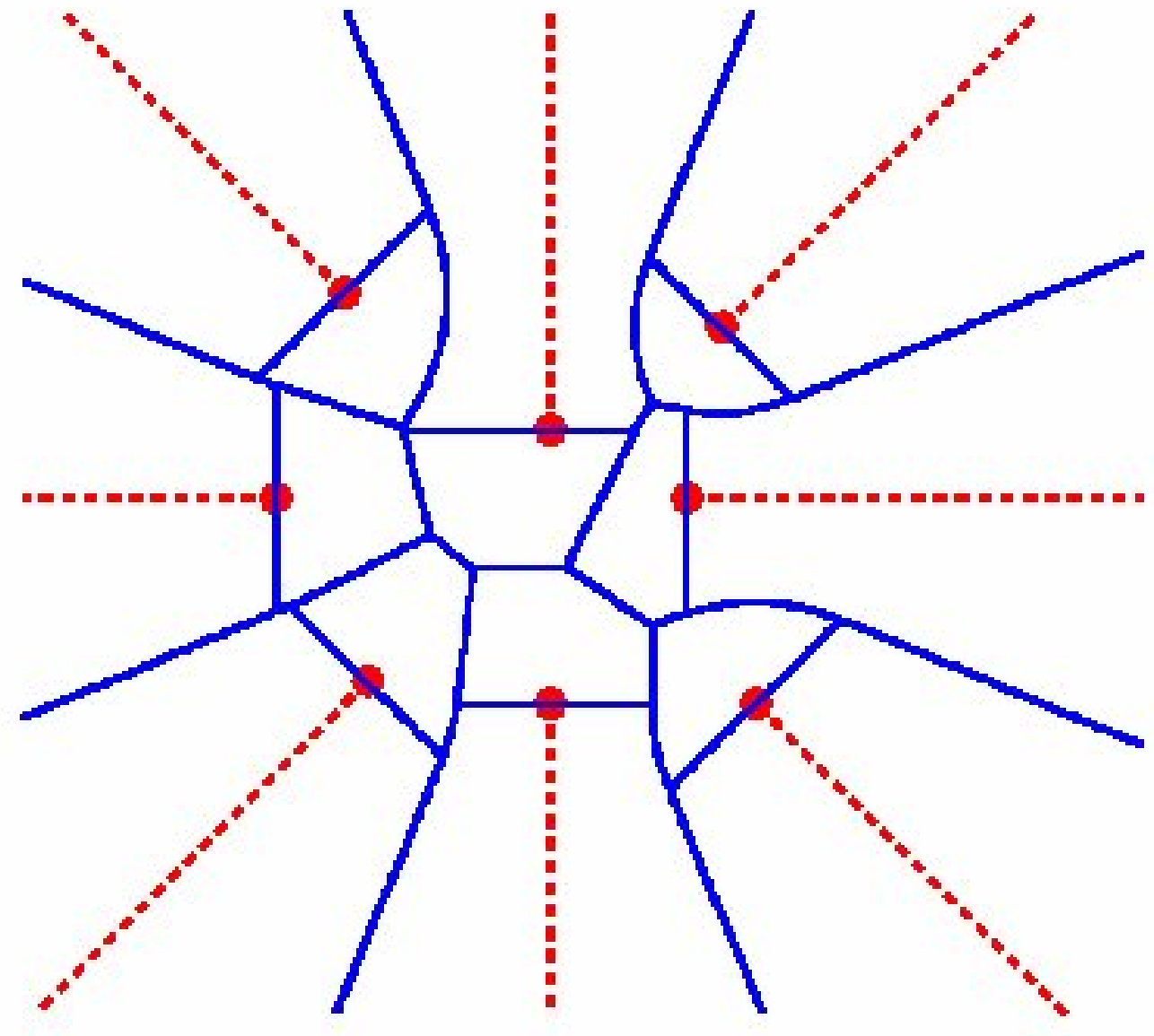}}
  \subfigure[]{\label{fig:linear:star}
    \includegraphics*[angle=-90,totalheight=\subfigwidth]{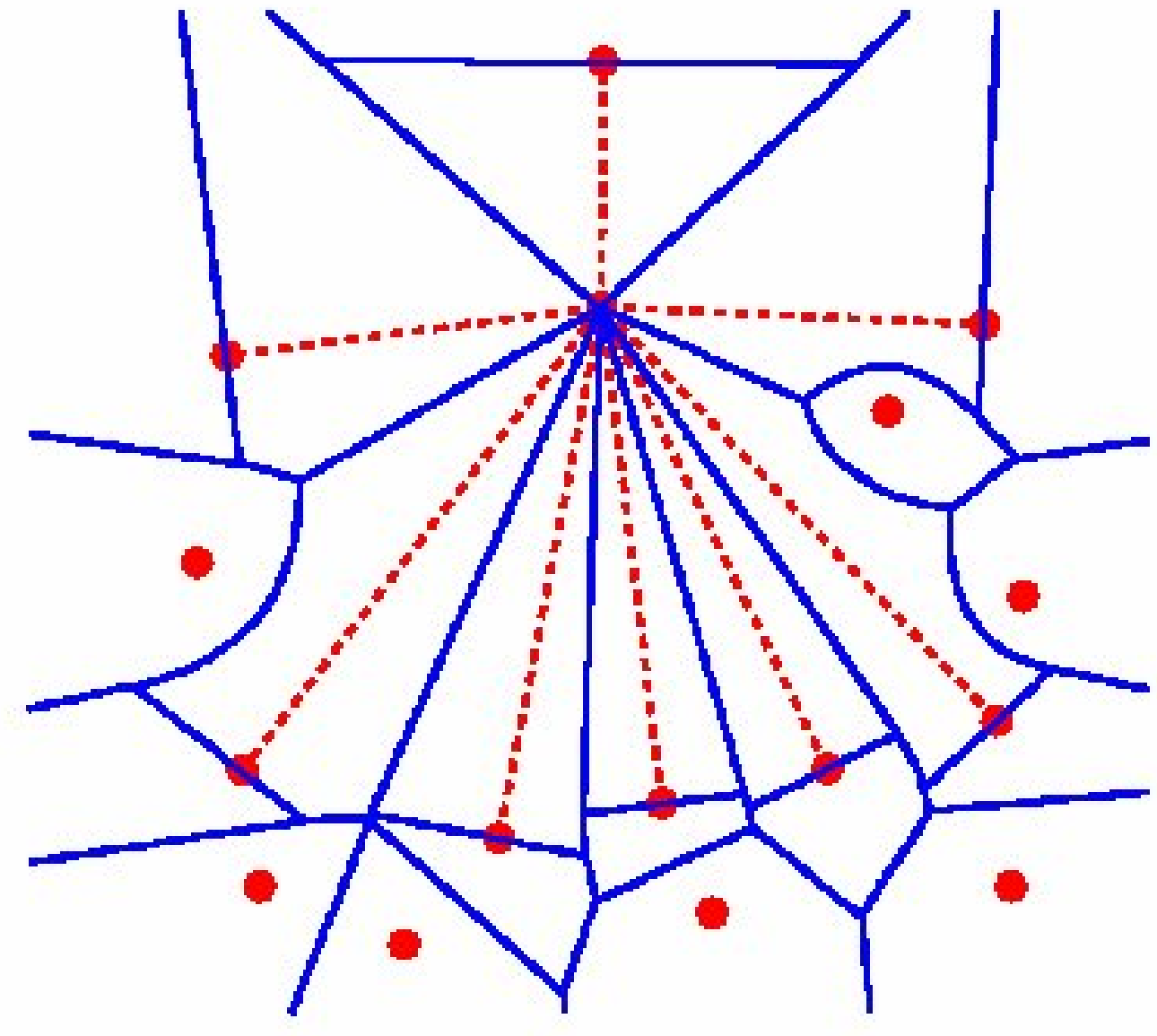}}

  \caption[Voronoi diagrams of linear objects]{Voronoi diagrams of
    various sets of linear objects as induced by the Euclidean metric
    computed with our framework. The sites are illustrated with dashed
    curves.
    \subref{fig:linear:grid}~The Voronoi diagram of 4$\times$4 grid with
    three line segments connecting 6~grid points.
    \subref{fig:linear:grid-and-bound}~The same set of sites as
    in~\subref{fig:linear:grid} with a line, a segment, and two rays
    around the grid points.
    \subref{fig:linear:ray}~The Voronoi diagrams of 8 rays in the plane.
    \subref{fig:linear:star}~The Voronoi diagram of 8~segments and
    7~isolated points. All segments intersect at one point, which is
    one of their endpoints.
  }
  \label{fig:linear}
\end{figure}

The Euclidean Voronoi diagram of segments in the plane
is probably the most frequently used Voronoi diagram after the Voronoi
diagram of points.
We present an exact implementation for a traits class for the computation
of the Voronoi diagram of a set of interior-disjoint linear geometric
objects (lines, rays, and line segments) and points.
Figure~\ref{fig:linear} shows diverse cases of Voronoi diagrams of
linear objects computed with our traits class.
Other exact implementations for the Voronoi diagram of segments
include the one provided by \cgal~\cite{k-reisvd-04,cgal:k-sdg2-08}
and the divide-and-conquer algorithm by
Aichholzer~\etal~\cite{aaahjp-dcvdr-09a}, which uses \cgal predicates
as basic building-blocks.
Both implementations do not support ray sites and line sites.
The algorithm by Aichholzer~\etal uses a circle to bound the Voronoi
edge-graph. A bounding circle cannot contain unbounded linear
entities, such as lines and rays.
Our implementation here is aimed at demonstrating the generality of
our framework.

As mentioned before, the bisectors of two points or two lines in the plane 
are composed of linear curves.
The bisector of a point and a line is a parabola.
As a result, the bisector of two sites in the plane is composed of
linear objects (segments, rays, and lines) and parabolic arcs.
The bisector of two segments can be composed of up to 
7~arcs~\cite{y-avdssc-87}.
It should be noted that even computing just the linear bisectors of 
the diagram is not a trivial task.
Indeed, the bisector of two lines is two perpendicular lines, but those 
lines do not necessarily possess rational coefficients.
We support segments with intersecting vertices, which can induce
two-dimensional bisectors as the region dominated by the joint vertex
can be two-dimensional.

The traits class \ccode{Linear\_objects\_Voronoi\_traits\_2} uses the following
technique to deal with the two-dimensional bisectors and the relative 
complex structure of the bisector curves.
The traits class treats each segment as a set of 3~different Voronoi sites: 
the two vertices of the segment and the open set that is the interior of 
the segment.\footnote{A ray is treated as 2~different sites.}
The handling of all predicates, specifically predicates related to 
bisector construction, is made much simpler
at the cost of an enlarged set of sites.
The splitting of a site into several sites is implemented in the
\ccode{Make\_xy\_monotone\_3} functor of the \envelopetraits
concept (this functor is not part of the \vorconcept concept).
The class \ccode{Linear\_objects\_Voronoi\_traits\_2} models the
\envelopetraits concept rather than the \vorconcept concept.

The interior of a segment behaves like a full straight line inside the region
between the two perpendicular lines at its vertices.
Moreover, it has no effect on any other point outside that region 
as the vertices of the segment dominate the rest of the Euclidean plane.
The domain of the distance function to the interior of 
the segment is, therefore, defined to be the aforementioned region.
The domain of a ray is the half-plane that contains it and bounded by
the perpendicular line at its source.
The domain of a distance function to a point site and a line site
remains the whole plane.
We implement this by providing the functor 
\ccode{Construct\_projected\_boundary\_2} defined in the \envelopetraits 
concept.

After the splitting we only need to consider three types of bisectors,
namely, the bisector between two points, the bisector between two
lines (where a line can also relate to the interior of a segment or a
ray) that is a two-line bisector, and the bisector between a point and
a line, which is a full parabola. 

Comparing the distances of a point in the plane to two points or two 
linear objects is trivial:
we just compare the ``regular'' squared distance from that point to
the sites.
When comparing the distances to a point and to a linear object we have
to be more careful, as the interiors of segments and rays are not
closed sets.
We compare the squared distances to the point and to the linear
object.
If an ``equal'' result is the outcome of considering an endpoint of the 
linear object, we decide that the point site is closer, as the
endpoint of a linear object (a segment or a ray) is not contained in
its interior.

The final problem we have to face is that the bisector of two lines in 
the plane is two lines that do not have rational coefficients in the
general case.
We observe that these two lines constitutes a degenerate hyperbola
with rational coefficients.
We use the algebraic traits class to construct the degenerate hyperbola that 
represents the bisector of the two input lines.

\section{Spherical Voronoi Diagrams}
\label{sec:impl:sphere-vd}

The computation of Voronoi diagrams makes use of two main operations 
supported by the \aos package:
\begin{inparaenum}[(i)]
\item sweep-based overlay operation, which is used to overlay two minimization
  diagrams, and
\item zone computation-based insertion operation, which is used to
  insert bisector curves that partition cells of the refined
  arrangement;
\end{inparaenum}
see Chapter~\ref{chap:env-to-vd}.

The new \aos{} package extends the aforementioned operations, that is,
the sweep-line and zone-computation, to support two-dimensional
parametric surfaces.
Thus, we utilize the \envelope{} code to handle minimization diagrams
that are embedded on two-dimensional parametric surfaces with little
effort.

This section demonstrates how the ability of constructing arrangements on 
two-dimensional parametric orientable surfaces is exploited to compute
Voronoi diagrams on the sphere
We compute Voronoi diagrams, the bisectors of which are composed of
geodesic arcs, namely, arcs of great circles which are created by 
intersecting the sphere with planes passing through the origin.
Section~\ref{sec:impl:sphere-vd:geodesic} gives some details about
the traits class we have created for the arrangement package that 
enables us to construct arrangements of geodesic arcs on the sphere
using rational arithmetic only.
Section~\ref{sec:impl:sphere-vd:voronoi} describes how we use the
above traits class to create a traits class for our Voronoi framework
for computing Voronoi diagrams embedded on the sphere, the bisectors of
which are geodesic arcs --- the Voronoi diagram of points on the
sphere and the power diagram of circles on the sphere. The section
focuses on the implementation details involved in the development of a
Voronoi diagram embedded on a two dimensional parametric surface.

Further details on the requirements from a traits class that enables the 
computation of Voronoi diagrams on surfaces can be found in 
Section~\ref{sec:env-to-vd:impl}.
Details on constructing arrangements on surfaces in general, and on 
constructing geodesic arrangements on the sphere in particular, 
can be found in~\cite{bfhmw-smtdas-07, fsh-eiaga-08} and in Efi Fogel's 
Ph.\,D.\ thesis~\cite{f-mscaag-08}.

\subsection{Arrangements of Geodesic Arcs on the Sphere}
\label{sec:impl:sphere-vd:geodesic}

\newcommand{\parms}{{\Phi}}                            %
\newcommand{\parmf}{{\phi}}                            %

Complying with the specifications in~\cite{bfhmw-smtdas-07},
we use the following parameterization of the unit sphere:
$\Phi = [-\pi + \alpha, \pi + \alpha] \times [-\frac{\pi}{2},
\frac{\pi}{2}]$,
$\phi_S(u, v) = (\cos u \cos v, \sin u \cos v, \sin v)$, where
$\alpha$ must be substituted for a value, which yields rational $\cos
\alpha$ and rational $\sin \alpha$ and defaults to $0$, when the class is
instantiated (at compile time).
The equator curve, for example, is given by
$\gamma(t) = (\pi(2t - 1) + \alpha, 0)$, for $t \in [0,1]$.
This parameterization induces two contraction points
$p_s = (0, 0, -1) = \phi_S(u,-\frac{\pi}{2})$ and
$p_n = (0, 0, 1) = \phi_S(u,\frac{\pi}{2})$, referred to as the south
and north poles, respectively, and an identification curve
$\{\phi_S(\pi + \alpha,v)\,|\,-\frac{\pi}{2} \leq v \leq \frac{\pi}{2}\}$,
as $\phi_S(-\pi + \alpha,v) = \phi_S(+\pi + \alpha,v)$ for all $v$ (which
coincides with the opposite Prime (Greenwich) Meridian when $\alpha = 0$).
We developed the topology traits to support any type of curves embedded 
on the sphere parameterized as above, without compromising the 
performance of the operations gathered in the traits class.
We concentrate on the details of the geometric traits class as it is most
relevant in this section.

The geometry-traits class for geodesic arcs on the sphere is
parameterized with a geometric kernel~\cite{hhkps-aegk-07}; see also
Section~\ref{ssec:impl:plane-vd:linear}.
The implementation handles all degeneracies, and is exact as long
as the underlying number type supports the arithmetic operations
$+$, $-$, $*$, and $/$ only in unlimited precision over the rationals,
such as the one provided by \gmp, the {\sc Gnu} Multi-Precision bignum
library\citelinks{gmp-link},
even though the embedding surface is a sphere. We are able to use
high-performance kernel models instantiated with exact rational
number-types for the implementation of this geometry-traits class,
as exact rational arithmetic suffices to carry out all necessary
algebraic operations.

The geometry-traits class defines the point type to be an unnormalized
vector in \R{3}, representing the place where the ray emanating
from the origin in the relevant direction pierces the sphere. An arc
of a great circle is represented by its two endpoints, and by the plane
that contains the endpoint directions and goes through the origin. The
orientation of the plane and the source and target points determine
which one of the two arcs of the great circle is considered.
This representation enables an exact, yet efficient, implementation of
all geometric operations required by the geometry-traits concept using
exact rational arithmetic, as normalizing directions and plane normals is
completely avoided.

We describe in details two predicates:
\textbf{Compare~\boldmath$u$} and \textbf{Compare~\boldmath$uv$};
see~\cite{f-mscaag-08}
for the 
\begin{wrapfigure}[12]{r}{7cm}
  \vspace{-15pt}
  \centerline{
    \subfigure[]{
      \includegraphics[width=3.5cm]{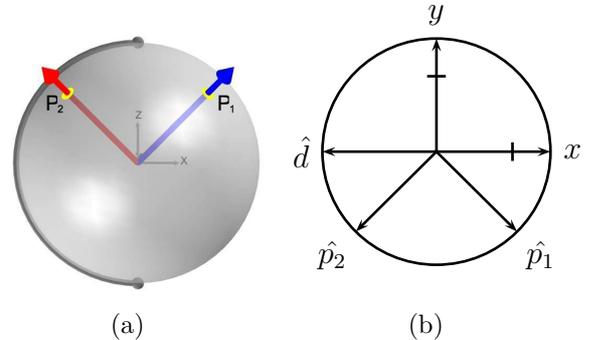}}
    \subfigure[]{
      \pspicture[](-1.9,-1.9)(1.9,1.9)
      \psset{unit=1cm,linewidth=1pt}
      \pscircle[linewidth=1pt](0,0){1.5}
      \psaxes[labels=none]{->}(0,0)(1.5,1.5)
      \uput[0]{0}(1.5,0){$x$}
      \uput[90]{0}(0,1.5){$y$}
      \psline{->}(0,0)(-1.5,0)
      \uput[180]{0}(-1.5,0){$\hat{d}$}
      \psline{->}(0,0)(1.06,-1.06)
      \uput[-45]{0}(1.06,-1.06){$\hat{p_1}$}
      \psline{->}(0,0)(-1.06,-1.06)
      \uput[-135]{0}(-1.06,-1.06){$\hat{p_2}$}
      \endpspicture
    }}
  \vspace{-10pt}
  \caption[Compare-$u$ predicate implementation]
          {The implementation of the Compare-$u$ predicate in the traits 
            class for geodesic arcs on the sphere}
\end{wrapfigure}
complete set of the concept requirements. The former compares two points $p_1$
and $p_2$ by their $u$-coordinates. 
The concept admits the assumption that the input points do not
coincide with the contraction points and do not lie on the
identification arc. 
Recall that points are in fact unnormalized vectors in \R{3}. We
project $p_1$ and $p_2$ onto the $xy$-plane to obtain two-dimensional
unnormalized vectors $\hat{p_1}$ and $\hat{p_2}$, respectively. We
compute the intersection between the identification arc and the
$xy$-plane to obtain a third two-dimensional unnormalized vector
$\hat{d}$. Finally, we test whether $\hat{d}$ is reached strictly
before $\hat{p_2}$ is reached, while rotating counterclockwise
starting at $\hat{p_1}$. This geometric operation is supported by
every geometric kernel of \cgal{}. In the figure on the right $\hat{d}$
is reached strictly before $\hat{p_2}$ is reached. Therefore, the
$u$-coordinate of $p_1$ is larger than the $u$-coordinate of $p_2$.

The predicate \textbf{Compare~\boldmath$uv$} compares two points
$p_1$ and $p_2$ lexicographically. It first applies
\textbf{Compare~\boldmath$u$} to compare the $u$-coordinates of the
two points. If the $u$-coordinates are equal, it applies a predicate
that compares the $v$-coordinates of two points with identical
$u$-coordinates, referred to as \textbf{Compare~\boldmath$v$}. This
predicate first compares the signs of the $z$-coordinates of the two
unnormalized input vectors. If they are identical, it compares the
squares of their normalized $z$-coordinates, essentially avoiding
the square-root operation.

All the required geometric types and geometric operations listed in
the geometric traits concept are implemented using rational
arithmetic only.
Degeneracies, such as overlapping arcs that occur during intersection
computation, are properly handled. The end result is a robust, yet
efficient, implementation.
\subsection{Power Diagrams on the Sphere}
\label{sec:impl:sphere-vd:voronoi}
The spherical Voronoi diagram and its generalization, the power
diagram of circles on the sphere (both defined below) are composed of
geodesic arcs.
Both diagrams have applications similar to the applications of Voronoi
diagrams of points and power diagrams in the plane.
For example, determining whether a point is included in the union of
circles on the sphere, and finding the boundary of the union of
circles on the sphere~\cite{iim-vdlgia-85,s-lvds-02}.
Both also have unique applications, for
example, the properties of the spherical Voronoi diagram can be used
to prove the \emph{the thirteen spheres} theorem~\cite{a-tsnp-04}.

Following are definition for the geodesic distance on the sphere (which is
the analog of the Euclidean distance between two points in the plane)
and the spherical Voronoi diagram
(which is the analog of the Voronoi diagram of points in the plane).

\begin{definition}[Geodesic distance]
  Given two points $p, q \in \unitsphere$, the \emph{geodesic distance}
  between them \distance{p}{q} is defined to be the shortest distance
  measured along a path on the surface of the sphere.
  The geodesic distance \distance{p}{q} is equal to the length of
  a geodesic arc that connects $p$ and $q$.
\end{definition}

\begin{wrapfigure}[13]{r}{7.5cm}
  \vspace{-10pt}
  \centerline{
    \includegraphics[width=6.5cm]{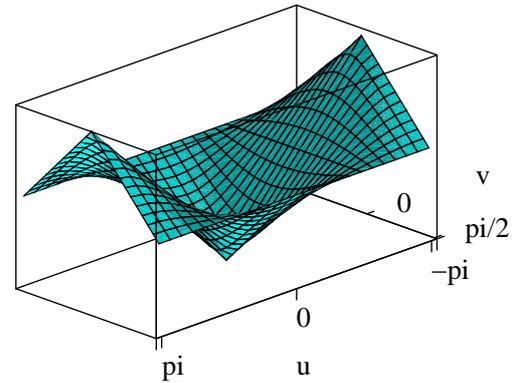}
  }
  \vspace{-5pt}
  \caption{Geodesic distance function on the sphere.}
  \label{fig:sphere-distance}
\end{wrapfigure}
Figure~\ref{fig:sphere-distance} to the right illustrates the distance
function from 
$(0, 0) \in [-\pi, \pi] \times [-\frac{\pi}{2}, \frac{\pi}{2}]$
in the parameter space (defined in
Section~\ref{sec:impl:sphere-vd:geodesic} above) to any other point in
the parameter space.
As expected from the topology of the surface, we get a function whose
values are identified at the boundaries of the parameter space.
Fortunately for us, we do not have to explicitly construct and handle these
types of functions. We only have to correctly answer the predicates
that are required by the algorithm.
The central projection of the intersection of two such functions onto the
sphere is a great circle, as the bisector of two points on the sphere
as induced by the geodesic distance is a great circle.

\begin{definition}[Spherical Voronoi diagram]
  Let $P = \{p_1, \ldots, p_n\}$ be a set of $n$ points in~\unitsphere.
  The \emph{spherical Voronoi diagram} of the set $P$ is the Voronoi 
  diagram as induced by the geodesic distance function.
\end{definition}

\begin{figure}[htbp]
  \addtolength{\subfigcapskip}{5pt}
  \addtolength{\subfigcapmargin}{3pt}

  \centering
  \subfigure[]{\label{sfig:vos:random}
    \includegraphics[width=3.85cm]{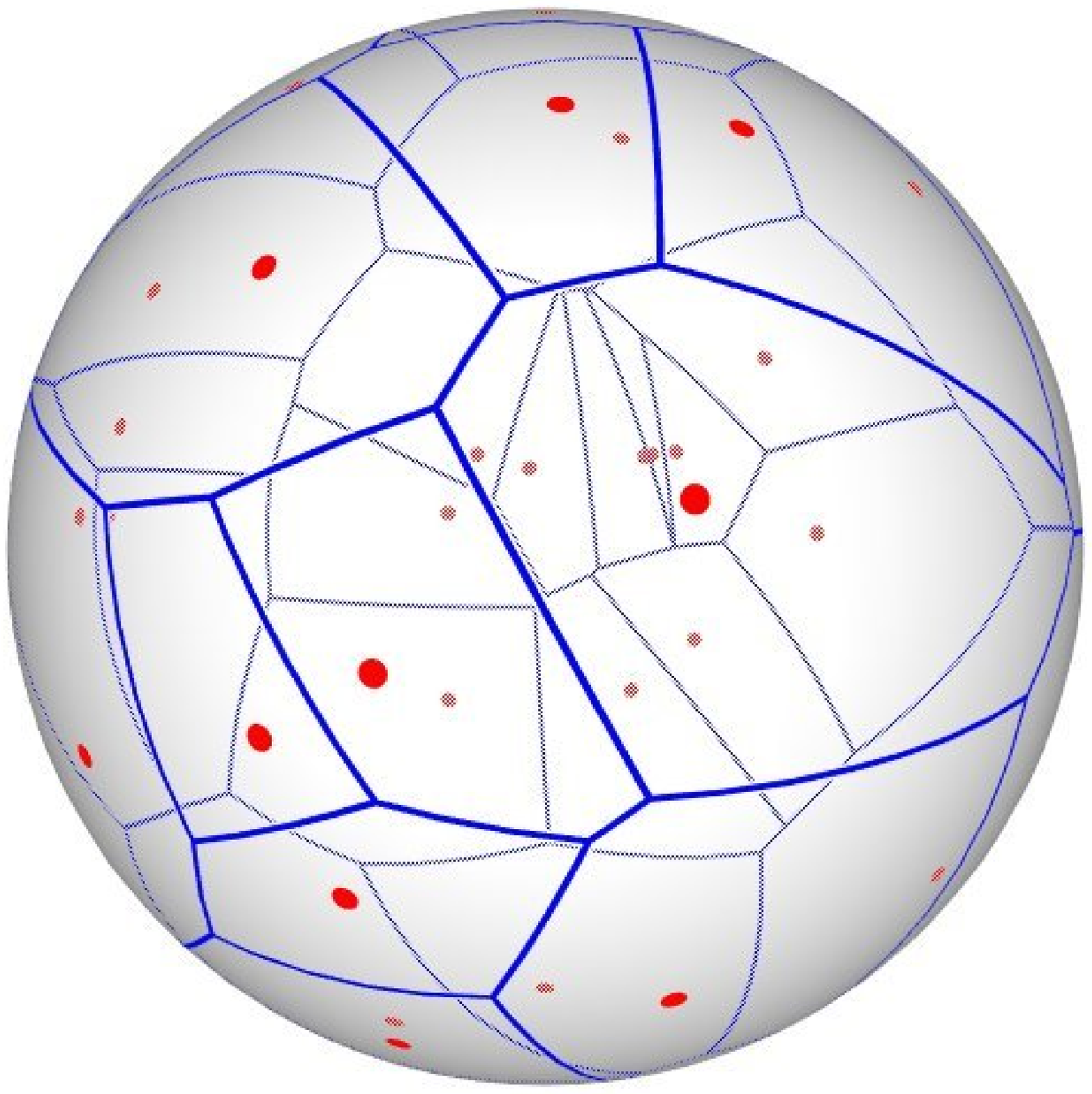}}
  \subfigure[]{\label{sfig:vos:degen}
    \includegraphics[width=3.85cm]{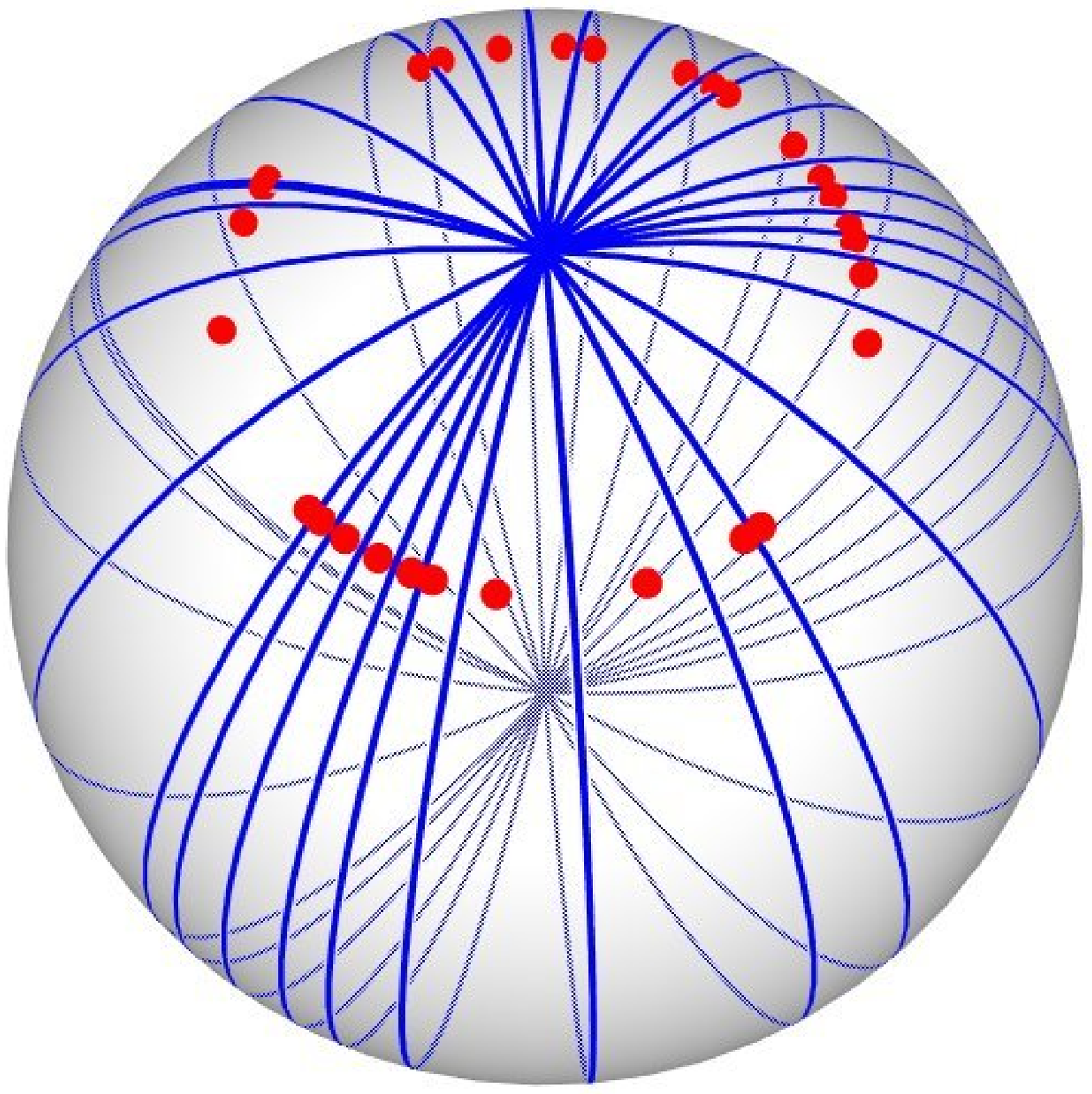}}
  \subfigure[]{\label{sfig:vos:power}
    \includegraphics[width=3.85cm]{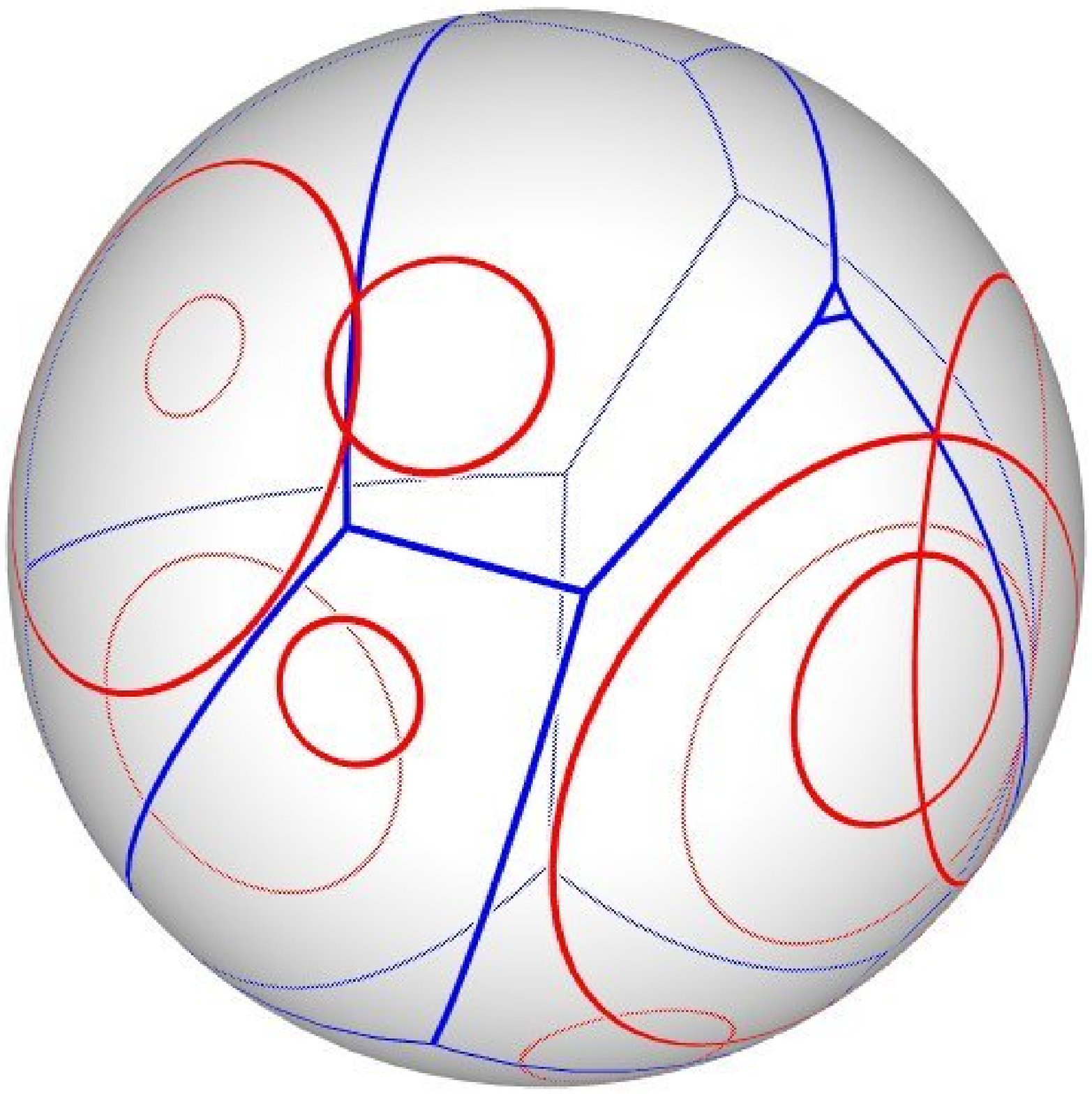}}
  \subfigure[]{\label{sfig:vos:degen-power}
    \includegraphics[width=3.85cm]{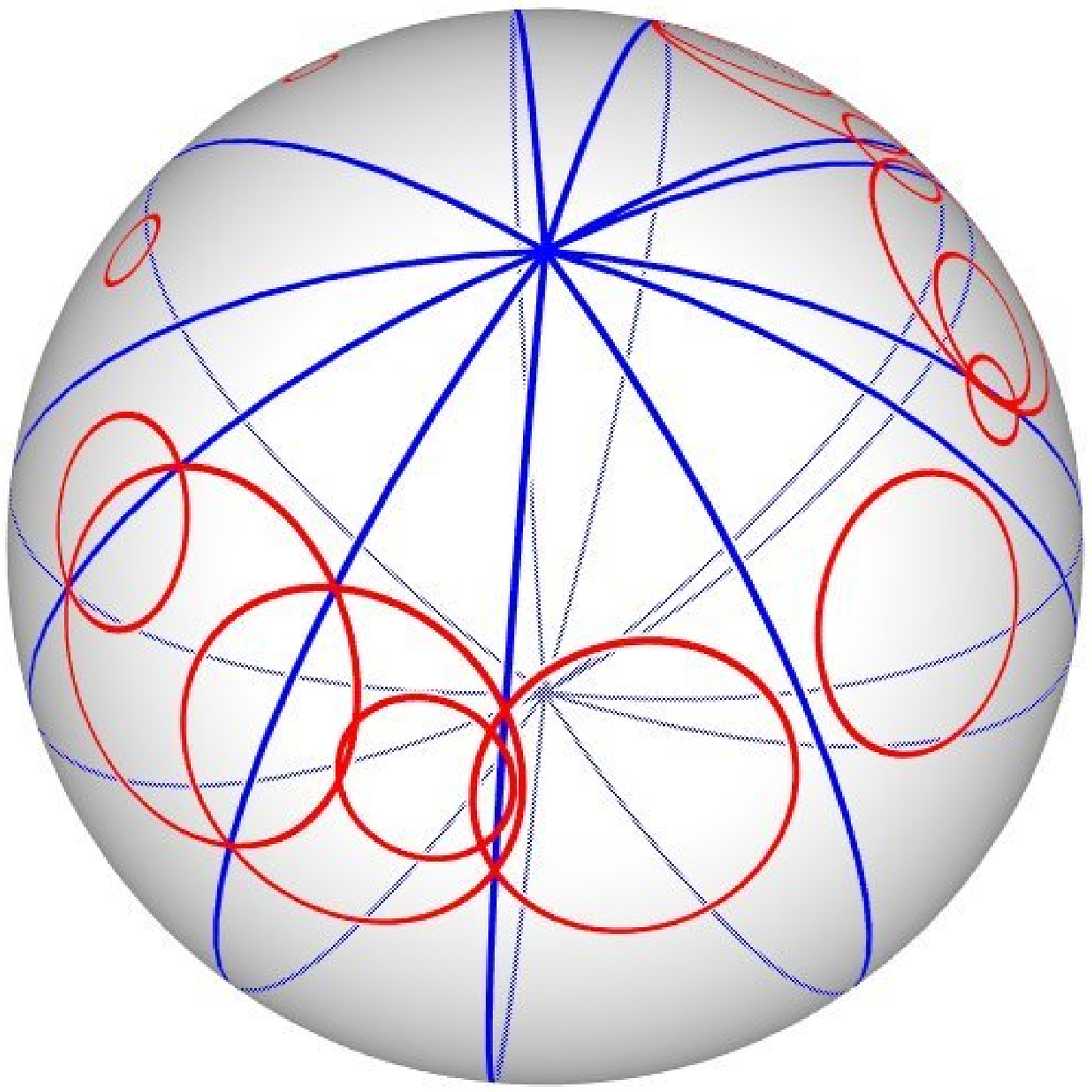}}

  \caption[Voronoi diagrams on the sphere]{Voronoi diagrams on the
    sphere. Sites are drawn in red and Voronoi edges are drawn in
    blue. \subref{sfig:vos:random} The Voronoi diagram of 32 random
    points.
    \subref{sfig:vos:degen} A highly degenerate case of Voronoi
    diagram of 30 point sites on the sphere.
    \subref{sfig:vos:power} The power diagram of 10 random circles.
    \subref{sfig:vos:degen-power} A degenerate power diagram of
    14 sites on the sphere.}
  \label{fig:vos}
  \addtolength{\subfigcapskip}{-5pt}
  \addtolength{\subfigcapmargin}{-5pt}
\end{figure}

The bisector of two point sites on the sphere is a great circle that
is the intersection of the sphere and the bisector plane of the points
in \R{3}, as imposed by the Euclidean
metric~\cite{nlc-vds-02, obsc-stcavd-00}.
The \emph{power diagram} of circles on the sphere, also known as the
\emph{Laguerre Voronoi diagram on the sphere}, is also 
composed of arcs of great circles~\cite{s-lvds-02}; see
Figures~\ref{sfig:vos:power} and~\ref{sfig:vos:degen-power}.

Given two circles on the sphere $c_1$ and $c_2$, let $\pi_1$ and $\pi_2$
be the planes containing $c_1$ and $c_2$, respectively. The bisector
of $c_1$ and $c_2$ is the intersection of the sphere and the plane that
contains the intersection line of $\pi_1$ and $\pi_2$ and the center
of the sphere.
If $\pi_1$ and $\pi_2$ are parallel planes, then the bisector is the
intersection of the sphere and the plane that contains the center of
the sphere and is parallel to both $\pi_1$ and $\pi_2$.

We believe that the following argument, similar to arguments in
previous sections, is true;
more details are supplied in Chapter~\ref{chap:conclusion}.
\begin{conjecture} 
The class of all Voronoi diagrams with great circles as bisectors is
identical to the class of power diagrams of circles on the sphere.
\end{conjecture}

The \ccode{Arr\_geodesic\_arc\_on\_sphere\_traits\_2} traits class for
computing arrangements of geodesic arcs on the sphere provides
predicates and operations needed for
the computation of Voronoi diagrams on the sphere, the bisectors of
which are great circles or piecewise curves composed of geodesic
arcs.
It is the basis of the \ccode{Spherical\_power\_diagram\_traits\_2}
class that enables the construction of power diagrams of sets of
circles on the sphere using our framework.
The traits class applies only rational arithmetic, and so, only
Voronoi diagrams of circles defined by the intersection of
\emph{rational} planes with the sphere (and rational points on the sphere)
are supported.
This is not a very strong limitation as every input point can be
approximated with a rational point up to any desired
precision~\cite{cdr-rrmrg-92}.

The \ccode{Spherical\_power\_diagram\_traits\_2} is a model of the
\vorconcept concept and defines all required types and functors.
It is parameterized by a geometric kernel of \cgal that is passed also
to the \ccode{Arr\_geodesic\_arc\_on\_sphere\_traits\_2} base class.
The kernel should support exact rational number-types to carry all
necessary operation in an exact manner.
The traits class handles all degenerate situations efficiently.

A \site in the \ccode{Spherical\_power\_diagram\_traits\_2} is an
object of type \ccode{Kernel::Plane\_3} --- a 3D plane object of the
geometric kernel.
The plane represents the Voronoi site that is the circle intersection
of the plane and the embedding sphere.
Not all planes in three dimensions can represent a site in our traits
class.
Recall that we are constructing the arrangement on the sphere, so only 
planes whose distance from the origin is $\leq 1$ are used.
The \ccode{Construct\_bisector\_2} functor is responsible for
constructing the bisector of two circle sites on the sphere.
It intersects the two underlying planes of the Voronoi sites, and
constructs the great circle that is defined by the plane containing
the intersection line and the origin.
In the case that both planes are parallel, we construct the great circle
that is defined by the plane parallel to both planes and passing
through the origin.

The functor \ccode{Construct\_point\_on\_x\_monotone\_2} constructs a
point in the interior of a $u$-monotone curve.
Every $u$-monotone arc has a source point and a target point.
Recall that points are represented as unnormalized vectors in
three-dimensions.
If the geodesic arc is smaller than~$\pi$, adding the source and
target vectors results in an unnormalized vector that represents a
point in the interior of the arc.
If the geodesic arc is bigger than~$\pi$ we construct the opposite
vector.
The case where the geodesic arc is equal to $\pi$, in which the source
and target are antipodal, is handled by constructing an orthogonal
vector to both the target vector and the normal of the plane.
The resulting vector represents a point in the interior of the
$u$-monotone arc.

The remaining functors are the proximity predicates
\ccode{Compare\_distance\_at\_point\_2} and
\ccode{Compare\_distance\_above\_2}.
The former predicate is given two circle sites (defined by
two planes) and a point on the sphere (defined by a vector) and is
to decide which of the sites the point is ``closer'' to.
The circle that is closer to the point is the one defined by the plane
encountered first when shooting a ray in the direction of the point
from the origin.
The predicate intersects the underlying planes with the line passing
through the origin and the input point.
Then, it compares the distances of the intersection points to the
origin.
An intersection point could be in the same direction as the input
point from the origin, or in the opposite direction.
A site whose intersection point is in the opposite direction is
considered farther from the input point than a site whose intersection
point is in the same direction.
If the intersection points of both circles are in the opposite
direction, the opposite result is returned.
If one of the planes is parallel to the vector that represents the
input point then, if the intersection point of the other site is in
the same direction, the other site is closer, otherwise the site with
the underlying parallel plane is closer.
If both planes are parallel to the vector of the point we have an
equality.

The \ccode{Compare\_distance\_above\_2} is implemented by
considering the geodesic arc itself and the planes that define the two sites.
We observe the $u$ and $v$ coordinates of the normals to the planes and
deduce an answer.
This is similar to the implementation of the same functor in the power
diagram of disks in the plane; see Section~\ref{ssec:impl:plane-vd:linear}.
The main difference is that on the sphere, circles with the same center
but with different radii still have a bisector,
whereas in the plane the disk with the larger radius dominates the
whole plane.
In other words, two sites always have a bisector.
If the normals to the planes are in the same direction (circles having
the same center) then the distance of the planes to the origin (radii
of the circles) is also considered while deducing an answer.

Figure~\ref{fig:vos} shows some Voronoi diagrams on the
sphere. Figures~\ref{sfig:vos:degen} and~\ref{sfig:vos:degen-power}
are of specific interest as they constitute highly degenerate
scenarios.

\section{Speeding-up the Computation}
\label{sec:exper:speed}
The main performance hit during the execution of our algorithm (and
exact geometric algorithms in general) is caused while trying to
exactly evaluate geometric predicates, \ie using exact arithmetic
operations.
There is a conceptual efficiency scale of operations, ranging from the
most efficient operations to the less efficient operations.
The most efficient operations are the ones using machine-supported
arithmetic, such as integers and floating-point numbers.
Using exact rational-arithmetic is less efficient than the
machine-supported arithmetic, but is still much faster than using
exact higher-degree algebraic computation.
As the algebraic degree goes up, the performance of operations goes down.

There are several optimizations that can be applied to a geometric
algorithm. The first is trying to use the more efficient arithmetic
according the scale described above. If the used arithmetic fails we
go down the scale of efficiency and use a higher-precision or higher
algebraic-degree arithmetic.
Alternatively, one can try and use simpler (approximated) operations
to the reduce the use of the exact and complicated operations (\eg
when performing an intersection test between two
polylines, the intersection of their bounding boxes can be tested
first).
Another technique is to reduce the number of (redundant) geometric
operations used.
The technique of using a simpler approximated version of a predicate
to alleviate the use of the more expensive operation is referred to as
\emph{filtering}. When the filtering exploits the specific geometry of
the problem at hand it is referred to as \emph{geometric filtering}.
We describe below techniques we have applied to reduce the running
time of our algorithm in the context of affine Voronoi diagrams;
Section~\ref{ssec:impl:plane-vd:linear}.
We use fast geometric filters and apply specific techniques at a
high level for (affine) Voronoi diagrams that reduce lower-level
filter-failures.

Most implementations of traits classes for our framework are based on
traits classes for the arrangement package.
The first step in optimizing the code is to pick the most efficient
traits class to handle the bisector curves of the specific diagram.
The \arrlineartraits class is parameterized with a kernel object, the
choice of which has a drastic effect on the performance of our code.

\cgal provides basic geometric kernels with built-in
filtering mechanisms.
The \lazyker class is an efficient geometric kernel of
\cgal~\cite{fp-gleseg-06}, which supplies lazy construction and
evaluation of the geometric predicates and operations that delay the
exact geometric computations at run-time.
Every geometric object of the \lazyker class holds two other instances of
different classes; one instance is approximated using machine
floating-point interval-arithmetic, and the other instance is exact 
using rational arithmetic.
The kernel tries to use the fast machine arithmetic, and only if it
fails, it performs the required operation using the exact version of
the object.
Every geometric object that was created using kernel operations keeps
the tree of operations that created it.
This way, when an approximated operation fails, the tree of operations
can be retrieved and the operations are performed all over again using
the exact version.
However, if an operation does not fail, no costly operation is
performed (a lazy behavior).
This filtering mechanism is applied at the geometric level, which
proved to be more efficient than the regular interval-arithmetic
filtering mechanisms that are applied at the arithmetic (number-type)
level.
We use the \lazyker as the base for our traits class in the case of
Voronoi diagrams with linear bisectors.
A future work is to implement a traits class for the \mobius diagram
using the \circker with other geometric filtering
optimizations~\cite{ptt-gfpca-06}; see Chapter~\ref{chap:conclusion}.

The aforementioned geometric kernels and traits classes for the
arrangement package provide a very general approach
for accelerating geometric computations.
However, the filtering techniques applied at this low level are not
aware of certain properties of the specific problem at hand, which can
be exploited to improve performance.
Specifically, the lower envelope algorithm applies some geometric
predicates that incur geometric filter-failures due to the
combinatorial structure of Voronoi diagrams.
With a simple observation, we can avoid those filter-failures during
the execution of the divide-and-conquer algorithm, increasing the
efficiency of our code.

\begin{observation}
  Given three Voronoi sites and their pairwise bisectors, if two of
  the bisectors intersect then the third bisector passes through their
  intersection point(s).
  \label{obs:3bis}
\end{observation}

Indeed, denoting the sites by $o_1$, $o_2$, and $o_3$, then at every
intersection point~$x$ of any two of the three corresponding
bisectors, we have 
$\rho(x, o_1) = \rho(x, o_2) = \rho(x, o_3)$.

The merge step of the divide-and-conquer of algorithm
(see Section~\ref{sec:env-to-vd:dc-vd}) consists of 
determining the structure of the Voronoi diagram over each face~$f$ of
the overlay of \vor{S_1} and \vor{S_2}.
The face~$f$ has a fixed nearest \neighbor~$s_1$ from~$S_1$ and a
fixed nearest \neighbor~$s_2$ from~$S_2$.
We partition~$f$ into subfaces consisting
of points nearer to $s_1$ and subfaces consisting of points nearer to
$s_2$ by inserting their bisector into the arrangement.

If $f$ is split into at least two regions, then the bisector of
$s_1$ and $s_2$ generally intersects some edges of $f$.\footnote{This
  is not always true as a bisector can form a closed loop within the
  face.}
Let us assume, without loss of generality, that the bisector
intersects an edge $e$ that originally belonged to \vor{S_1}.
The edge $e$ is part of the bisector between $s_1 \in S_1$ and
another site $t \in S_1$.
Assuming general position, when examining the neighboring face $f'$ of
$f$ on the other side of $e$, the two fixed nearest sites are $t \in
S_1$ and $s_2 \in S_2$.
According to Observation~\ref{obs:3bis} the bisector of $t$ and $s_2$
will pass through the intersection points of the bisector of $s_1$ and
$s_2$, and $e$.

\begin{figure}%
  \providelength{\subfigwidth}\setlength{\subfigwidth}{140pt}

  \centerline{
    \subfigure[]{\label{fig:3bisectors-a}\includegraphics[width=\subfigwidth]{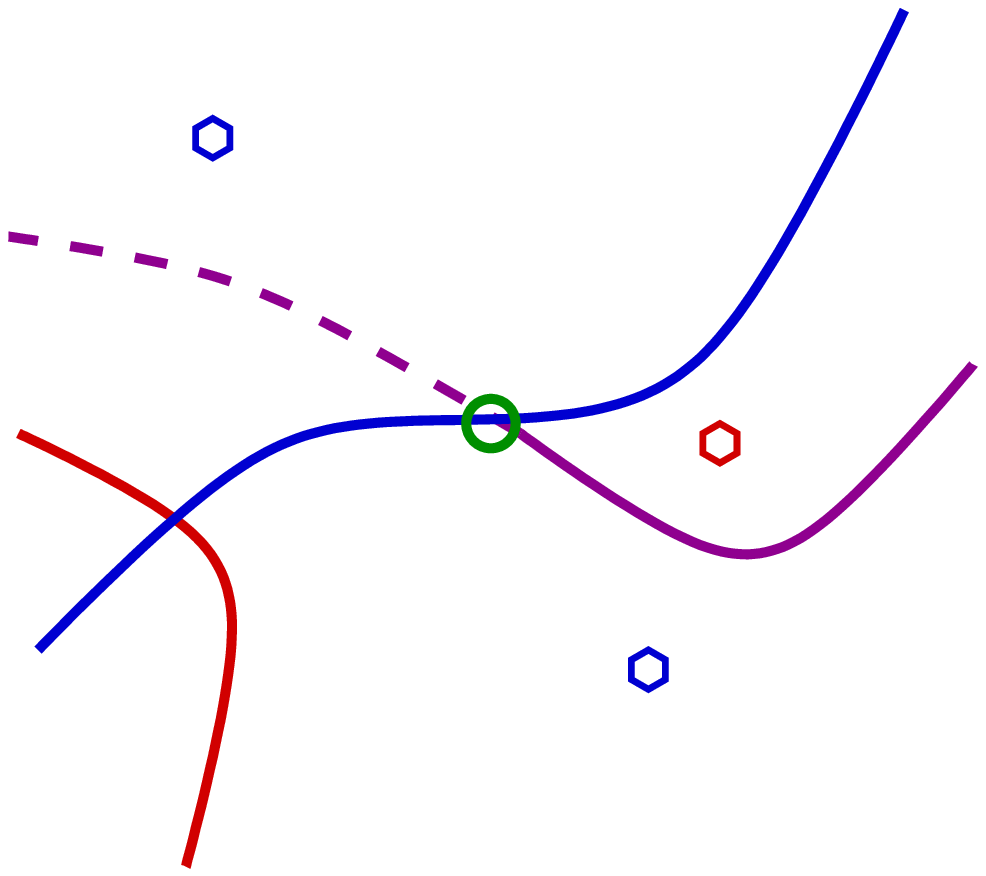}}
    \hspace{\spacewidth}
    \subfigure[]{\label{fig:3bisectors-b}\includegraphics[width=\subfigwidth]{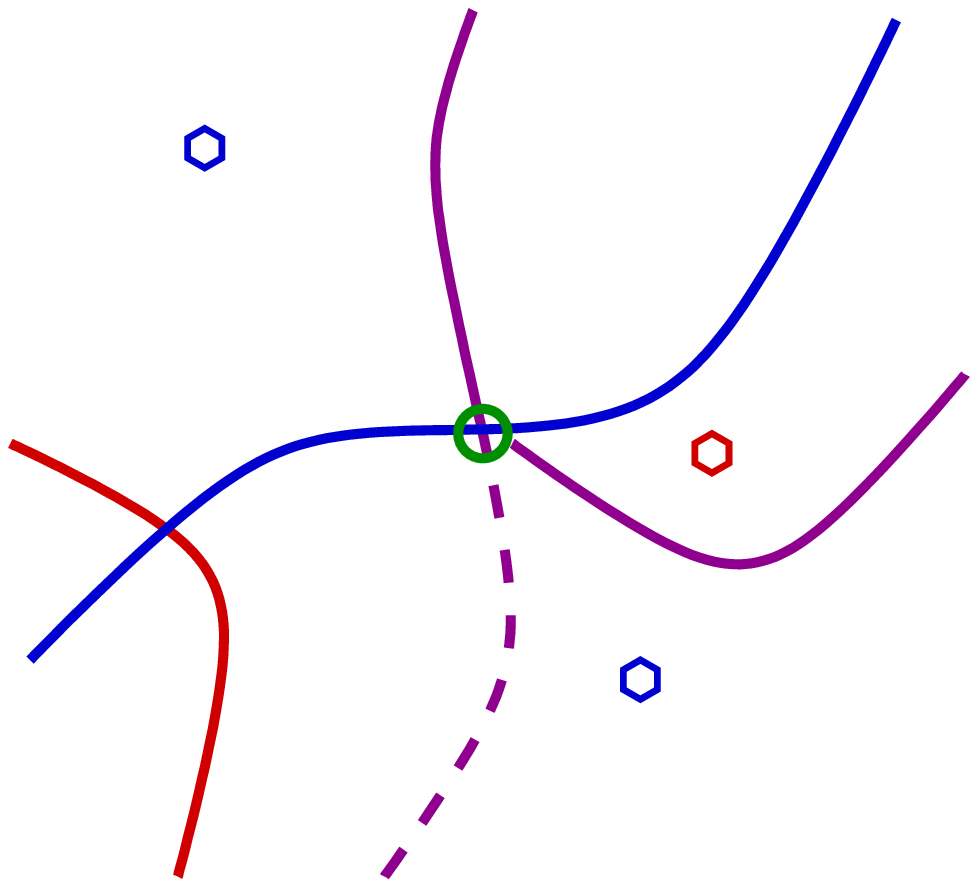}}
  }
  
  \vspace{-10pt}
  \caption[Three-bisectors optimization]{Three-bisectors optimization.
    During the merge step of the blue and red Voronoi diagrams each
    face is split by the bisector of the two nearest sites of both
    diagrams.
    \subref{fig:3bisectors-a}~The bisector of the right blue site and
    the red site splits the right face.
    \subref{fig:3bisectors-b}~The bisector of the left blue site and
    the red site splits the left face.
    Notice that the bisector passes through the intersection point of
    the former bisector and the boundary of the right face 
    (the intersection point is circled in green).
  }
  \label{fig:3bisectors}
\end{figure}

\begin{table}[h] %
  \caption[Three-bisector and simplified-zone optimizations]{Effect of
    the three-bisector and simplified zone optimizations on random
    points. Time is measured in seconds.
    FF --- the number of filter-failures that occurred during the
    computation,
    V --- the number of vertices of the diagram,
    E --- the number of edges of the diagram,
    F --- the number of faces of the diagram,
    3-Bis. --- the three-bisector optimization,
    S-Z --- the simplified zone optimization.
  }
  \label{tab:3bisec}

  \begin{center}
  \begin{tabular}{|l||r|r|r||r|r||r|r||r|r|}
    \hline

    \multirow{2}*{\textbf{Input}} &
    \multicolumn{3}{c||}{\textbf{Diagram Size}} & 
    \multicolumn{2}{c||}{\textbf{No opt.}} &
    \multicolumn{2}{c||}{\textbf{3-Bis. opt.}} &
    \multicolumn{2}{c|}{\textbf{3-Bis. + S-Z}} \\ \cline{2-10}

    &
    \multicolumn{1}{c|}{\textbf{V}} &
    \multicolumn{1}{c|}{\textbf{E}} &
    \multicolumn{1}{c||}{\textbf{F}} &
    
    \multicolumn{1}{c|}{\textbf{Time}} &
    \multicolumn{1}{c||}{\textbf{FF}} &

    \multicolumn{1}{c|}{\textbf{Time}} &
    \multicolumn{1}{c||}{\textbf{FF}} &

    \multicolumn{1}{c|}{\textbf{Time}} &
    \multicolumn{1}{c|}{\textbf{FF}} \\
    
    \hline
    \hline

    \textit{rnd\_1000} & 1979  & 2978  & 1000  & 5.52  & 75105  &
                         3.58  & 11408 & 2.72  & 599             \\ \hline
    \textit{rnd\_2000} & 3983  & 5982  & 2000  & 12.57 & 168411 &
                         8.30  & 25099 & 6.58  & 1285            \\ \hline
    \textit{rnd\_4000} & 7979  & 11978 & 4000  & 28.54 & 370848 &
                         19.66 & 58098 & 15.41 & 2807            \\ \hline
  \end{tabular}
  \end{center}
\end{table}

Each feature of the overlay is extended with the nearest Voronoi sites
to it.
When inserting a bisector into the overlay and creating a new
vertex at the intersection point, we update the information of the new
vertex.
When inserting another bisector to the overlay and trying to decide if
a point is on the bisector curve, we first check this information
instead of using a geometric predicate, which is redundant and would
have caused a filter to fail.
We call this optimization the \emph{three-bisectors optimization}; see
Figure~\ref{fig:3bisectors} for an illustration.

A bisector curve is inserted into the underlying arrangement using a
\emph{zone-construction} algorithm. The zone-construction algorithm of
the arrangement package is most general and can handle any input,
including a face that contains inner-holes, isolated vertices, or
``antennas''~\cite{wfzh-aptaca-07}.
In the case of a Voronoi diagram with affine bisectors we always
insert the bisector curve into a \emph{convex} face (with no inner
holes). 
We use a simplified version of the zone algorithm, which gives us
additional performance improvement.

The simple version works in the case of inserting a curve (line) into
a bounded convex face (that is composed of linear segments).
First, we try to detect if one of the vertices of the faces was
created by inserting a bisector (a vertex that will invoke the
three-bisector optimizations).
This avoids the costly operations performed by the regular zone
algorithm for finding the vertex by traversing the edges of the face.
After we have found a maximum of two intersection points we stop, as
this is the maximum number of intersections.
Additionally, we do not check intersections with inner holes and do
not consider other degenerate situations.
For the other cases, \ie inserting a line into an unbounded face or
a zone-insertion that admits an overlap, we use the standard zone
algorithm supplied by the arrangement package.
Table~\ref{tab:3bisec} shows the improvements in running time achieved
by our optimizations.

\chapter{Discussion: Advantages and Limitations}
\label{chap:exper}

Our framework consists of several fundamental concepts:
\begin{inparaenum}[(i)]
\item robust computation through the adherence to the exact
  computation paradigm, utilizing components from \cgal,
\item using randomization on the input to the divide-and-conquer
  construction algorithm, yielding an efficient algorithm, and
\item the representation of Voronoi diagrams as arrangements, \cgal's
  arrangements to be precise.
\end{inparaenum}

Following is a comprehensive discussion on both the advantages and
the limitations of our framework.
The advantages are discussed in Section~\ref{sec:advantages} and
the limitations are discussed in Section~\ref{sec:disadvantages}.

\section{Advantages}
\label{sec:advantages}
\label{sec:advantages:optimality}

The major strength of our approach is its completeness, robustness,
and generality, that is, the ability to handle degenerate input, the
agility to produce exact results, and the capability to construct
diverse types of Voronoi diagrams with a relatively small effort.
The code is designed to successfully handle degenerate input, while
exploiting the synergy between generic programming and exact geometric
computing, and the divide-and-conquer framework to construct Voronoi
diagrams.

\subsection*{Generality}
\label{sec:advantages:general}

A prime advantage of our framework is the ability to compute new
types of Voronoi diagrams with relative ease.
The diagrams are represented as arrangements of \cgal, and the
implementation of a new type of diagrams is based on a traits class
for the arrangement package, which supplies the handling of bisector
curves of the sites.
A Voronoi diagram whose bisectors can be represented with an existing
traits class for the arrangement package can be easily
developed. Existing traits classes support bounded and unbounded
linear curves (line segments, rays, and lines), circles and linear
segments, conic arcs, \bez{} curves, and algebraic curves of arbitrary
degree.

The framework supports a vast variety of Voronoi diagrams with
different properties.
The entire collection of supported Voronoi diagrams cannot be
supported by most other frameworks.
Existing frameworks for computing Voronoi diagrams usually support a
specific family of Voronoi diagrams, \eg planar Voronoi diagrams of
points under the $L_p$ metric, planar Voronoi diagrams of general
sites under the Euclidean metric etc.
Using an envelope-construction based algorithm allows our framework to
support two-dimensional diagrams with almost no restrictions;
we can implement linear Voronoi diagrams as well as Voronoi diagrams
with quadratic complexity, and Voronoi diagrams with two-dimensional
bisectors.
The diagrams do not have to conform with the abstract Voronoi diagrams
definition, for example anisotropic diagrams: we construct such
diagrams whereas the theoretical requirements for an abstract Voronoi
diagram include that all bisectors will divide the plane into two
unbounded regions.
As mentioned before, we can also compute Voronoi diagrams on
two-dimensional orientable parametric surfaces.
Existing topology-traits classes in the arrangement package include a
topology-traits class for the plane, the sphere, elliptic quadrics, and
ring Dupin cyclides that generalize tori.

More information on realizations of the framework can be found in
Chapter~\ref{chap:impl}.

\subsection*{Theoretical Optimality}

\begin{figure}[h]
  \providelength{\subfigwidth}\setlength{\subfigwidth}{160pt}
  \providelength{\spacewidth}\setlength{\spacewidth}{40pt}

  \centering
  \subfigure[]{\label{fig:exper:randomization-dani} \hspace{20pt}
    \includegraphics[width=\subfigwidth - 30pt]{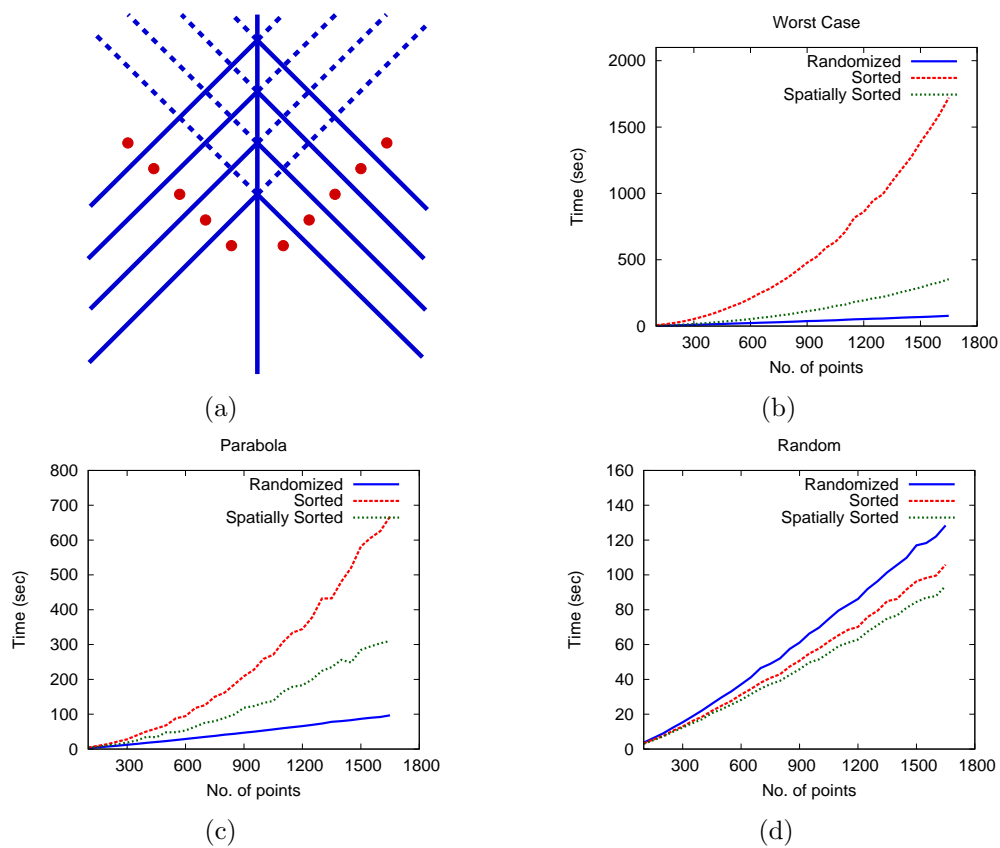} \hspace{0pt}}
  \hspace{\spacewidth}
  \subfigure[]{\label{fig:exper:randomization-wc}\includegraphics[width=\subfigwidth]{worst_case_randomization.epsi}}
  \hspace{\spacewidth}
  \subfigure[]{\label{fig:exper:randomization-p}\includegraphics[width=\subfigwidth]{parabola.epsi}}
  \hspace{\spacewidth}
  \subfigure[]{\label{fig:exper:randomization-r}\includegraphics[width=\subfigwidth]{random_input.epsi}}
  \caption[Effect of randomization]{Effect of the random partitioning
    of Voronoi sites.
    The figure shows the running time in seconds as a function of the
    number of sites in various cases of the standard Voronoi
    diagram (point sites with $L_2$ metric)
    using different partitioning strategies.
    \subref{fig:exper:randomization-dani} A worst-case example of
    10~point sites in the case of the standard Voronoi diagram. 
    The dashed segments are segments that exist in the overlay
    (if we use the ``Sorted'' partition strategy) but will not be
    present in the final diagram.
    \subref{fig:exper:randomization-wc} The running time
    graph in the worst-case constellation where the point sites are
    arranged as in \subref{fig:exper:randomization-dani}.
    \subref{fig:exper:randomization-p} The running time
    graph in the case where the point sites are on a single parabola.
    \subref{fig:exper:randomization-r} The running time
    graph in the case of random point sites inside a square.
  }
  \label{fig:exper:randomization}

  \addtolength{\subfigcapmargin}{-20pt}
\end{figure}

The random partitioning of the Voronoi sites into two sets in the
divide step of the algorithm yields a near-optimal expected running
time in the size of the output diagram
(Section~\ref{sec:micha-proof}).
This optimality result is general, and true for every type of diagrams
supported by the framework.

Figure~\ref{fig:exper:randomization} demonstrates the effect of
randomly partitioning the sites on the standard Voronoi diagram of
points with diverse sets of inputs.
We apply the following three different partitioning strategies:
\begin{inparaenum}[(i)]
\item the partitioning based on lexicographic sort (Sorted) --- the
  first set comprises the smaller $\ceil{n/2}$~points and the second
  set of points comprises the larger $\floor{n/2}$~points,
\item a partitioning based on 
  \ccode{CGAL::spatial\_sort} function (Spatial Sorted), and
\item a randomized partitioning (Randomized).  
\end{inparaenum}
In all cases, if we randomly partition the sites, we get a nearly-linear
running time in the number of sites.
In the case of a random point-set
(Figure~\ref{fig:exper:randomization-r}) we get a slightly larger,
though still nearly-linear, running time.

Figure~\ref{fig:exper:randomization-wc} uses the following set of
inputs.
Each input set of $n$ points is defined to be \daniex.
If we partition the set into two subsets, to the left and to the right
of the $y$-axis (the ``Sorted'' partitioning strategy), then
the overlay of the two Voronoi diagrams has $\Theta(n^2)$ complexity; see
Figure~\ref{fig:exper:randomization-dani} for an illustration.
Hence the algorithm runs in $\Omega(n^2)$ time, whereas the complexity
of the final diagram is only $\Theta(n)$.

\subsection*{Handling Degeneracies}
\label{sec:advantages:degeneracies}

\begin{figure}[h]
    \setlength{\spacewidth}{0pt}
    \providelength{\subfigwidth}
    \setlength{\subfigwidth}{110pt}
    
    \centering
    \subfigure[]{\label{fig:degen-a}\includegraphics[width=\subfigwidth]{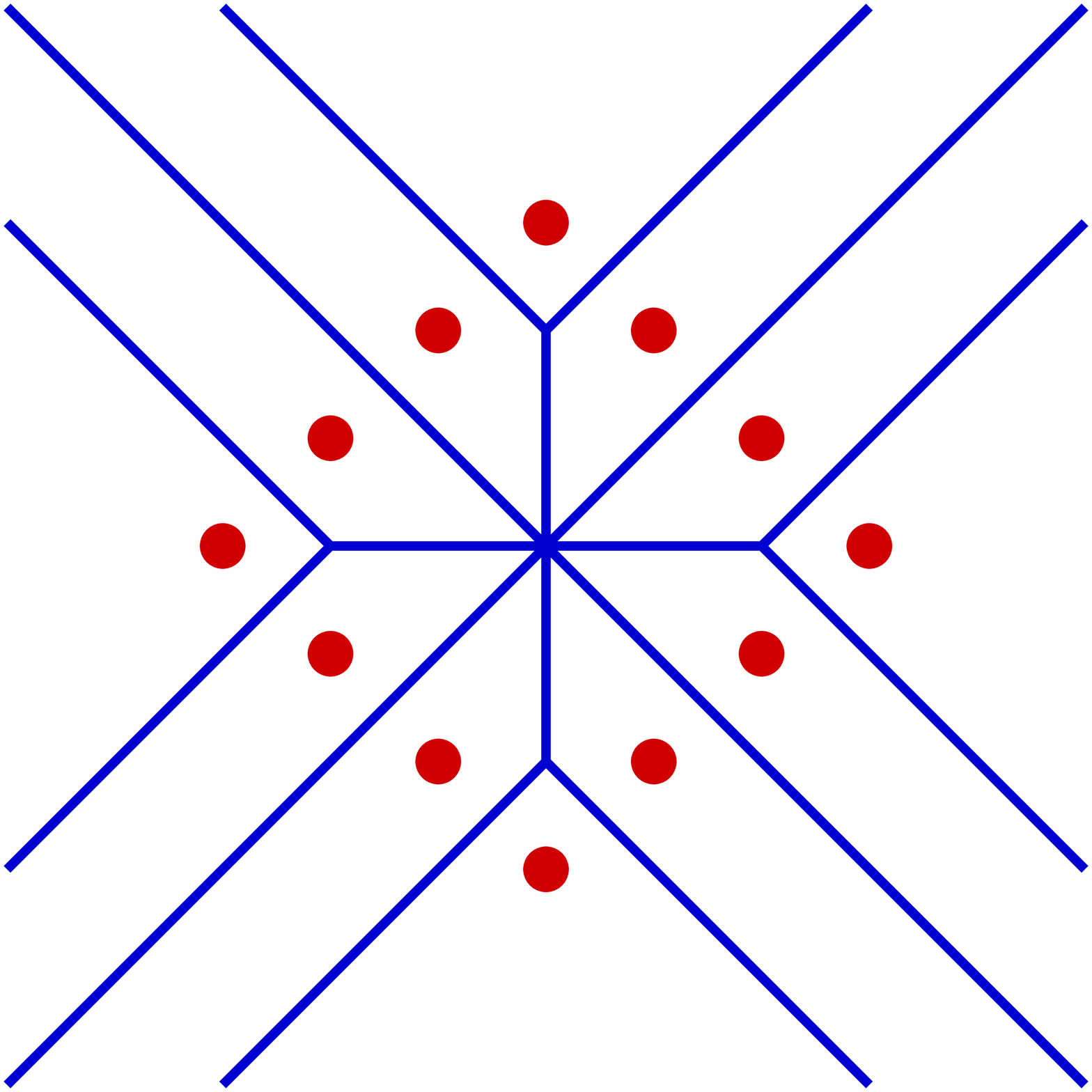}}
    \hspace{\spacewidth}
    \subfigure[]{\label{fig:degen-b}\includegraphics[width=\subfigwidth]{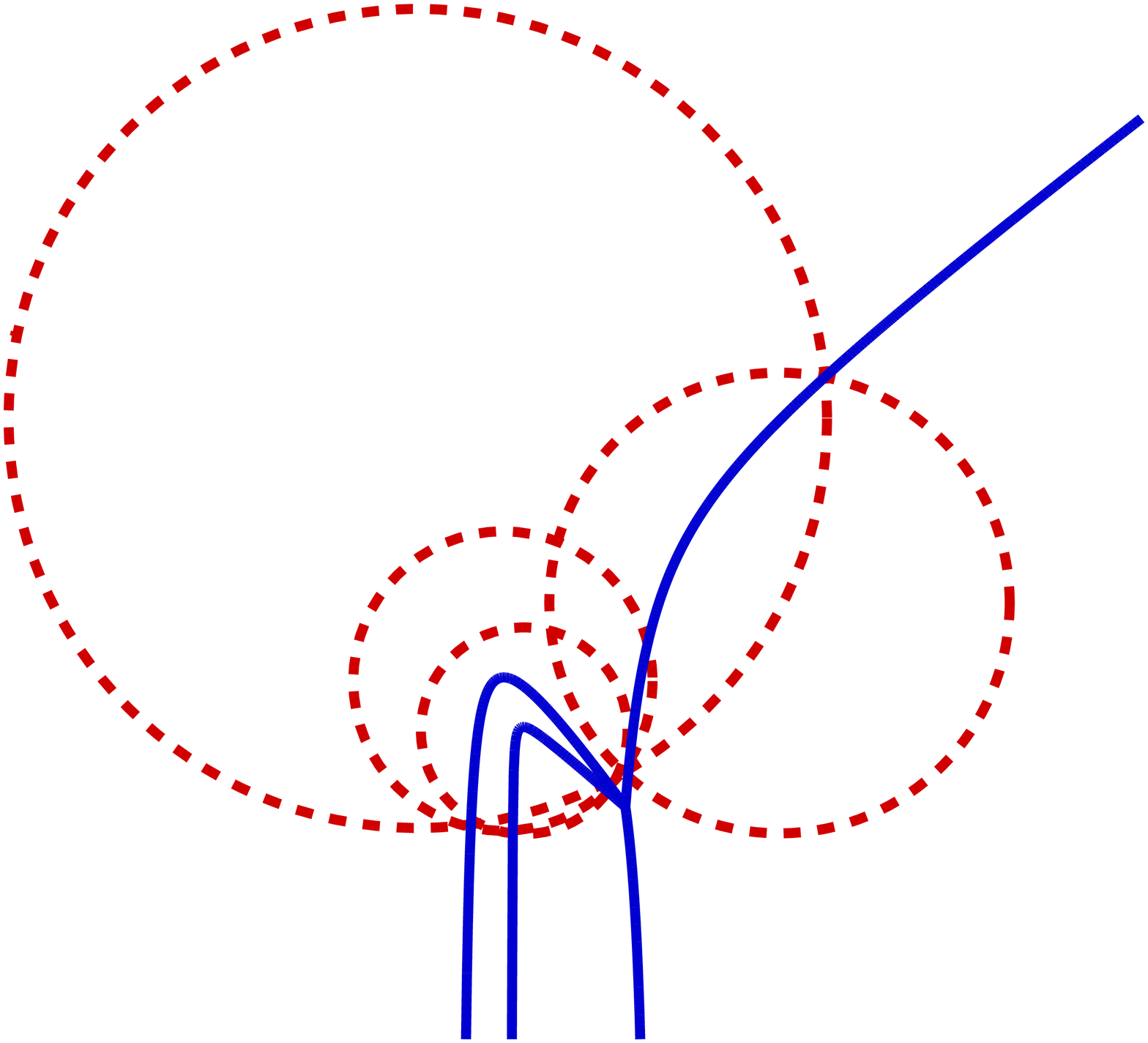}}
    \hspace{\spacewidth}
    \subfigure[]{\label{fig:degen-c}\includegraphics[width=\subfigwidth]{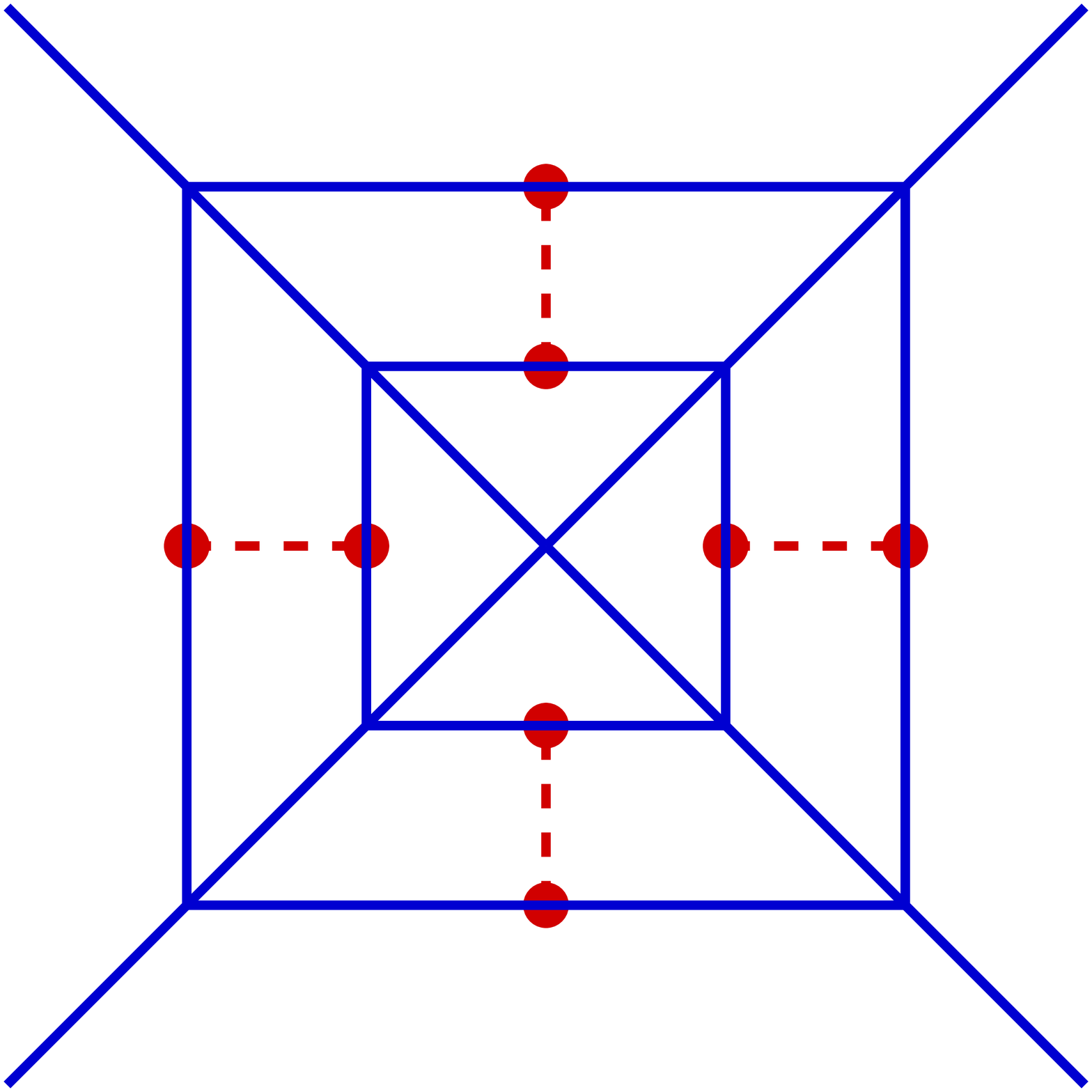}}
    \hspace{\spacewidth}
    \subfigure[]{\label{fig:degen-d}\includegraphics[width=\subfigwidth]{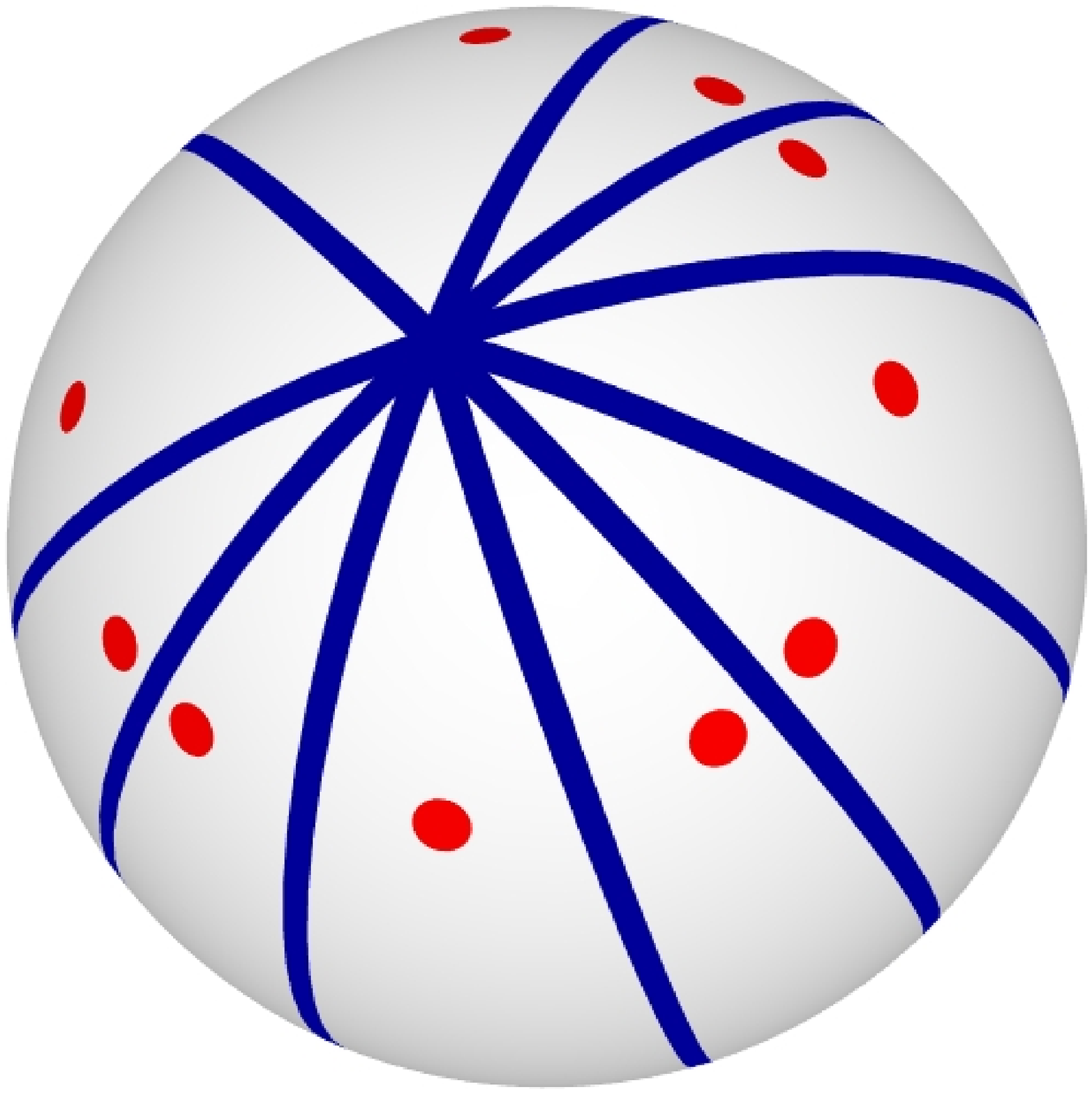}}
  
  \caption[Degenerate Voronoi diagrams]{Degenerate Voronoi diagrams
    computed with our software.
    \subref{fig:degen-a}~A degenerate standard Voronoi diagram.
    \subref{fig:degen-b}~A degenerate additively-weighted Voronoi
    (Apollonius) diagram.
    \subref{fig:degen-c}~A degenerate Voronoi diagram of segments.
    \subref{fig:degen-d}~A degenerate spherical Voronoi diagram.
    The sites in \subref{fig:degen-b} and \subref{fig:degen-c} are
    illustrated with dashed curves.}
  \label{fig:degen}
\end{figure}

Implementing robust geometric algorithms is known to be a hard
task~\cite{kmpsy-cerpg-08}. Inherent problems in designing
geometric algorithms are the coping with the limited precision of
machine-supported arithmetic and the handling of complicated boundary
conditions that arise in various degenerate situations (see
Section~\ref{sec:background:cgal}).
Our framework, like the \aos package of \cgal, handles all
degenerate situations while computing Voronoi diagrams, as long as the
supplied traits class handles a small number of degenerate cases.
For example, the traits class should detect if two $x$-monotone curves
overlap when trying to intersect them.
All implemented traits classes described in Chapter~\ref{chap:impl}
handle all degenerate situations.
Figure~\ref{fig:degen} depicts various degenerate situations that are
properly handled by our framework and the traits classes.

\subsection*{Farthest-\neighbor Voronoi diagrams}
\label{sec:advantages:farthest}
A useful meta-type of Voronoi diagrams is the farthest-\neighbor
Voronoi diagram.
A farthest Voronoi diagram can be formalized as a nearest-\neighbor
Voronoi diagram.
A farthest Voronoi diagram with positive distances comprises the
nearest Voronoi diagram induced by the following distance functions:
$\rho_F(x, o) = \nicefrac{1}{\distance{x}{o}}$.
In addition, reversing the assignment of dominance regions of each
pair of sites constitutes the definition for farthest Voronoi diagrams
using the abstract Voronoi diagram terminology.

\begin{figure}[h]
    \setlength{\spacewidth}{0pt}
    \setlength{\subfigwidth}{110pt}

    \centering
    \subfigure[]{\label{fig:farthest-a}\includegraphics[width=\subfigwidth]{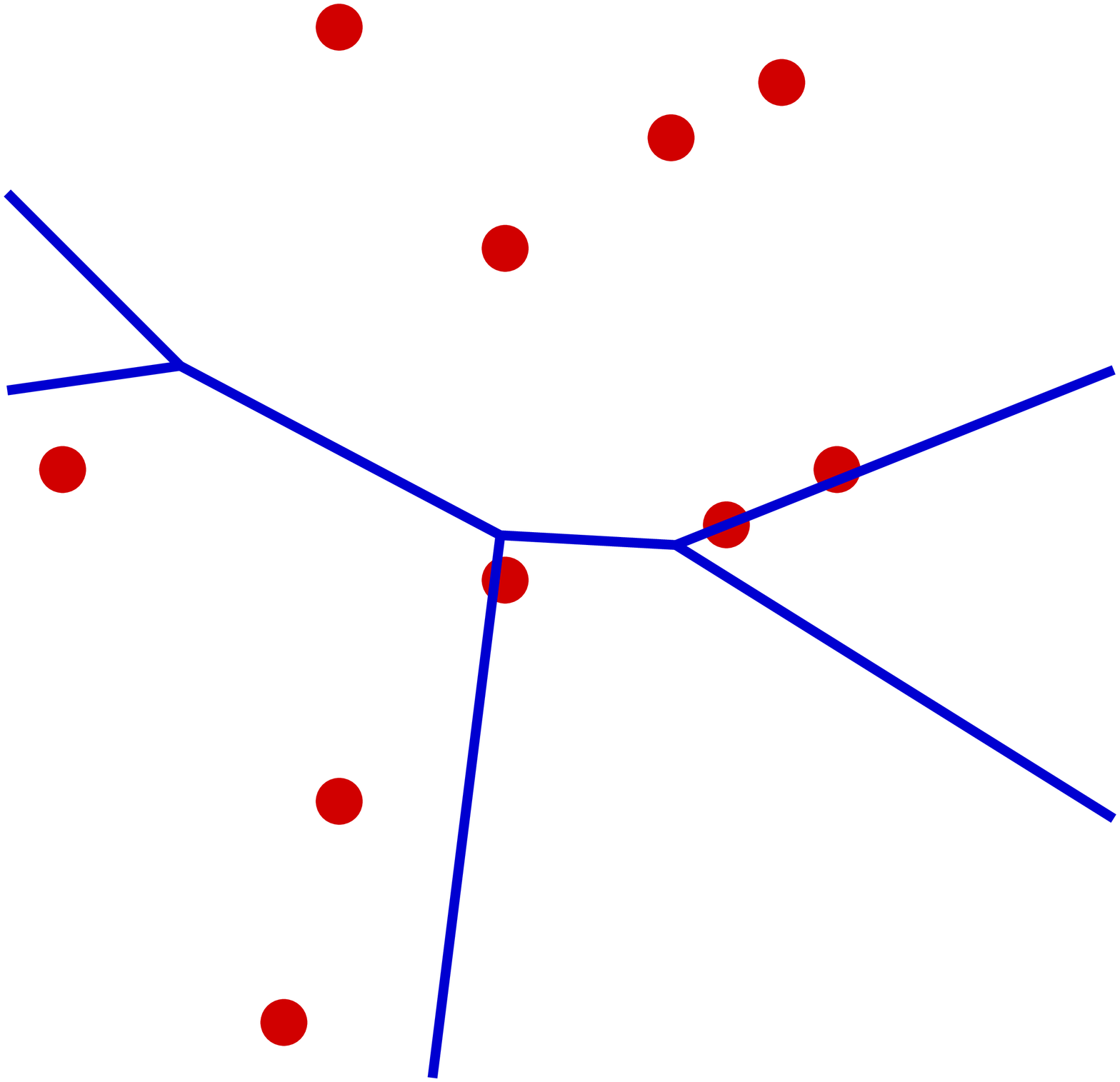}}
    \hspace{\spacewidth}
    \subfigure[]{\label{fig:farthest-b}\includegraphics[width=\subfigwidth]{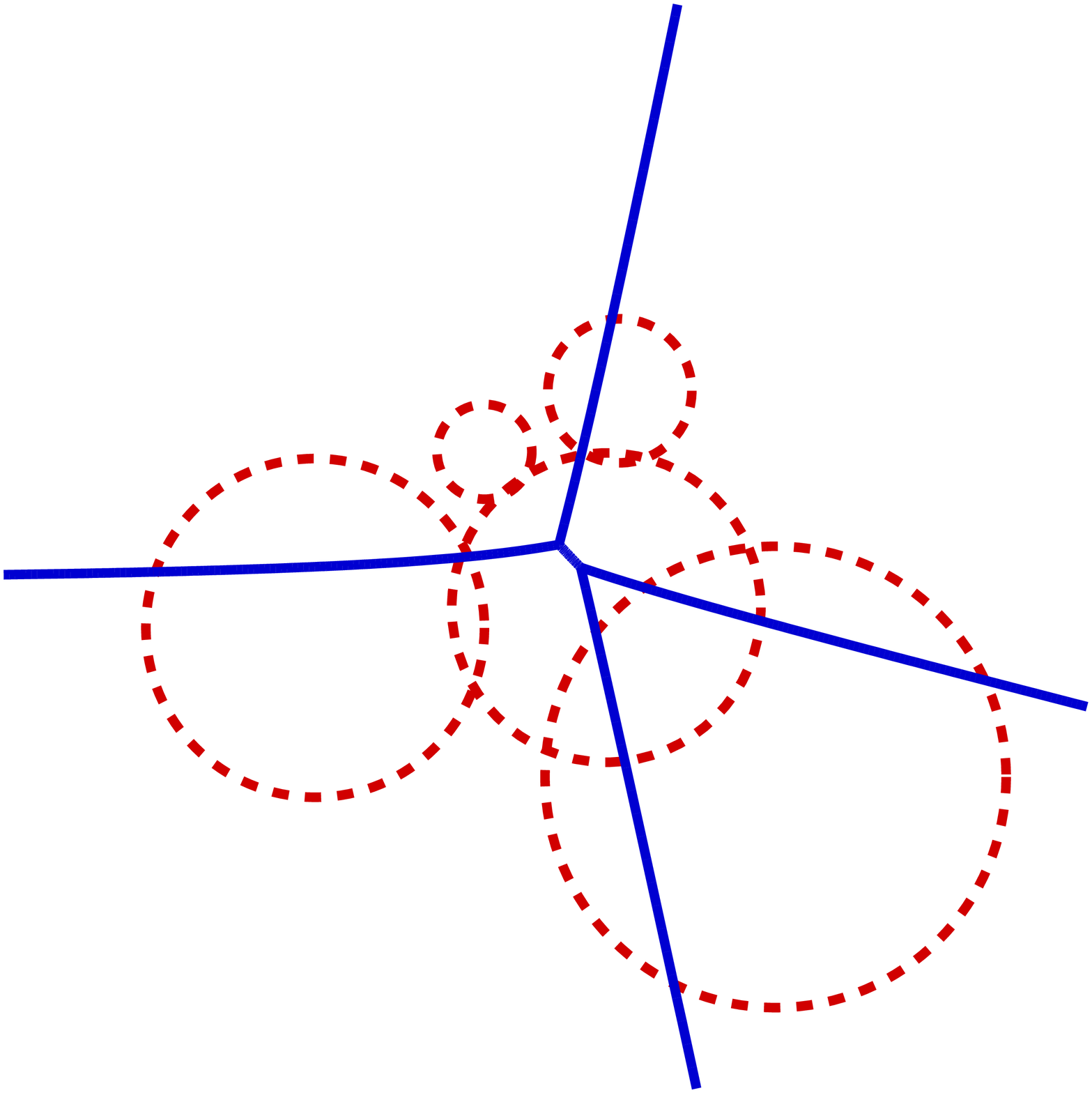}}
    \hspace{\spacewidth}
    \subfigure[]{\label{fig:farthest-c}\includegraphics[width=\subfigwidth]{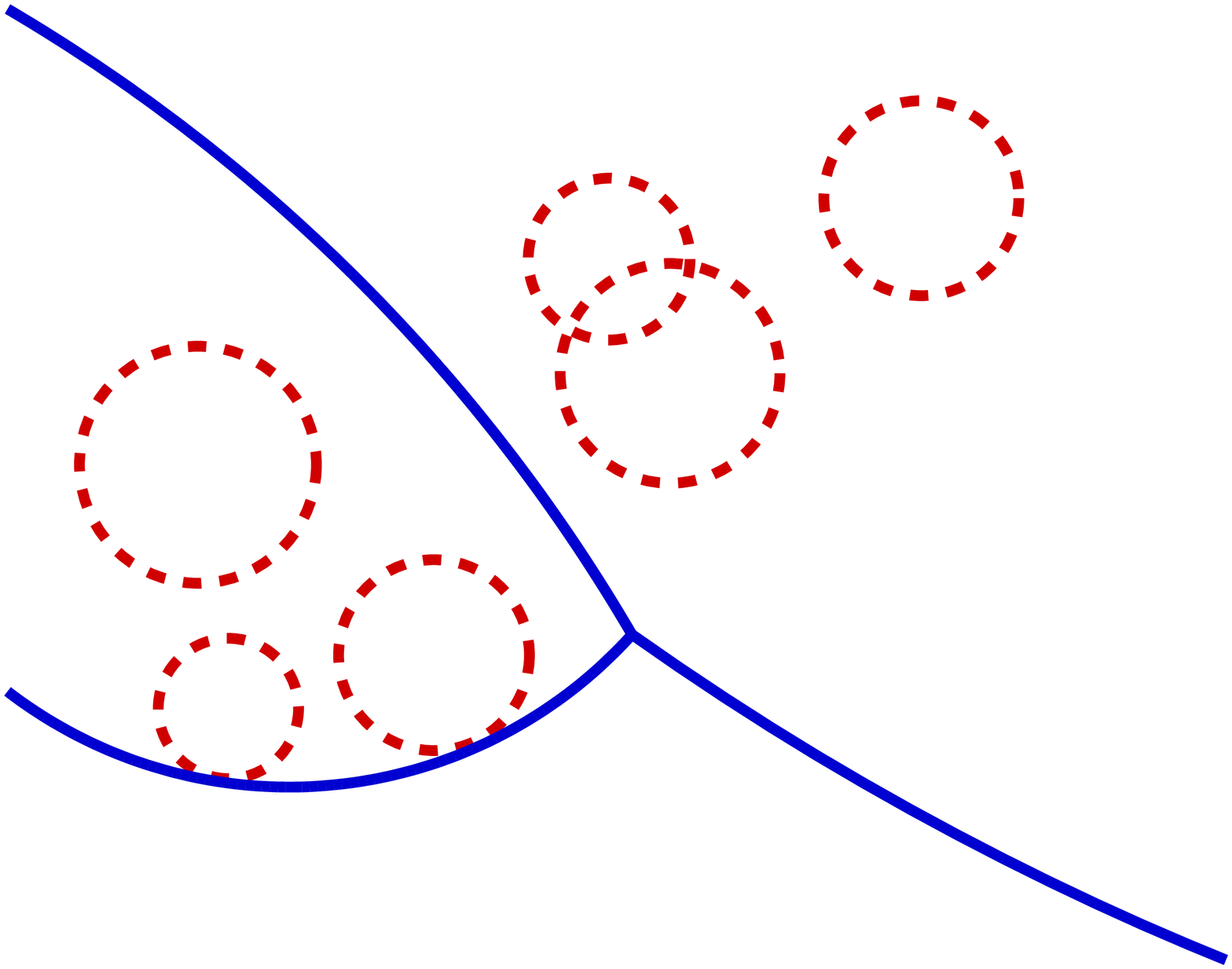}}
    \hspace{\spacewidth}
    \subfigure[]{\label{fig:farthest-d}\includegraphics[width=\subfigwidth]{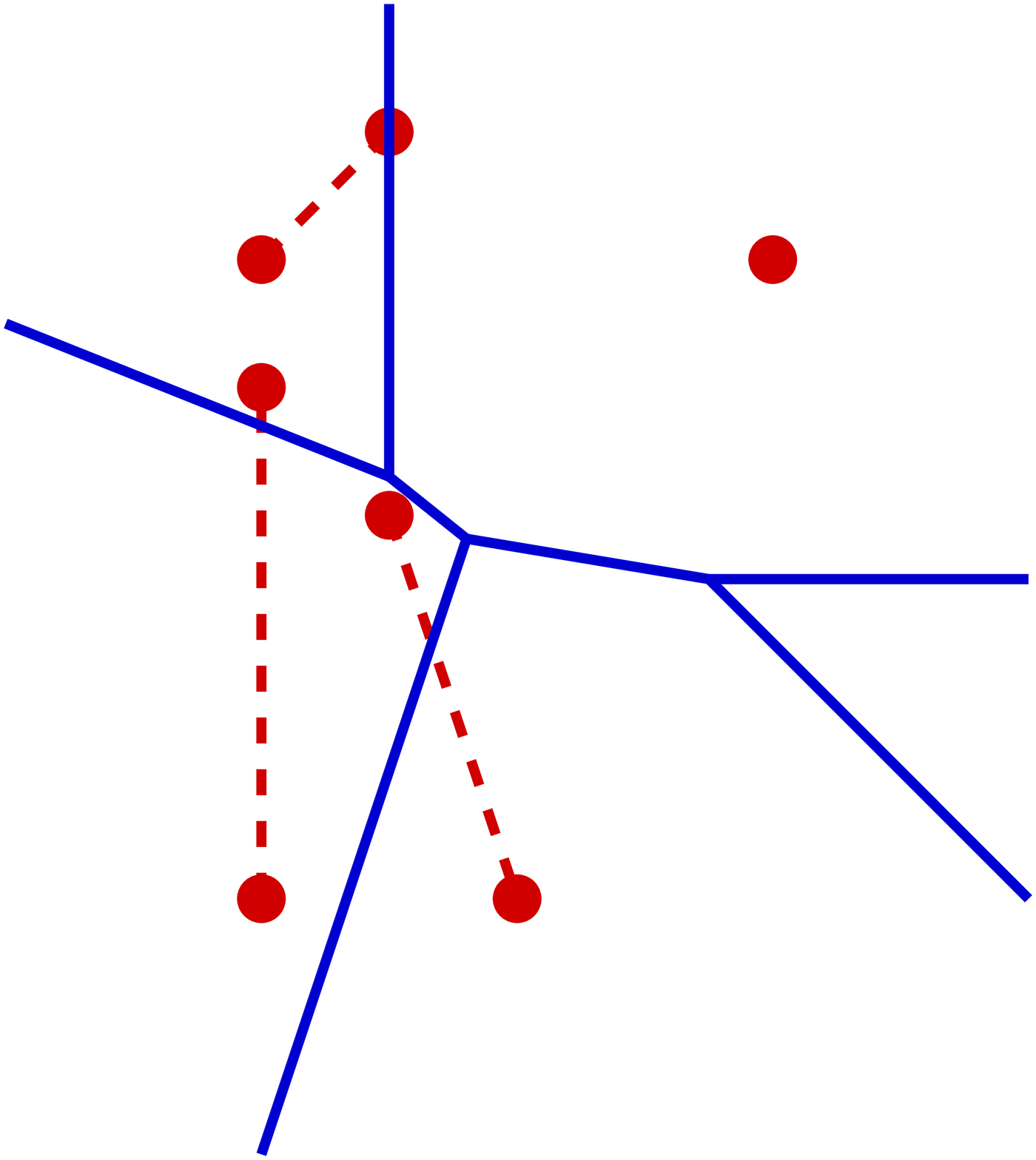}}

  \caption[Farthest-\neighbor Voronoi diagrams]{Farthest-\neighbor Voronoi
    diagrams of the same sites as in Figure~\ref{fig:various}, computed
    with our software.
    \subref{fig:farthest-a}~A farthest point Voronoi diagram.
    \subref{fig:farthest-b}~A farthest additively-weighted Voronoi diagram.
    \subref{fig:farthest-c}~A farthest \mobius{} diagram.
    \subref{fig:farthest-d}~A farthest Voronoi diagram of segments.
    The sites in \subref{fig:farthest-b}, \subref{fig:farthest-c}, and
    \subref{fig:farthest-d} are illustrated with dashed curves.}
  \label{fig:farthest}
\end{figure}

Given a traits class for nearest Voronoi diagrams, our framework
enables the immediate computation of the respective farthest Voronoi
diagrams (using the \ccode{CGAL::farthest\_voronoi\_2}
function --- see Section~\ref{sec:env-to-vd:impl}).
The construction of the upper envelope of the distance functions from
the sites yields the farthest Voronoi diagram.
Figure~\ref{fig:farthest} shows various farthest-\neighbor Voronoi
diagrams computed with our framework.

\subsection*{Post-computation Operations}
\label{sec:advantages:post-op}
Voronoi diagrams are represented as exact \cgal arrangements, and
their vertices, edges, and faces can easily be traversed while obtaining
the coordinates of the vertices of the diagram to any desired
precision, if the situation requires.

Most applications require additional operations
to be performed on the resulting Voronoi diagrams.
The computed diagrams can be passed as input to consecutive
operations supported by the \aos{} package and its derivatives.
We get plenty of additional functionalities for free.
Among those are
\begin{inparaenum}[(i)]
  \item point-location queries and vertical-ray shooting,
  \item the ability to perform aggregated insertion of additinal curves,
  \item computing the zone of a curve inside a Voronoi diagram and
    inserting a curve in an incremental fashion to an existing
    diagram,
  \item the ability to remove existing edge of the diagram,
  \item overlaying two (or more) Voronoi diagrams or arrangements, 
  \item the ability to attach user-defined data to all of the
    geometric and topological primitives, and
  \item the ability to use the \boost graph library to run various
    graph algorithms on the diagram and its dual
    structure~\cite{cgal:wfzh-a2-08}.
\end{inparaenum}
There are specific applications for the overlay of Voronoi diagrams,
for example, representing the local zones of two competing
telecommunication operators~\cite{bgz-spvt-00}.
We describe an application of the overlay operation for computing a
minimum-width annulus of a set of disks in the plane in
Chapter~\ref{chap:min-annulus}.

\begin{figure}[h] %
  \setlength{\subfigwidth}{150pt}

  \centerline{
    \subfigure[]{\label{fig:world:world}\includegraphics[width=\subfigwidth]{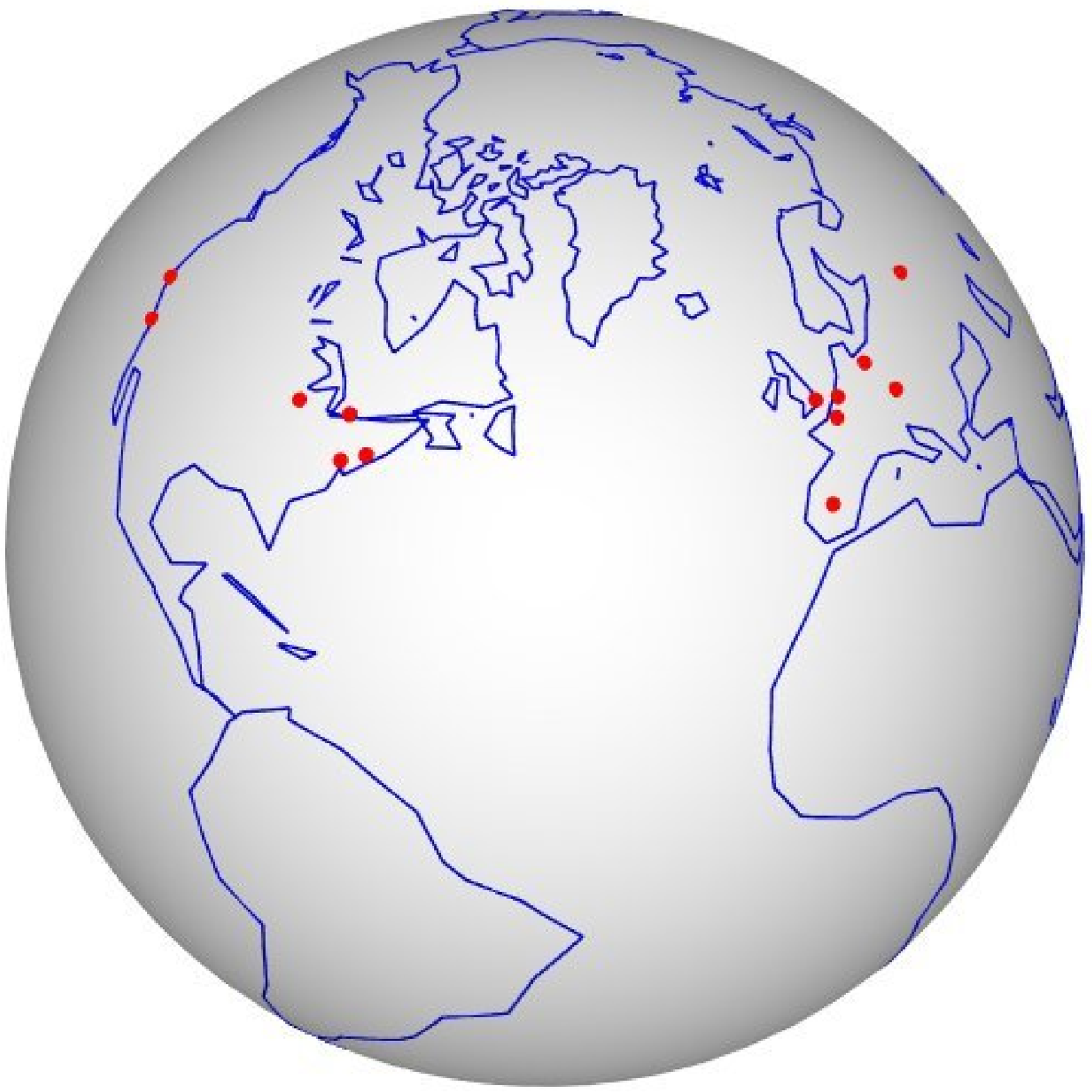}}
    \hspace{\spacewidth}
    \subfigure[]{\label{fig:world:cities}\includegraphics[width=\subfigwidth]{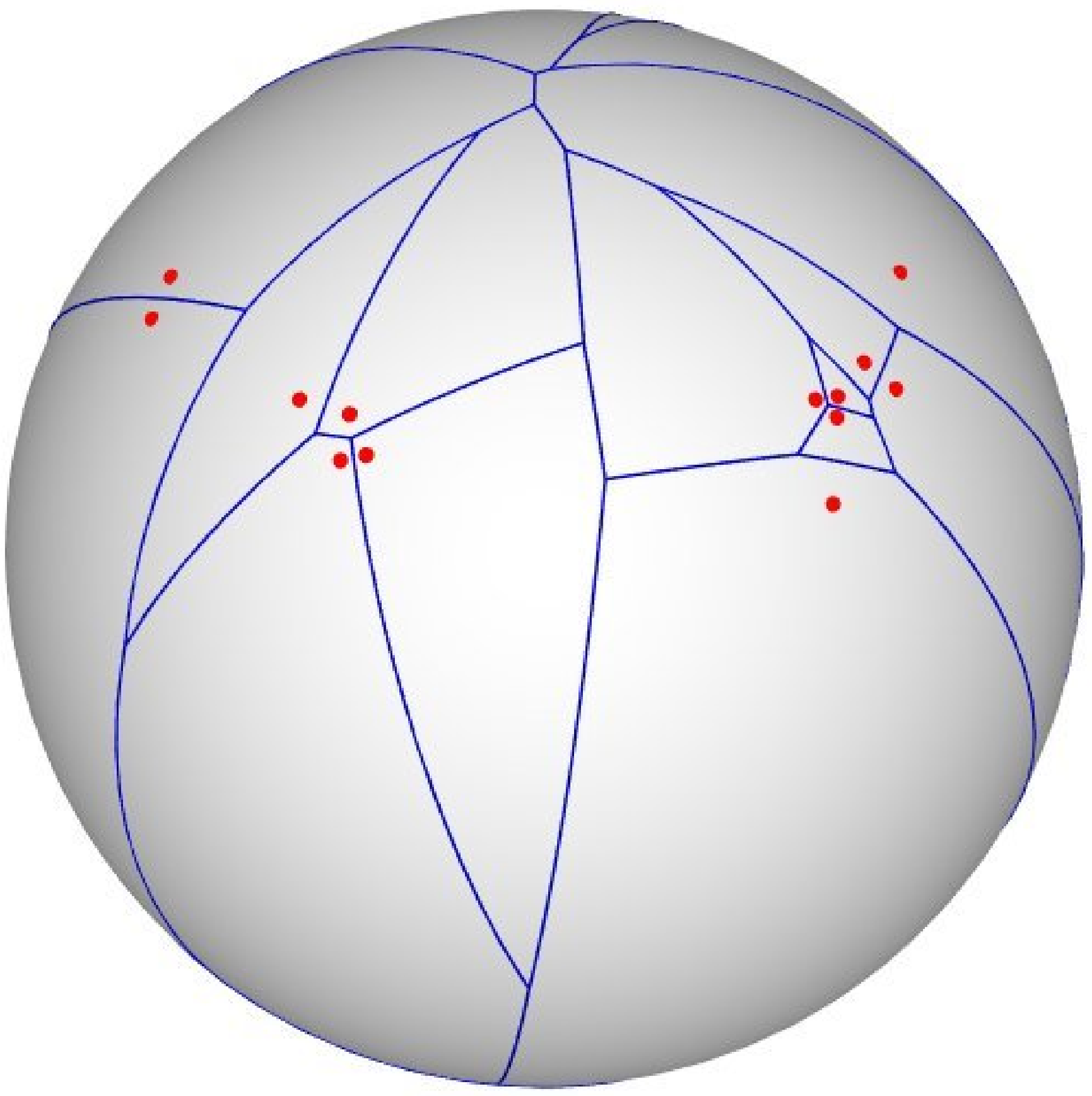}}
    \hspace{\spacewidth}
    \subfigure[]{\label{fig:world:overlay}\includegraphics[width=\subfigwidth]{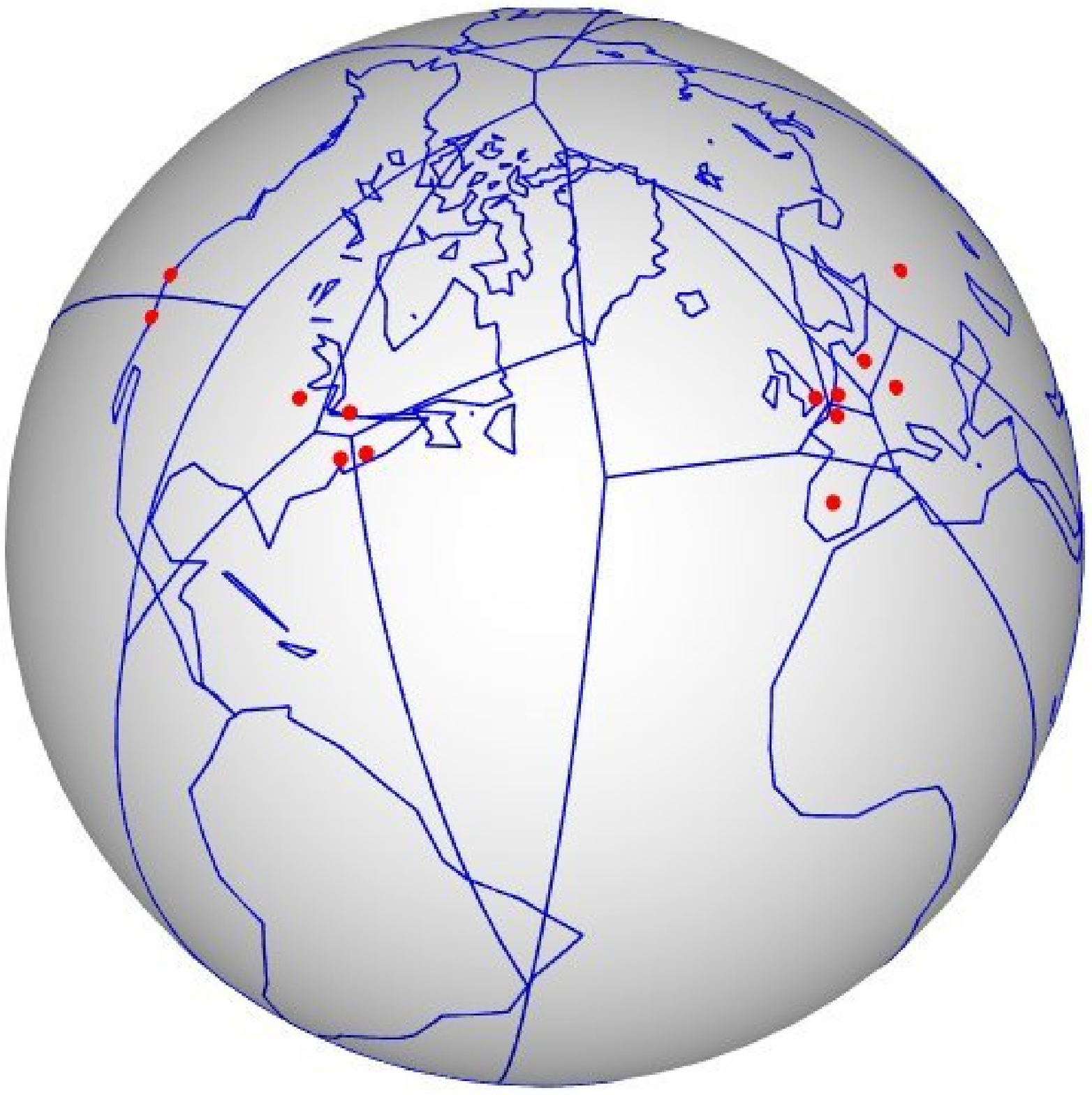}}
  }
  
  \caption[Overlaying an arrangement and a Voronoi diagram on the
    sphere]{Overlaying an arrangement and a Voronoi diagram on the
    sphere.
    \subref{fig:world:world} Arrangement induced by the continents and
    some of the islands on earth.
    \subref{fig:world:cities} Voronoi diagram of 20~points on the
    sphere representing 20~major cities on the planet.
    \subref{fig:world:overlay} The overlay of the arrangement from
    \subref{fig:world:world} and the Voronoi diagram
    from~\subref{fig:world:cities}.
  }
  \label{fig:world}
\end{figure}

Figure~\ref{fig:world} shows an arrangement on the sphere induced by 
\begin{inparaenum}[(i)]
\item the continents and some of the islands on earth, and 
\item 20 major cities in the world,
\end{inparaenum}
which appear as isolated vertices.
The arrangement consists of 1065 vertices, 1081 edges, and 117 faces.
The initial data was taken from gnuplot~\citelinks{gnuplot-link} and
google maps~\citelinks{googlemaps-link}.
The middle sphere embeds an arrangement that represents the Voronoi
diagram of the 20 cities above.
The right figure shows the overlay of the two aforementioned
arrangements computed with the generic overlay function from the \aos
package.

\section{Limitations}
\label{sec:disadvantages}
\subsection*{Practical Computation Time}
\label{sec:desadvantages:practical}

Theoretically, the randomized divide-and-conquer envelope approach for
computing Voronoi diagrams is efficient and is asymptotically
comparable to other (near-)optimal methods.
Practically, however, the concrete running time of our implementation
is inferior to those of existing implementations of various dedicated
implementations (for specific type of diagrams).
Following are possible explanations for the gap.
\begin{enumerate}
\item The method uses constructions of bisectors and Voronoi vertices
  as elementary building-blocks and they must be exact.
  Geometric algorithms based on constructors (rather than
  predicates) are usually more time and space consuming, especially
  when the constructed objects are exact, like in our
  case.\footnote{Recall, for example, that every constructed
    geometric object of the lazy kernel holds the entire construction
    tree.}
\item Each face of the overlay is inspected during the merge step.
  Though randomization ensures us that the expected number of faces is
  proportional to the complexity of the final diagram, a large number
  of redundant faces is being examined.
  Other algorithms do not examine all the faces of the overlay, for
  example, the classic algorithm by Shamos and Hoey~\cite{sh-cpp-75}
  suggests that by sorting the input point sites only a small number
  of faces has to be handled.
  Our general framework is unaware of these kind of optimizations that
  relate to the specific type of the Voronoi diagram.
\item In many dedicated Voronoi diagrams algorithms implementations,
  various optimizations are employed to reduce the running time of the
  algorithm.
  Most of these optimizations cannot be applied by the general
  algorithm.
  For example, some Voronoi diagrams algorithms harness the fact that
  sites usually affect only their surroundings to optimize the
  code. Our general algorithm cannot assume such a thing.
\end{enumerate}

Our work can be probably improved further, but it is reasonable to
assume that the general implementation will continue to be inferior to
other implementations.

\subsection*{Bisector Construction}
\label{sec:desadvantages:bisector}

Besides affecting the time performance of the algorithm, constructing
and manipulating bisectors have other limitations.
In certain cases, it is hard to construct the bisector of two sites.
The bisector could be two-dimensional or composed of several arcs,
and the efficient construction of it may not be trivial.
The user does not need to know the details of constructing lower
envelopes of distance functions, indeed, but he/she
may have to posses fairly advanced knowledge in computational algebra.

In other cases, there may not be an existing traits class for the
arrangement package that supports the construction of an arrangement
induced by the bisector curves. The user may have to create a traits
class for the arrangement package themselves.

One can also imagine  diagrams that will not be supported by the
framework due to the hardness of the computation of their bisectors.
For example, it is not clear what is the type of the curves that
constitute the boundaries of the cells of the \emph{zone
  diagram}~\cite{amt-zdeuac-07}.

\subsection*{Incremental Construction and Removal}
\label{sec:desadvantages:incremental}

Incremental insertion of Voronoi sites could be very
useful~\cite{obsc-stcavd-00,mgd-diovdm-03}, for example, for
developing online Voronoi diagrams' construction algorithms.
Our framework can support the insertion of a single site by
partitioning the input to one set that consists of $n-1$ sites and
another set that consists of a single site.
Unfortunately, this insertion procedure is expected to be most
inefficient.
The current implementation will try and bisect each of the faces with
the bisector of the new site and the dominant site on the face, which
will result in $\Omega(n)$ time per insertion operation.

A deletion operation of sites is also
useful~\cite{d-ddt-02,mgd-diovdm-03} but is not supported by the
framework.
Adding this operation to the framework is not trivial.

\chapter{Application: Minimum-Width Annulus of Disks}
\label{chap:min-annulus}

In this chapter we describe the application of our framework to solve
the problem of computing a minimum-width annulus of a set of disks in
the plane.
Our algorithm is a generalization of a known algorithm for computing a
minimum-width annulus of a set of points in the plane, and its
implementation exploits the generality and the flexibility of our
framework.
Section~\ref{sec:min-annulus:intro} gives a short introduction and
background on the problem.
Section~\ref{sec:min-annulus:proofs} presents our generalized
algorithm.
Finally, Section~\ref{sec:min-annulus:impl} gives specific
implementation details and experimental results for the application.

\section{Introduction}
\label{sec:min-annulus:intro}

An {\it annulus}\index{annulus|see{minimum-width annulus}} is the bounded
area between two concentric circles. The width
of an annulus is the difference between the radii of the outer circle
and the inner circle.
Given a set of objects in the plane the objective is to find a
\Index{minimum-width annulus} containing those objects.

Constructing a \Index{minimum-width annulus} of a set of points has
applications in diverse fields. We list below two examples, taken
from the areas of tolerancing metrology and \Index{facility
  location}; more applications can be found
in~\cite{bbbrw-ccmwap-98,cd-fdtcpr-09} and other papers.

In the field of mechanical design, assessing the deviation of a
manufactured object from its ideal design is a key problem.
If the manufactured object is round then this deviation is called
the \emph{roundness error} of the object.
When assessing the roundness error of a manufactured object, four
types of roundness errors are most commonly used, that is Maximum
Inscribed Circle, Minimum Circumscribing Circle, Least Squares Center,
and Minimum Radial Separation~(MRS).
The MRS roundness error is defined to be the minimum difference
between the radii of two concentric circles (one circumscribing and
one inscribed). 
Therefore, computing the MRS roundness error amounts to computing a
minimum-width annulus containing the points sampled from the
manufactured object~\cite{y-ecgtm-94}.

Facility location generally deals with the placement of facilities for
minimizing costs under various restrictions.
When considering a location for a new facility among existing
facilities, in many cases the most ``fair''
location is the location where the difference between the maximal and
the minimal effects on existing facilities is brought to a
minimum~\cite{ms-emflar-94}.
In other cases, facilities may have both desirable and obnoxious
properties. For example, the service area of a cell site\footnote{The
  location where antennas and other network communication equipment
  are placed in order to provide wireless service in a geographic area.}
should cover the locations of its clients, and due to aesthetic and
health constraints, should preferably be in the farthest location
possible from all clients.\footnote{See
  \url{http://cgm.cs.mcgill.ca/~athens/cs507/Projects/2004/Emory-Merryman}
  for more details.}
In both cases the center of a \Index{minimum-width annulus}
containing the locations of the existing facilities and the locations
of the clients, respectively, is a good location.

A \Index{minimum-width annulus} does not always exist. If the
width of the set of objects is smaller than the width of any
containing annulus, then there is no minimum-width
annulus.\footnote{The width of a set is defined to be the width of the
  thinest strip (\ie the area bounded between two parallel lines)
  containing it.}
Consider, for example, four collinear points in the plane. For every
annulus containing the four points, we can always find another annulus with
smaller width by moving the center of the original annulus away from
the points on the line perpendicular to the line passing through
them.

In the case of point sets, the minimum-width annulus must have (at
least) two points on its outer circle and (at least) two points on its
inner circle~\cite{r-ac-79,sr-wrspp-99}, which results in the
following observation:
\begin{observation}
  The center of a minimum-width annulus of points in the plane  must
  lie on an intersection point of the nearest-\neighbor Voronoi diagram
  and the farthest-\neighbor Voronoi diagram of the points.
\end{observation}
The center of an empty circle touching (at least) two points is on an
edge of the nearest-\neighbor Voronoi diagram of the points, and the
center of a circle containing all the points and touching (at least)
two points is on an edge of the farthest-\neighbor Voronoi diagram.

Using the above observation, Ebara~\etal~\cite{efnn-ravd-89} gave a
worst-case $O(n^2)$ time algorithm for finding a \Index{minimum-width annulus} 
of points in the plane.
The algorithm was reintroduced to the manufacturing community
by Roy and Zhang~\cite{rz-epccm-92}.

Similar methods were used to solve different variations of the problem.
de~Berg~\etal \cite{bbbrw-ccmwap-98}\ used nearest-\neighbor and 
farthest-\neighbor Voronoi diagrams to compute a minimum-width
annulus of point sets with various constraints on its radii (\eg fixed
inner radius).
Le and Lee~\cite{ll-orpr-91} showed how to use the \Index{overlay} of the medial
axis of a simple polygon (which is the Voronoi diagram of its edges 
restricted to the interior of the polygon) and the 
\Index{farthest Voronoi diagram} of
the vertices of the polygon to compute a \Index{minimum-width annulus}
 bounding the polygon.
Barequet~\etal considered several problems and applied similar
techniques for the case of offset-polygon annuli
containing a set of points~\cite{bbdg-opapp-98,bbdg-occpa-05}.

For special cases there are deterministic sub-quadratic
algorithms. 
Garcia-Lopez~\etal \cite{lrs-fspc-98} showed that given
the final circular order of the points around the center of the
annulus (not a far-fetched scenario in real-world applications), a
\Index{minimum-width annulus} can be found in worst-case time complexity of
$O(n \log n)$.
Swanson~\etal~\cite{slw-oardc-95} introduced an optimal $O(n)$ time
algorithm for computing the roundness of convex polygons.
Devillers and Ramos~\cite{dr-cresar-02} gave an algorithm that
exhibits linear running time in practice for point-sets that are
almost round.\footnote{By their terminology, a set is \emph{almost
    round} if it is contained inside an annulus which satisfies $W /
  R_M \leq 0.1$, where $W$ is the width of the annulus and $R_M$ is
  the average of the inner and outer radii.}

Several randomized sub-quadratic algorithms exist for finding a
\Index{minimum-width annulus} of points in the
plane~\cite{aas-cefda-97,as-erasgo-96,tas-apsgo-94}.
The algorithm by Agarwal and Sharir~\cite{as-erasgo-96} achieves an
expected running time of $O(n^{3/2 + \varepsilon})$ which is the best
known algorithm to date.
Using \Index{core-sets} Chan~\cite{c-fcscd-06} presented an
$(1+\varepsilon)$-factor \Index{approximation} algorithm for the
\Index{minimum-width annulus}, improving previously known
results~\cite{aaps-aamwas-00,ahv-aemp-04,c-adwsec-02}.

In this chapter we describe an application of our framework that
demonstrates its usefulness.
We use our tools to find a minimum-width annulus containing a set
of \emph{disks} in the plane.
Section~\ref{sec:min-annulus:proofs} deals with the theoretical
aspect of our algorithm. Section~\ref{sec:min-annulus:impl}
describes major implementation details and some experimental results.

\section{Algorithm for Minimum-Width Annulus of Disks}
\label{sec:min-annulus:proofs}

\index{farthest Voronoi diagram}
As mentioned above, when dealing with  point sets, a 
\Index{minimum-width annulus}
can be found by computing the \Index{overlay} of the nearest-\neighbor and 
farthest-\neighbor Voronoi
diagrams of the points.
A similar approach applies to the case of disk sites but
the diagrams have to be defined more carefully.

\index{farthest Voronoi diagram}
\index{\FPFS{} Voronoi diagram}
Consider the following farthest-point \Index{distance function} from a point
$p \in \R{2}$ to a set of points $S \subset \R{2}$: 
\begin{displaymath}
  f(p, S) = \sup_{x \in S}{||p - x||},
\end{displaymath}
which measures the farthest distance from the point
$p$ to the set $S$. 
Consider the farthest-\neighbor Voronoi diagram with respect to this distance
function. We call this diagram the ``\FPFS{}'' Voronoi diagram.
The distance function $f(p, S)$ becomes the
Euclidean distance when the set $S$ consists of a single point.
However, this is not the case when the set $S$ is, say, a disk in the
plane.

The following observation establishes a connection between the \FPFS{} 
Voronoi diagram of a set of disks and an additively-weighted Voronoi 
diagram, and comes in handy below.

\begin{observation}
  \label{obs:min-annulus}
  Let \disks be a set of disk in \R{2}.
  The \FPFS Voronoi diagram of \disks is identical to the Apollonius
  diagram of the disks with negated radii.
\end{observation}

For a disk~$d$ in the plane with center~$c$ and radius~$r$ the 
farthest-point distance function is $\rho(p, d) = ||p - c|| + r$.
Recall that the Apollonius distance function is defined to be
$\rho(p, d) = ||p - c|| - r$. 
Next, observe that the farthest-point distance function is the same as 
the Apollonius distance function with negative radii.

We prove that there is a minimum-width annulus
(in case one exists) whose center is a vertex of the overlay of 
the Apollonius diagram and the \FPFS{} Voronoi diagram of the disks.

Recall that in the case of point sets the center of a minimum-width
annulus is an intersection point of the nearest-\neighbor Voronoi
diagram and the farthest-\neighbor Voronoi diagram of the points.
In the case of disks, this is not true. Take for example
a set of disks located at $(2,0)$, $(0,3)$, $(-4,0)$, and $(0,-5)$
with radii~$1$, $2$, $3$, and~$4$, respectively.
In this setting, the annulus centered at the origin with radii~$1$
and~$9$ is a minimum-width annulus, as its width is equal to a
diameter of one of the disks.
The origin does not lie on an edge of the \FPFS{} Voronoi diagram, as
the outer circle touches only one disk.

Let $\disks{} = \{d_1, \ldots, d_n\}$ be a collection of disks in the
plane such that for all $i$, $d_i \not\subseteq \bigcup_{j \neq i}d_j$.
For simplicity of exposition, we assume here that $n \geq 3$; the case of 
$n < 3$ is simple to handle.
We show that there
is a minimum-width annulus whose bounding circles intersect the disks of
\disks{} in at least 4 points.

\begin{observation} 
  \label{lem:min-annulus:tri}
  
  If \annulus{} is a \Index{minimum-width annulus} containing \disks{}
  then each of the inner circle and the outer circle of \annulus{}
  touches a disk of \disks{}.
\end{observation}

Each minimum-width annulus must touch the disks of \disks{} in at
least two points, as we can shrink the outer circle of the annulus and
expand the inner circle until each of them touches a disk.
Thus, fixing a point~$p$ in the plane as the center of a containing
annulus fixes an annulus bounding \disks for this center. We call this
annulus the \emph{tight annulus} fixed at~$p$.

Let \annulus{p} denote the tight annulus fixed at point~$p$ and let
\anwid{p} denote its width.
Let $\aninner{p}, \anouter{p} \subseteq \disks$ denote the set of disks
that are touched by the inner and outer circles of \annulus{p}, respectively.

\begin{definition}[\anpointorg{} in direction~\andirection]
  For a point~$p$ and a direction~\andirection in the plane, consider
  the ray~$r$ emanating from~$p$ in direction~\andirection.
  $r$ can be divided into maximally-connected cells, such that every
  point~$x$ in a cell ($x$~is the center of a tight annulus) obtains
  the same $\aninner{x}$ and $\anouter{x}$.
  Let~$p'$ be an interior point of the cell of~$p$, or, if~$p$
  comprises a single-point cell, 
  an interior point of the next cell in the direction of the ray.
  We call~$p'$ a \anpoint{} of~$p$ in direction~\andirection.
\end{definition}

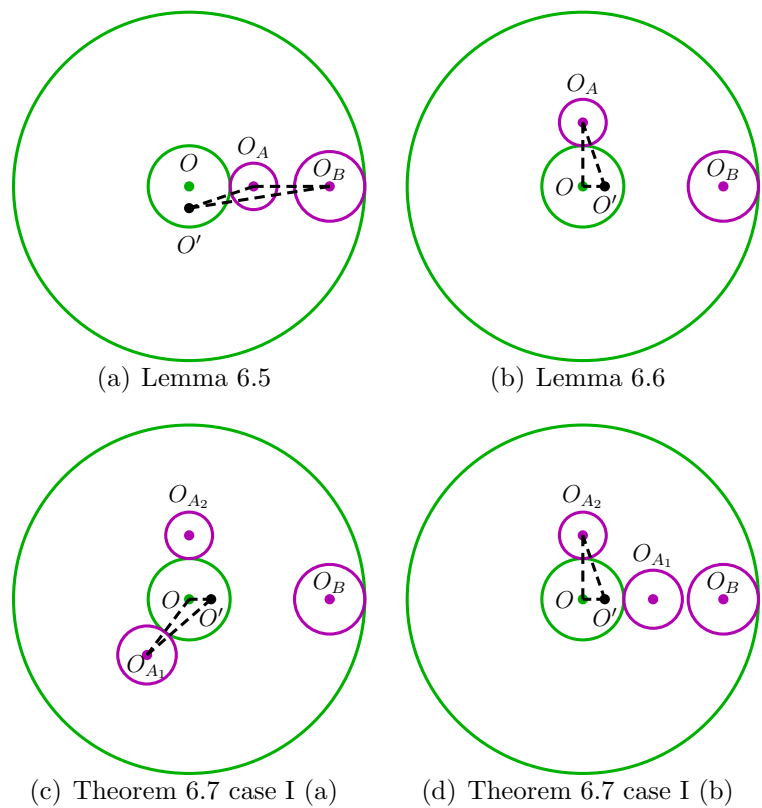
\begin{figure} %
  \vspace{-10pt}
  \centering
  \begin{tabular}{cc}
    \subfigure[Lemma~\ref{lem:min-annulus:col}]{\label{fig:min-annulus:col}
      \scalebox{0.77}{%
        \begin{pspicture}(0,-3.1)(6, 3.1)
          \psset{unit=1cm,linewidth=\annuluslinewidth{}}
          \pstGeonode[PosAngle=90, PointSymbol=*,linecolor=annulus-color,](3,0){O}
          \pstGeonode[PosAngle={90,90}, PointNameSep={.7em,1.6em},
            linecolor=circles-color](5.4,0){O_B}(4.1,0){O_A}
          \pstGeonode[PointSymbol={none,none}, 
            PointName={none,none}](6,0){OUT}(3.7,0){IN}
          \pstGeonode[PosAngle=-90,PointNameSep=1.5em](3,-.375){O'}
          \pstCircleOA[linecolor=annulus-color]{O}{OUT}
          \pstCircleOA[linecolor=annulus-color]{O}{IN}
          \pstCircleOA[linecolor=circles-color]{O_A}{IN}
          \pstCircleOA[linecolor=circles-color]{O_B}{OUT}
          \pstLineAB[linestyle=dashed]{O'}{O_A}
          \pstLineAB[linestyle=dashed]{O'}{O_B}
          \pstLineAB[linestyle=dashed]{O_B}{O_A}
    \end{pspicture}}} &
    \subfigure[Lemma~\ref{lem:min-annulus:three-points}]{\label{fig:min-annulus:three-points}
      \scalebox{0.77}{%
        \begin{pspicture}(0,-3.1)(6, 3.1)
          \psset{unit=1cm,linewidth=\annuluslinewidth{}}
          \pstGeonode[PosAngle=180, PointSymbol=*,linecolor=annulus-color,
            PointNameSep=.75em](3,0){O}
          \pstGeonode[PosAngle=90, PointNameSep=.7em,
            linecolor=circles-color](5.4,0){O_B}
          \pstGeonode[PosAngle=90, PointNameSep=1.6em,
            linecolor=circles-color](3, 1.1){O_A}
          \pstGeonode[PointSymbol={none,none}, 
            PointName={none,none}](6,0){OUT}(3,0.7){IN}
          \pstGeonode[PosAngle=-90,PointNameSep=.75em,PointName=O'](3.375,0){O'}
          \pstCircleOA[linecolor=annulus-color]{O}{OUT}
          \pstCircleOA[linecolor=annulus-color]{O}{IN}
          \pstCircleOA[linecolor=circles-color]{O_A}{IN}
          \pstCircleOA[linecolor=circles-color]{O_B}{OUT}
          \pstLineAB[linestyle=dashed]{O'}{O_A}
          \pstLineAB[linestyle=dashed]{O'}{O}
          \pstLineAB[linestyle=dashed]{O}{O_A}
    \end{pspicture}}} \\
    \subfigure[Theorem~\ref{the:min-annulus:annulus} 
      case~\ref{the:min-annulus:annulus:caseI}~(\ref{the:min-annulus:annulus:case1})]
              {\label{fig:min-annulus:annulus:case1}
      \scalebox{0.77}{%
        \begin{pspicture}(0,-3.1)(6, 3.1)
          \psset{unit=1cm,linewidth=\annuluslinewidth{}}
          \pstGeonode[PosAngle=180, PointSymbol=*,linecolor=annulus-color,
            PointNameSep=.75em](3,0){O}
          \pstGeonode[PosAngle=90, PointNameSep=.7em,
            linecolor=circles-color](5.4,0){O_B}
          \pstGeonode[PosAngle=-90, PointNameSep={0.5em},
            PointName=O_{A_1}, linecolor=circles-color](2.28,-0.96){O_A1}
          \pstGeonode[PosAngle=90, PointNameSep=1.6em,
            PointName=O_{A_2}, linecolor=circles-color](3, 1.1){O_A2}
          \pstGeonode[PointSymbol={none}, 
            PointName={none}](2.58,-0.56){IN1}
          \pstGeonode[PointSymbol={none,none}, 
            PointName={none,none}](6,0){OUT}(3,0.7){IN}
          \pstGeonode[PosAngle=-90,PointNameSep=.75em,PointName=O'](3.375,0){O'}
          \pstCircleOA[linecolor=annulus-color]{O}{OUT}
          \pstCircleOA[linecolor=annulus-color]{O}{IN}
          \pstCircleOA[linecolor=circles-color]{O_A1}{IN1}
          \pstCircleOA[linecolor=circles-color]{O_A2}{IN}
          \pstCircleOA[linecolor=circles-color]{O_B}{OUT}
          \pstLineAB[linestyle=dashed]{O'}{O_A1}
          \pstLineAB[linestyle=dashed]{O'}{O}
          \pstLineAB[linestyle=dashed]{O}{O_A1}
    \end{pspicture}}} &
    \subfigure[Theorem~\ref{the:min-annulus:annulus} 
      case~\ref{the:min-annulus:annulus:caseI}~(\ref{the:min-annulus:annulus:case2})]
              {\label{fig:min-annulus:annulus:case2}
      \scalebox{0.77}{%
        \begin{pspicture}(0,-3.1)(6, 3.1)
          \psset{unit=1cm,linewidth=\annuluslinewidth{}}
          \pstGeonode[PosAngle=180, PointSymbol=*,linecolor=annulus-color,
            PointNameSep=.75em](3,0){O}
          \pstGeonode[PosAngle=90, PointNameSep=.7em,
            linecolor=circles-color](5.4,0){O_B}
          \pstGeonode[PosAngle={90,90}, PointNameSep={1.8em},
            PointName=O_{A_1}, linecolor=circles-color](4.2,0){O_A1}
          \pstGeonode[PosAngle=90, PointNameSep=1.6em,
            PointName=O_{A_2}, linecolor=circles-color](3, 1.1){O_A2}
          \pstGeonode[PointSymbol={none}, 
            PointName={none}](3.7,0){IN1}
          \pstGeonode[PointSymbol={none,none}, 
            PointName={none,none}](6,0){OUT}(3,0.7){IN}
          \pstGeonode[PosAngle=-90,PointNameSep=.75em,PointName=O'](3.375,0){O'}
          \pstCircleOA[linecolor=annulus-color]{O}{OUT}
          \pstCircleOA[linecolor=annulus-color]{O}{IN}
          \pstCircleOA[linecolor=circles-color]{O_A1}{IN1}
          \pstCircleOA[linecolor=circles-color]{O_A2}{IN}
          \pstCircleOA[linecolor=circles-color]{O_B}{OUT}
          \pstLineAB[linestyle=dashed]{O'}{O_A2}
          \pstLineAB[linestyle=dashed]{O'}{O}
          \pstLineAB[linestyle=dashed]{O}{O_A2}
    \end{pspicture}}}
  \end{tabular}
  \caption[Theorem~\ref{the:min-annulus:annulus}]{Illustrations for Theorem~\ref{the:min-annulus:annulus}}
  \label{fig:min-annulus:the}
\end{figure}

\begin{lemma} \label{lem:min-annulus:col}
  There is no minimum-width annulus \annulus{O} and two different
  disks $A$ and $B$ such that 
  $\aninner{O} = \{A\}$, $\anouter{O} = \{B\}$, and the centers of
  $A$, $B$ and \annulus{O} are collinear.
\end{lemma}

\begin{proof}
  Suppose to the contrary that such a minimum-width annulus \annulus{O}
  exists. Denote by $O_A$ and $O_B$ the respective centers of $A$ and $B$
  and by $R_A$ and $R_B$ their respective radii. $O_A$ is on the segment
  $OO_B$; see Figure~\ref{fig:min-annulus:col} for an illustration.
  Choose $O'$ to be a \anpoint{} of~$O$ in a direction
  perpendicular to $OO_B$. Then from triangle inequality $|O_AO_B| >
  |O'O_B| - |O'O_A|$ and 
  \begin{displaymath}
    \anwid{O} = (|OO_B| + R_B) - (|OO_A| - R_A) =
    |O_AO_B| + R_B + R_A > |O'O_B| - |O'O_A|  + R_B + R_A.
  \end{displaymath}
  Thus, \annulus{O'} has a smaller width which is a contradiction.
\end{proof}

\begin{lemma}\label{lem:min-annulus:three-points}
  If \annulus{O} is a minimum-width annulus containing \disks{} then
  either $|\aninner{O}| > 1$, or $|\anouter{O}| > 1$, or there is
  another annulus $\annulus{O'}$ of minimum-width containing \disks{}
  such that $|\aninner{O'}| >~1$ or $|\anouter{O'}| > 1$.
\end{lemma}

\begin{proof}
  Suppose to that $\aninner{O} = \{A\}$, $\anouter{O} = \{B\}$. 
  Let $R_A$ and $R_B$ denote the respective radii of $A$ and $B$, and
  let $O_A$ and $O_B$ denote their respective centers.

  In case that $A = B$ then we can move $O$ on the line that goes
  through $O$ and $O_A$ away from $O_A$. The new annulus, centered at
  $O'$, has the same width but the radii of its bounding circles are larger.
  We keep moving $O$ until one of the circles intersects another
  disk.

  If $A \neq B$ then from Lemma~\ref{lem:min-annulus:col}, $O_A$, $O_B
  \text{, and } O$ are not collinear (see
  Figure~\ref{fig:min-annulus:three-points} for an illustration).
  Choose $O'$ to be a \anpoint{} of~$O$ in the direction of
  $O_B$. Then by triangle inequality $|OO_A| < |OO'| + |O'O_A|$ and 
  {\setlength\arraycolsep{2pt}
  \begin{eqnarray*}
    \anwid{O} & = & |OO_B| + R_B - (|OO_A| - R_A) = |OO'| + |O'O_B| +
    R_B - |OO_A| + R_A \\
    & & {} > |O'O_B| - |O'O_A| + R_B + R_A. 
  \end{eqnarray*}}
  Thus the
  annulus $\annulus{O'}$ has a smaller width then \annulus{O}, which
  is a contradiction.
\end{proof}

\begin{theorem}
  \label{the:min-annulus:annulus}
  If there is a \Index{minimum-width annulus} containing \disks{}, then there
  is a minimum-width annulus \annulus{O} such that $|\aninner{O}| +
  |\anouter{O}| \geq 4$.
\end{theorem}

\begin{proof}
  Suppose to the contrary that there is a minimum-width annulus
  \annulus{O} containing the disks of \disks{}, but there is no
  minimum-width annulus whose total number of intersection points with
  the disks of \disks{} $\geq 4$.

  From Lemma~\ref{lem:min-annulus:three-points}, we may assume that
  either $|\aninner{O}| > 1$, or $|\anouter{O}| > 1$.

  \begin{enumerate}[leftmargin=*,labelsep=5pt,itemindent=30pt,labelindent=-50pt,label={\bf Case~\Roman*:},ref={\Roman*}]

  \item\label{the:min-annulus:annulus:caseI}
    Assume that $\aninner{O} = \{A_1, A_2\}$ and $\anouter{O} = \{B\}$
    Let $R_{A_1}$, $R_{A_2}$, and $R_B$ denote the respective radii of
    $A_1$, $A_2$, and $B$, and let $O_{A_1}$, $O_{A_2}$, and $O_B$
    denote their respective centers.

    If $B \in \aninner{O}$ then we can move $O$ along the bisector
    of $A_1$ and $A_2$ (which is one branch of a
    hyperbola), increasing the distance from
    $A_1$ and $A_2$.
    \anwid{O} (which equals to
    $2 \cdot R_B$) will stay the same. As the radius of the bounding
    circles grows one of the circles will eventually intersect another disk.

    If $B \not\in \aninner{O}$ then there are two cases:
    
    \begin{enumerate}[leftmargin=*,labelsep=5pt,itemindent=30pt,labelindent=-35pt,label={\bf Case~\alph*:},ref={\alph*}]
    \item\label{the:min-annulus:annulus:case1} 
      $O_{A_1}$ and $O_{A_2}$ are not on the segment $OO_B$ (see
      Figure~\ref{fig:min-annulus:annulus:case1} for an illustration).
      Choose $O'$ to be the \anpoint{} near $O$ in the direction of
      $O_B$.
      Then again from triangle inequality we get 
      $|OO'| > |OO_{A_i}| - |O'O_{A_i}|$ and
      \begin{align*}
        & \anwid{O} = |OO_B| + R_B - (|OO_{A_i}|  - R_{A_i}) = \\ & |OO'| +
        |O'O_B| - |OO_{A_i}| + R_B + R_{A_i} > 
        |O'O_B| - |O'O_{A_i}| + R_B + R_{A_i} = \anwid{O'},
      \end{align*}
      which is a contradiction.
      
    \item\label{the:min-annulus:annulus:case2} 
      Without loss of generality, assume that $O_{A_1}$ is on the segment
      $OO_B$ (see Figure~\ref{fig:min-annulus:annulus:case2} for an
      illustration). 
      For every point $P$ on the open segment $OO_{A_1}$ we get
      \begin{displaymath}
        |OP| + |PO_{A_2}| - R_{A_2} > |OO_{A_2}| - R_{A_2} =
        |OO_{A_1}| - R_{A_1} = |OP| + |O'O_{A_1}| - R_{A_1},
      \end{displaymath}
      and therefore
      \begin{displaymath}
        |PO_{A_2}| - R_{A_2} > |PO_{A_1}| - R_{A_1}.
      \end{displaymath}
      
      This means that an \anpoint{} $O'$ in the direction of~$O_{A_1}$
      from $O$ satisfies $\anouter{O'} = \anouter{O}$ and 
      $\aninner{O'} = \aninner{O} \setminus \{A_2\}$.
      The annulus \annulus{O'} has the same width as
      \annulus{O} and therefore is also a minimum-width annulus whose
      center is collinear with $O_{A_1}$ and $O_B$.
      This is a contradiction to Lemma~\ref{lem:min-annulus:col}.
    \end{enumerate}

  \item \label{the:min-annulus:annulus:caseII}
    $|\aninner{O}| = 1$ and $|\anouter{O}| = 2$.
    The proof of this case is similar to the first case and therefore omitted.
  \end{enumerate}
\end{proof}

\begin{figure}[h]
  \vspace{-5pt}
  \centering
    \subfigure[]{\label{fig:min-annulus:annulus-a}\includegraphics[width=115pt]{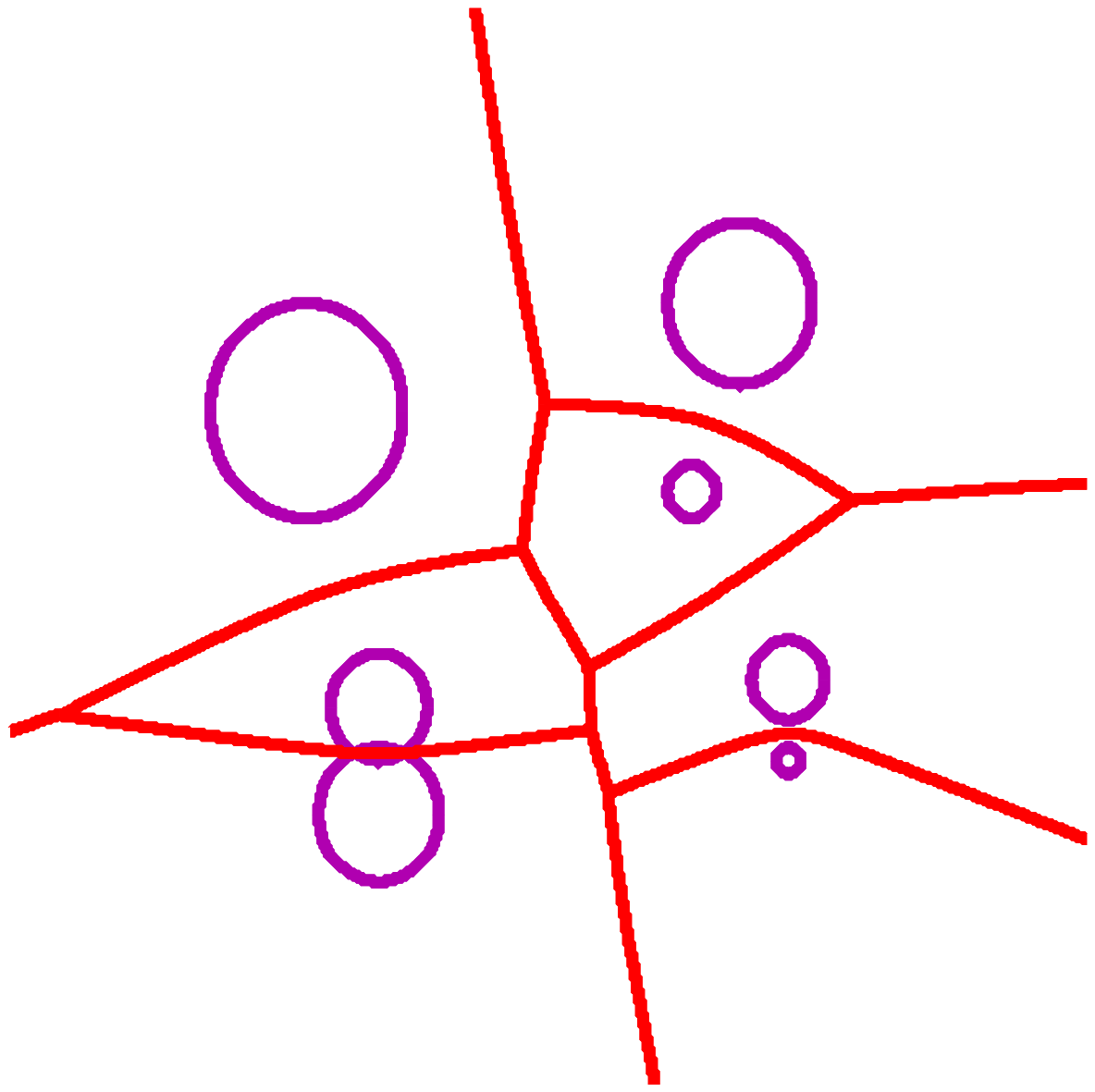}}
    \subfigure[]{\label{fig:min-annulus:annulus-b}\includegraphics[width=115pt]{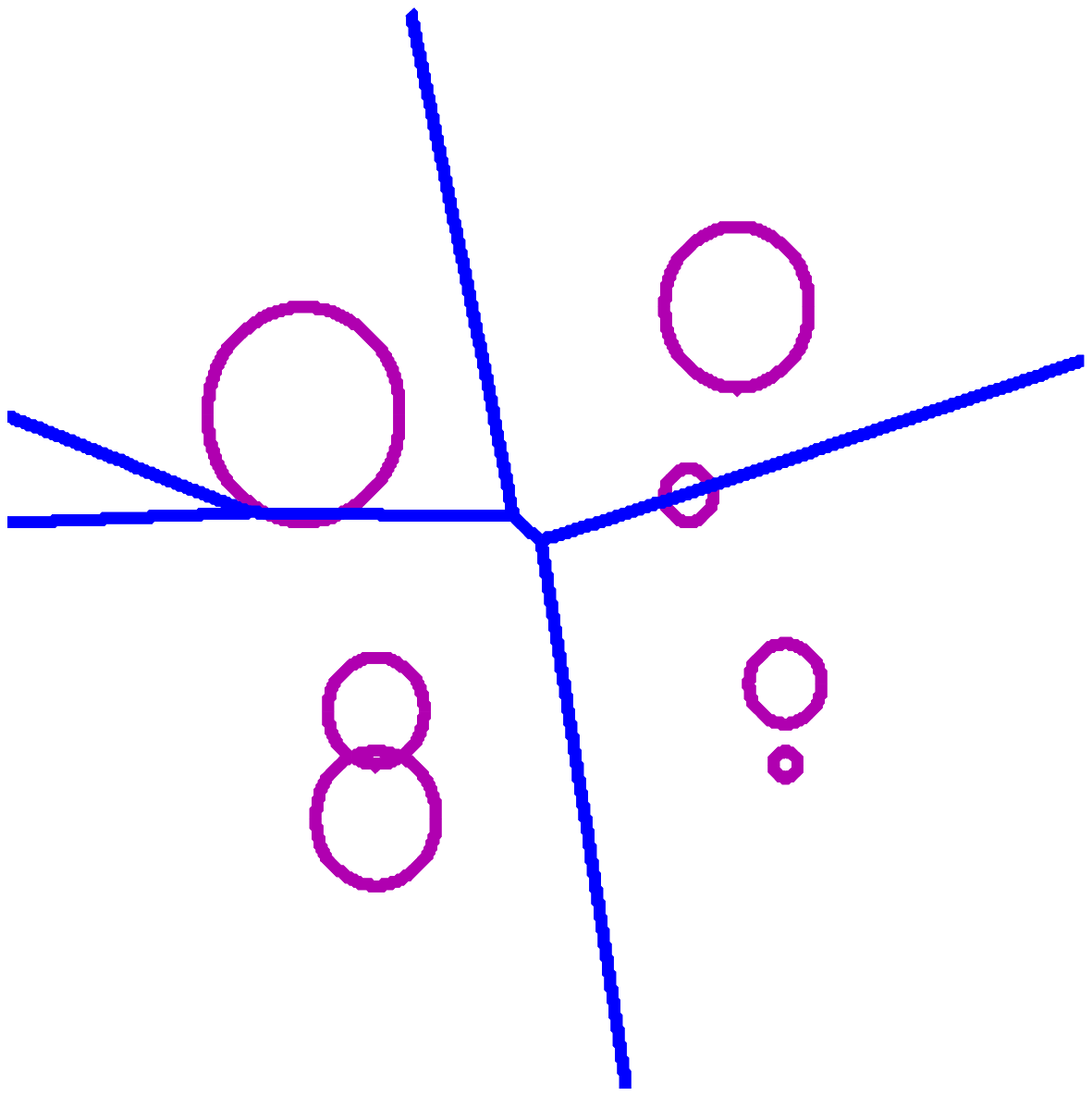}}
    \subfigure[]{\label{fig:min-annulus:annulus-c}\includegraphics[width=115pt,height=100pt]{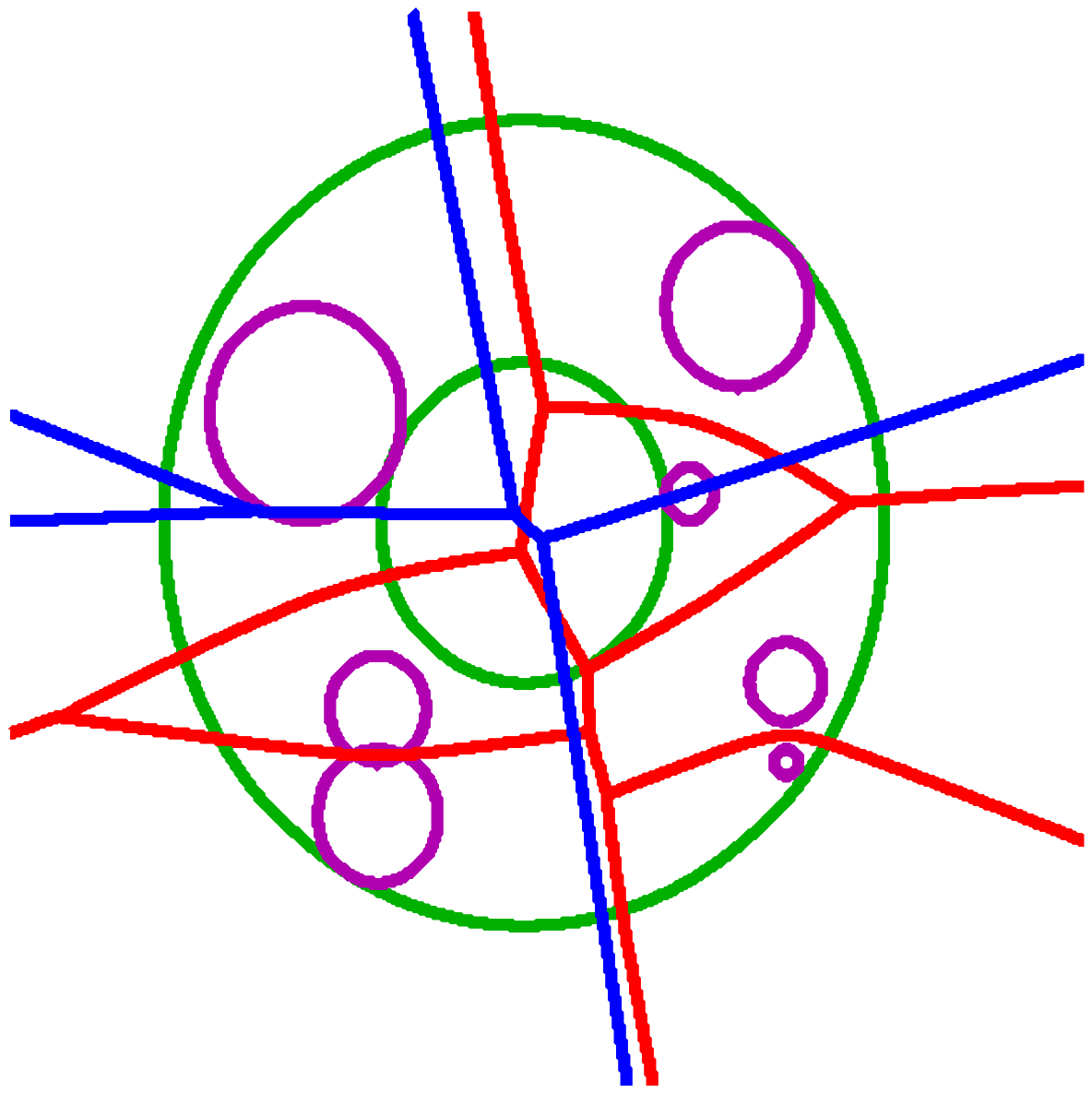}}
    \subfigure[]{\label{fig:min-annulus:annulus-d}\includegraphics[width=104pt,height=100pt]{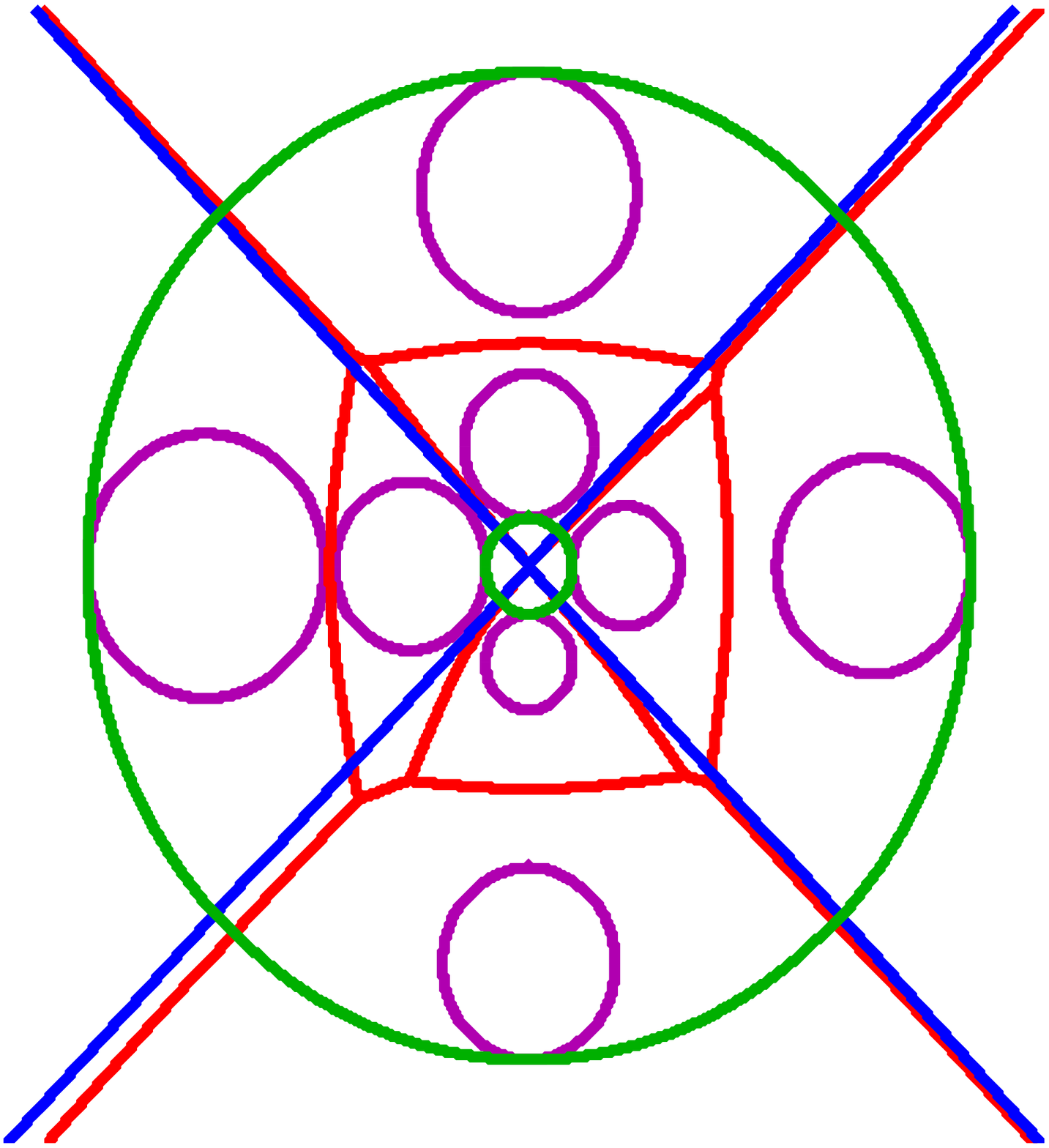}}

  \caption[Computing a minimum-width annulus of disks]{Computing a 
    minimum-width annulus of a set of disks.
    (a) The Apollonius diagram of the set of disks.
    (b) The \FPFS{} Voronoi diagram of the set of disks.
    (c) A minimum-width annulus of the set of disks. The center of
    the annulus is a vertex in
    the overlay of the Apollonius and the \FPFS{} Voronoi diagrams.
    (d) A highly degenerate scenario for computing a minimum-width
    annulus of a set of disks.}
  \label{fig:min-annulus:annulus}
\end{figure}

From Theorem~\ref{the:min-annulus:annulus} it is clear that if there
is a minimum-width annulus then there is a minimum-width annulus
centered at a point $O$ that falls into one of the following three
cases:
\begin{compactenum}
\item\label{itm:min-annulus:iiio}
  $|\aninner{O}| \geq 3$ and $|\anouter{O}| = 1$.
  This means that $O$ coincides with a vertex of the Apollonius
  diagram of the disks.
\item\label{itm:min-annulus:oooi}
  $|\aninner{O}| = 1$ and $|\anouter{O}| \geq 3$.
  This means that $O$ coincides with a vertex of the \FPFS{} Voronoi
  diagram of the disks.
\item\label{itm:min-annulus:iioo} 
  $|\aninner{O}| \geq 2$ and $|\anouter{O}| \geq 2$.
  In this case $O$ lies on a Voronoi edge of the Apollonius diagram of
  the disks and on a Voronoi edge of the \FPFS{} Voronoi diagram of
  the disks.
\end{compactenum}
\index{Apollonius diagram}\index{\FPFS{} Voronoi diagram}\index{overlay}
We therefore construct each of the diagrams (the Apollonius and the \FPFS{} 
Voronoi), and then overlay them. 
For each vertex $O$ of the overlay, we retrieve four relevant disks
(either three touching the inner circle and the one touching the
outer circle, in case~\ref{itm:min-annulus:iiio}, or three touching the
outer circle and the one touching the inner circle, in
case~\ref{itm:min-annulus:oooi}, or two pairs of disks touching
respectively the inner and outer circles), and compute the width of the
resulting annulus. We output the annulus of the smallest width. 
Figure~\ref{fig:min-annulus:annulus} illustrates the algorithm for computing a
minimum-width annulus of disks and a highly degenerate input that is
being handled properly by our implementation.
More implementation details can be found in Section~\ref{sec:min-annulus:impl}.

The Apollonius diagram (additively-weighted Voronoi diagram) is of
linear complexity and constitutes an
\emph{abstract Voronoi diagram}~\cite{k-cavd-89}.
The farthest-\neighbor Apollonius diagram is a
\emph{farthest abstract Voronoi diagram}~\cite{mmr-fsavd-01}, and is,
therefore, of linear complexity too.
Observation~\ref{obs:min-annulus} suggests that the \FPFS Voronoi
diagram is actually a linear-sized farthest Apollonius diagram. 

The construction of each of the above types of Voronoi diagrams
using the divide-and-conquer algorithm described in
Chapter~\ref{chap:env-to-vd} yields a worst-case time complexity of
$O(n^{2 + \varepsilon})$.
Overlaying the two diagrams with a sweep based algorithm has 
$O((n + k) \log n)$ worst-case time complexity where $k$ is the number of
intersections between the diagrams. 
The total worst-case time complexity of the algorithm is therefore
$O(n^{2 + \varepsilon})$.

Though the worst-case time complexity of the algorithm is larger than
quadratic, it is reasonable to assume that the expected time
complexity is, in many cases, smaller.
Indeed, applying Corollary~\ref{cor:env-to-vd:exp-comp} to our case
results in expected construction time of $O(n\log^2n)$ for both
Apollonius and \FPFS{} Voronoi  diagrams, and in total, expected
running time of  $O(n\log^2 n + k \log n)$
where $k$ is the number of intersections between the diagrams.
It is true that we may still have $\Theta(n^2)$ intersections between the 
two diagrams, but, though not proved for disks, for random point
sites\footnote{Drawn independently from a density on a compact convex
  set of the plane.}
it is known that the expected number of intersections between the 
farthest and the nearest Voronoi diagrams is linear~\cite{bd-irgo-98}.

\section{Implementation Details and Experimental Results}
\label{sec:min-annulus:impl}

This section gives further, and more technical, information about the 
implementation details of the algorithm described in the previous
section.
The implementation is exact and robust, utilizing software components
described in Chapter~\ref{chap:impl} and other tools from
\cgal{}\citelinks{cgal-link}.

The first step of the algorithm is the construction of the Apollonius
diagram of a set of disks.
The construction is enabled through the implemented \apolloniustraits
traits class for constructing Apollonius diagrams of disks in the
plane; see Section~\ref{ssec:impl:plane-vd:apo}.

The second step of the algorithm is the construction of the \FPFS{}
Voronoi diagram of a set of disks.
Again, Observation~\ref{obs:min-annulus} comes in handy.
As the \FPFS Voronoi diagram is a farthest-\neighbor Apollonius
diagram, we easily produce a traits class for constructing \FPFS
Voronoi diagrams.
The new \fpfstraits traits class provides the user with a better
interface for constructing the \FPFS Voronoi diagram as
it does not require to negate the radii of the
disks.

We use the \overlay function from the \aos package to overlay the
Apollonius diagram and the \FPFS{} Voronoi diagram.
Given two arrangements, the overlay operation sweeps over them and
constructs the resulting arrangement, while updating its features
based on data associated with the input arrangements'
features~\cite{wfzh-aptaca-07}.
For example, a new face $f$ is created by the overlap of two faces
$f_1$ and $f_2$ of the two input arrangements, respectively, and its
data is updated with the data from $f_1$ and $f_2$.

The generic \overlay function is parameterized with an overlay-traits
class, which handles the merge operations of data associated with
any two features of the respective two diagrams
(there are 10~different cases to handle).
In our case, the \ccode{MWA\_overlay\_traits\_2} class is the model
of \cgal's \overlaytraits{} concept and updates the features of the
resulting arrangement with dominating sites from features of both
diagrams.
The sites of the Apollonius diagram are kept separate 
from the sites of the \FPFS{} Voronoi diagram in two sets associated
with a feature of the resulting arrangement.

\begin{table*}[t]
  \setlength{\tabcolsep}{3.77pt}
  
  \begin{center}
    \caption[Time consumption of the minimum-width annulus
      computation.]{\capStyle{
        Time consumption (in seconds) of minimum-width annuli 
        computation and sizes of the corresponding constructed
        diagrams.
        \FPFSabrv{} --- \FPFS,
        V --- the number of vertices of the diagram,
        E --- the number of edges of the diagram,
        F --- the number of faces of the diagram,
        T --- the time in seconds consumed by the
        respective phase,
        C --- the time in seconds consumed during
        the comparison of the candidates for the annulus center.
        ``dgn\_*'' are input files in a degenerate setting 
        (see Figure~\ref{fig:min-annulus:annulus-d} for an illustration), and
        ``rnd\_*'' are input files in a random setting.
    }}
    \label{sec:min-annulus:table}
    
    \vspace{5pt}
    \small
    \begin{tabular}{|l||r|r|r|r||r|r|r|r||r|r|r|r||c||c|}

      \hline
      \multirow{2}*{\textbf{Input}} &
      \multicolumn{4}{c||}{\textbf{Apollonius}} &
      \multicolumn{4}{c||}{\textbf{\FPFSabrv VD}} &
      \multicolumn{4}{c||}{\textbf{Overlay}} &
      \multirow{2}*{\textbf{C}} &
      \multirow{2}*{\parbox{30pt}{\textbf{Total Time}}} \\ \cline{2-13}
      & 
      \multicolumn{1}{c|}{\textbf{V}} &
      \multicolumn{1}{c|}{\textbf{E}} &
      \multicolumn{1}{c|}{\textbf{F}} &
      \multicolumn{1}{c||}{\textbf{T}}
      & 
      \multicolumn{1}{c|}{\textbf{V}} &
      \multicolumn{1}{c|}{\textbf{E}} &
      \multicolumn{1}{c|}{\textbf{F}} &
      \multicolumn{1}{c||}{\textbf{T}}
      & 
      \multicolumn{1}{c|}{\textbf{V}} &
      \multicolumn{1}{c|}{\textbf{E}} &
      \multicolumn{1}{c|}{\textbf{F}} &
      \multicolumn{1}{c||}{\textbf{T}}
      & & \\
      \hline
      \hline
      \textit{rnd\_50} & 84  & 128 & 45 & 6.74 & 
      10  & 21  & 12 & 2.40 & 
      126 & 213 & 88 & 0.44 &
      0.57 & \multicolumn{1}{r|}{10.16} \\
      \hline
      \textit{rnd\_100} & 162 & 242 & 81 & 17.47 &
      17 & 35 & 19 & 5.36 &
      238 & 395 & 158 & 0.81 &
      1.46 & \multicolumn{1}{r|}{25.12} \\
      \hline
      \textit{rnd\_200} & 317 & 460 & 144 & 44.07 &
      15 & 31 & 17 & 11.16 &
      416 & 659 & 244 & 1.28 &
      1.33 & \multicolumn{1}{r|}{57.86} \\
      \hline
      \textit{rnd\_500} & 672 & 967 & 296 & 136.46 &
      16 & 33 & 18 & 29.30 &
      775 & 1174 & 400 & 1.85 &
      1.97 & \multicolumn{1}{r|}{169.59} \\
      \hline \hline
      
      \textit{dgn\_50} & 59 & 106 & 48 & 5.74 &
      1 & 25 & 25 & 1.94 &
      100 & 213 & 114 & 0.84 &
      0.30 & \multicolumn{1}{r|}{8.83} \\
      \hline

      \textit{dgn\_100} & 
      134 & 232 & 99 & 16.57 &
      1 & 50 & 50 & 5.83 &
      239 & 492 & 254 & 2.93 &
      1.11 & \multicolumn{1}{r|}{26.46} \\
      \hline

      \textit{dgn\_200} & 
      304 & 502 & 199 & 42.60 &
      1 & 100 & 100 & 15.77 &
      581 & 1156 & 576 & 7.60 &
      2.78 & \multicolumn{1}{r|}{68.76} \\
      \hline

      \textit{dgn\_500} & 
      785 & 1262 & 478 & 154.37 &
      1 & 239 & 239 & 56.21 &
      1645 & 3221 & 1577 & 37.93 &
      7.94 & \multicolumn{1}{r|}{256.47} \\
      \hline

    \end{tabular}
    \vspace{-10pt}
  \end{center}
\end{table*}

The two geometry-traits classes (\apolloniustraits{} and
\fpfstraits{}) use the same (\cpp) types of geometric primitives and
operations, even though they themselves consist different \cpp types.
The overlay function previously supported only arrangements that were
instantiated with the same geometric-traits type --- even if the
underlying curves were of the same type.
We changed the interface of the \overlay function and parts of its
implementation to also support arrangements with different
geometry-traits types but the same types of geometric primitives
(points, curves, $x$-monotone curves, etc.).

We compare the widths of all annuli created at vertices of the overlay
to find the center of our minimum-width annulus (as described in 
Section~\ref{sec:min-annulus:proofs}).
We use rational interval-arithmetic optimization similar to the one
applied in the implementation of the proximity predicates of the
Apollonius traits class (Section~\ref{ssec:impl:plane-vd:apo}).

Table~\ref{sec:min-annulus:table} shows the sizes of the constructed
diagrams in each phase of the algorithm and the time consumption
(in seconds) of the execution of each phase.
The vertices of the overlay are the candidates for the center of the
annulus.
The experiments were carried out on an Intel\textregistered{}
Core\texttrademark{}2 Duo 2GHz processor with 1GB memory running
Linux operating system.

\chapter{Conclusions and Future Work}
\label{chap:conclusion}

Our framework together with the existing traits classes of the
arrangement package of \cgal provide the means to produce various
Voronoi diagrams embedded on two-dimensional surfaces in an exact and
robust way.
Moreover, the theoretical bound on the expected running time of the
algorithm is nearly optimal.

Future work is to enrich the variety of Voronoi diagrams computed with 
our software to include Voronoi diagrams of circular
arcs~\cite{y-avdssc-87}, Bregman Voronoi diagrams~\cite{nbn-bvd-07},
Voronoi diagrams in the hyperbolic Poincar\'e
half-plane~\cite{ot-cvduhp-96} or in the Poincar\'e hyperbolic
disk~\cite{nm-hvd-06,nn-hvdme-09}, and Hausdorff Voronoi
diagrams~\cite{obsc-stcavd-00}.
Other Voronoi diagrams embedded on different surfaces, for example,
cylinders or tori, can be developed.

Some of the above diagrams have bisectors that can be handled with
existing traits classes available in the arrangement package.
All two-sites Voronoi diagrams mentioned in~\cite{bdd-psvd-02} can be
developed by traits class from the arrangement package; the most
complicated diagrams there can use the algebraic traits class for the
arrangement package.

As the arrangement package evolves, more Voronoi diagrams will become
achievable.
For example, one of the future traits classes for the arrangement
package will probably be a traits class for the construction of
arrangements induced by general circular arcs, embedded on the sphere;
such an implementation can be based on the work by Cazals and
Loriot~\cite{cl-cacsa-09}.
This traits class will enable the construction of \mobius diagrams
on the sphere.

The class of all diagrams on the sphere with circle bisectors that
constitute a Voronoi diagram is identical to the class of \mobius
diagrams on the sphere~\cite[\S2.4.1]{bwy-cvd-06}.
A similar conjecture appears in
Section~\ref{sec:impl:sphere-vd:voronoi}, namely, the class of all
Voronoi diagrams with great circles as bisectors is identical
to the class of power diagrams on the sphere.
The proof for the \mobius case relies on a similar theorem for planar
power diagrams and affine bisectors, which is proved using advanced
tools from linear algebra.
Proving this conjecture remains an open problem.

Another major goal is to improve the performance of the code in
practice.
A possible direction to consider is to avoid overlaying the entire
arrangements in the merge step.
Some diagrams are known to merge in linear time by a deterministic
partitioning of the set of sites~\cite{k-cavd-89}, which could
alleviate the need to overlay large portions of the Voronoi diagrams
where portions from one diagram are known to always dominate portions
from the other.

The time consumption of our algorithm directly depends on the
implementation of the traits classes, as they supply the algorithm
with all predicates and constructors.
The performance of some of the implemented traits classes in this
thesis may also be improved.
For example, one can try and implement a traits class for the \mobius
diagram based on the \circker provided by \cgal~\cite{ptt-gfpca-06},
and apply other filtering techniques.
Such an implementation should be compared against our implementation.

As described in Chapter~\ref{chap:min-annulus} the problem of finding
a minimum-width annulus was addressed almost solely for the case of
point sets.
The case of a simple linear polygon can also be solved using an
overlay of its medial axis and the farthest Voronoi diagram of its
vertices.
The medial axis of the polygon corresponds to the nearest Voronoi
diagram of the polygon's edges restricted to the inside of the
polygon.
There is a strong reason to believe that similar techniques to the
ones applied in this thesis, to compute the minimum-width annulus of a
set of disks, can be applied to other types of objects, utilizing our
framework.

\bibliography{abrev,how_to_cite_cgal,thesis}
\bibliographystyle{alpha}

\bibliographystylelinks{plain}
\bibliographylinks{links}
\label{sec:linkbib}
\end{document}